\documentclass[12pt]{article}
\usepackage{amssymb}
\usepackage{graphicx}
\begin{document}
\newcommand{\beq}{\begin{equation}}
\newcommand{\eeq}{\end{equation}}
\newcommand{\beqa}{\begin{eqnarray}}
\newcommand{\eeqa}{\end{eqnarray}}
\newcommand{\beqar}{\begin{eqnarray*}}
\newcommand{\eeqar}{\end{eqnarray*}}
\newcommand{\al}{\alpha}
\newcommand{\be}{\beta}
\newcommand{\del}{\delta}
\newcommand{\D}{\Delta}
\newcommand{\eps}{\epsilon}
\newcommand{\ga}{\gamma}
\newcommand{\Ga}{\Gamma}
\newcommand{\ka}{\kappa}
\newcommand{\nn}{\nonumber}
\newcommand{\inn}{\!\cdot\!}
\newcommand{\h}{\eta}
\newcommand{\ii}{\iota}
\newcommand{\kk}{\varphi}
\newcommand\F{{}_3F_2}
\newcommand{\la}{\lambda}
\newcommand{\La}{\Lambda}
\newcommand{\na}{\prt}
\newcommand{\Om}{\Omega}
\newcommand{\om}{\omega}
\newcommand{\p}{\Phi}
\newcommand{\sig}{\sigma}
\renewcommand{\t}{\theta}
\newcommand{\z}{\zeta}
\newcommand{\ssc}{\scriptscriptstyle}
\newcommand{\eg}{{\it e.g.,}\ }
\newcommand{\ie}{{\it i.e.,}\ }
\newcommand{\labell}[1]{\label{#1}} 
\newcommand{\reef}[1]{(\ref{#1})}
\newcommand\prt{\partial}
\newcommand\veps{\varepsilon}
\newcommand{\pol}{\varepsilon}
\newcommand\vp{\varphi}
\newcommand\ls{\ell_s}
\newcommand\cF{{\cal F}}
\newcommand\cA{{\cal A}}
\newcommand\cS{{\cal S}}
\newcommand\cT{{\cal T}}
\newcommand\cV{{\cal V}}
\newcommand\cL{{\cal L}}
\newcommand\cM{{\cal M}}
\newcommand\cN{{\cal N}}
\newcommand\cG{{\cal G}}
\newcommand\cK{{\cal K}}
\newcommand\cH{{\cal H}}
\newcommand\cI{{\cal I}}
\newcommand\cJ{{\cal J}}
\newcommand\cl{{\iota}}
\newcommand\cP{{\cal P}}
\newcommand\cQ{{\cal Q}}
\newcommand\cg{{\it g}}
\newcommand\cR{{\cal R}}
\newcommand\cB{{\cal B}}
\newcommand\cO{{\cal O}}
\newcommand\tcO{{\tilde {{\cal O}}}}
\newcommand\bz{\bar{z}}
\newcommand\bb{\bar{b}}
\newcommand\ba{\bar{a}}
\newcommand\bg{\bar{g}}
\newcommand\bc{\bar{c}}
\newcommand\bomega{\bar{\omega}}
\newcommand\bH{\bar{H}}
\newcommand\bw{\bar{w}}
\newcommand\bX{\bar{X}}
\newcommand\bK{\bar{K}}
\newcommand\bA{\bar{A}}
\newcommand\bR{\bar{R}}
\newcommand\bZ{\bar{Z}}
\newcommand\bxi{\bar{\xi}}
\newcommand\bphi{\bar{\phi}}
\newcommand\bpsi{\bar{\psi}}
\newcommand\bprt{\bar{\prt}}
\newcommand\bet{\bar{\eta}}
\newcommand\btau{\bar{\tau}}
\newcommand\hF{\hat{F}}
\newcommand\hA{\hat{A}}
\newcommand\hT{\hat{T}}
\newcommand\htau{\hat{\tau}}
\newcommand\hD{\hat{D}}
\newcommand\hf{\hat{f}}
\newcommand\hK{\hat{K}}
\newcommand\hg{\hat{g}}
\newcommand\hp{\hat{\Phi}}
\newcommand\hi{\hat{i}}
\newcommand\ha{\hat{a}}
\newcommand\hb{\hat{b}}
\newcommand\hQ{\hat{Q}}
\newcommand\hP{\hat{\Phi}}
\newcommand\hS{\hat{S}}
\newcommand\hX{\hat{X}}
\newcommand\tL{\tilde{\cal L}}
\newcommand\hL{\hat{\cal L}}
\newcommand\MZ{\mathbb{Z}}
\newcommand\MR{\mathbb{R}}
\newcommand\tG{{\tilde G}}
\newcommand\tg{{\tilde g}}
\newcommand\tphi{{\widetilde \Phi}}
\newcommand\tPhi{{\widetilde \Phi}}
\newcommand\ti{{\tilde i}}
\newcommand\tj{{\tilde j}}
\newcommand\tk{{\tilde k}}
\newcommand\tl{{\tilde l}}
\newcommand\ttm{{\tilde m}}
\newcommand\tn{{\tilde n}}
\newcommand\ta{{\tilde a}}
\newcommand\tb{{\tilde b}}
\newcommand\tc{{\tilde c}}
\newcommand\td{{\tilde d}}
\newcommand\tm{{\tilde m}}
\newcommand\tmu{{\tilde \mu}}
\newcommand\tnu{{\tilde \nu}}
\newcommand\talpha{{\tilde \alpha}}
\newcommand\tbeta{{\tilde \beta}}
\newcommand\trho{{\tilde \rho}}
 \newcommand\tR{{\tilde R}}
\newcommand\teta{{\tilde \eta}}
\newcommand\tF{{\widetilde F}}
\newcommand\tK{{\widetilde K}}
\newcommand\tE{{\tilde E}}
\newcommand\tpsi{{\tilde \psi}}
\newcommand\tX{{\widetilde X}}
\newcommand\tD{{\widetilde D}}
\newcommand\tO{{\widetilde O}}
\newcommand\tS{{\tilde S}}
\newcommand\tB{{\tilde B}}
\newcommand\tA{{\widetilde A}}
\newcommand\tT{{\widetilde T}}
\newcommand\tC{{\widetilde C}}
\newcommand\tV{{\widetilde V}}
\newcommand\thF{{\widetilde {\hat {F}}}}
\newcommand\Tr{{\rm Tr}}
\newcommand\tr{{\rm tr}}
\newcommand\STr{{\rm STr}}
\newcommand\hR{\hat{R}}
\newcommand\M[2]{M^{#1}{}_{#2}}

\newcommand\bS{\textbf{ S}}
\newcommand\bI{\textbf{ I}}
\newcommand\bJ{\textbf{ J}}

\begin{titlepage}
\begin{center}

\vskip 2 cm
{\LARGE \bf  Effective action of heterotic string theory  \\  \vskip 0.25 cm at order $\alpha'^2$
 }\\
\vskip 1.25 cm
   Mohammad R. Garousi\footnote{garousi@um.ac.ir}

\vskip 1 cm
{{\it Department of Physics, Faculty of Science, Ferdowsi University of Mashhad\\}{\it P.O. Box 1436, Mashhad, Iran}\\}
\vskip .1 cm
 \end{center}

\begin{abstract}

Upon examining the effective action of the heterotic string theory at order $\alpha'^2$, an inconsistency between the Chern-Simons coupling $\Omega^2$ and T-duality has been discovered. To address this issue, we introduce 60 parity-even independent geometrical couplings involving the $B$-field, metric, and dilation at the same order, each with arbitrary coefficients.
To ensure the invariance of these couplings under T-duality, we consider the Miessner action for the couplings at order $\alpha'$ and rigorously determine the coefficients of the 60 couplings in terms of the Chern-Simons coupling. Notably, it is found that the coefficients of the Riemann cubed terms must be zero, which aligns with the results of S-matrix calculations. Additionally, the parity odd couplings at order $\alpha'^2$ are obtained through T-duality.
Overall, our analysis successfully resolves the inconsistency between the Chern-Simons coupling and T-duality, providing a more comprehensive understanding of the behavior of the heterotic string theory at order $\alpha'^2$.

\end{abstract}
\end{titlepage}

\section{Introduction}

String theory is a promising candidate for a consistent theory of quantum gravity. It postulates a finite number of massless fields and an infinite tower of massive fields, reflecting the underlying stringy nature of gravity.
To study physics within this framework, it is convenient to use an effective action that includes only the massless fields. The effects of the massive fields manifest as higher derivatives of the massless fields, commonly referred to as $\alpha'$-corrections. Both the bosonic string theory and the five superstring theories share the same massless NS-NS fields, which include the metric, Kalb-Ramond field, and dilaton. The superstring theories also feature additional massless fields  (see, for example, \cite{Becker:2007zj}), but these are not the focus of this paper.
The effective actions of string theory can be derived by imposing various symmetries and dualities within the theory. These symmetries and dualities dictate the form of the effective action and provide insight into the fundamental nature of the theory.

The effective actions of string theory exhibit gauge symmetries corresponding to their various massless fields. For instance, the metric is associated with diffeomorphism symmetry, while the Kalb-Ramond field $B$ is associated with gauge symmetry. The heterotic string theory, which is the focus of this paper, has an anomaly that can only be cancelled by assuming the gauge group to be $SO(32)$ or $E_8\times E_8$ and introducing nonstandard gauge transformations and nonstandard local Lorentz-transformations for the $B$-field \cite{Green:1984sg}.
By utilizing the diffeomorphism and $B$-field gauge symmetries, one can identify the independent gauge-invariant couplings, or geometrical couplings, at each order of $\alpha'$ up to field redefinitions. The number of independent couplings depends on the background topology, with closed spacetime manifolds having fewer independent couplings than open spacetime manifolds with boundaries. However, the coefficients of these independent couplings cannot be determined by the geometrical gauge symmetries alone.
Fortunately, using string field theory techniques, it has been shown in \cite{Sen:1991zi,Hohm:2014sxa} that the classical effective actions of string theories at all orders of $\alpha'$ possess $O(d,d,\MR)$ symmetry after reducing the theories on the torus $T^{(d)}$ and ignoring the massive Kaluza-Klein (KK) modes. The effective actions are invariant under the non-geometrical subgroups of the T-duality group $O(d,d,\MR)$, such as the Buscher rules \cite{Buscher:1987sk,Buscher:1987qj} and their $\alpha'$ corrections, which can be used to establish relations between the geometrical couplings. This method has been successfully employed to obtain the effective action of the bosonic string theory at order $\alpha'^2$ \cite{Garousi:2019mca}, as well as the NS-NS couplings in the effective action of the superstring theory at order $\alpha'^3$ \cite{Garousi:2020gio}. In this paper, we will apply this method to determine the NS-NS couplings in the effective action of the heterotic string theory at order $\alpha'^2$.

The heterotic string theory features 496 massless vector fields in the adjoint representation of the $SO(32)$ or $E_8\times E_8$ gauge group, as well as NS-NS fields that are scalar in these   groups. For the purposes of this paper, we consider zero vector gauge fields. In this case, the nonstandard local Lorentz-transformation for the $B$-field requires a specific field strength in the effective action, as described in \cite{Green:1984sg}:
\beqa
\hat{H}_{\mu\nu\alpha}&=& H_{\mu\nu\alpha}+\frac{3}{2}\alpha' \Omega_{\mu\nu\alpha}\,.
\labell{replace}
\eeqa
Here, $H_{\mu\nu\alpha}=3\partial_{[\mu }B_{\nu\alpha]}$, and the Chern-Simons three-form $\Omega$ is defined as
\beqa
\Omega_{\mu\nu\alpha}&=&\omega_{[\mu i}{}^j\partial_\nu\omega_{\alpha] j}{}^i+\frac{2}{3}\omega_{[\mu i}{}^j\omega_{\nu j}{}^k\omega_{\alpha]k}{}^i\,,
\eeqa
where $\omega_{\mu i}{}^j$ is the spin connection, defined in terms of the Christoffel connection $\Gamma_{\mu\nu}{}^\rho$ and the vielbein $e_\mu{}^i$ by $\omega_{\mu i}{}^j=\partial_\mu e_\nu{}^j e^\nu{}_i-\Gamma_{\mu\nu}{}^\rho e_\rho{}^j e^\nu{}_i$. The  vielbein is related to the metric $G_{\mu\nu}$ via $e_\mu{}^i e_\nu{}^j\eta_{ij}=G_{\mu\nu}$.
The above nonlinear field strength organizes the couplings at different orders of $\alpha'$ into a single action. For instance, the universal leading-order effective action includes couplings at orders $\alpha'^0, \alpha'$  and $\alpha'^2$, \ie 
\beqa
\bS^{(0,1,2)}\!\!+\partial\!\!\bS^{(0)}=-\frac{2}{\kappa^2}\left[\int d^{10}x \sqrt{-G} e^{-2\Phi}\! \left( R + 4\nabla_{\mu}\Phi \nabla^{\mu}\Phi-\frac{1}{12}\hat{H}^2\right) \!+ 2\int d^{9}\sigma\sqrt{|g|} e^{-2\Phi}K\right].\labell{baction}
\eeqa
Here, $\kappa$ is related to the 10-dimensional Newton's constant. The couplings at orders $\alpha'^0$ and $\alpha'^2$ are even parity, while the coupling at order $\alpha'$ is odd parity\footnote{It is important to note that the world-sheet action of a string in the presence of a background $B$-field incorporates the 2-dimensional antisymmetric Levi-Civita tensor. As a result, the $B$-field exhibits an odd behavior under world-sheet parity.}. The action also includes a Gibbons-Hawking boundary term \cite{Gibbons:1976ue} that depends on the extrinsic curvature $K$ and the induced metric on the boundary.

In addition to the odd parity couplings at order $\alpha'$, there is also a set of even parity couplings at the same order. The specific couplings depend on the chosen scheme for the geometrical couplings. In this study, we adopt the Meissner scheme \cite{Meissner:1996sa} for the even parity couplings at order $\alpha'$. Upon replacing $H$ in this action with $\hat{H}$, it yields the following couplings at order $\alpha'^n$, where $n = 1, \ldots, 5$:
\beqa
\bS^{(1,2,3,4,5)}&\!\!\!\!\!=\!\!\!\!\!&-\frac{2\alpha' c_1}{\kappa^2}\int d^{10} x\sqrt{-G} e^{-2\Phi}\Big[R_{GB}^2+\frac{1}{24} \hat{H}_{\alpha }{}^{\delta \epsilon } \hat{H}^{\alpha \beta
\gamma } \hat{H}_{\beta \delta }{}^{\varepsilon } \hat{H}_{\gamma \epsilon
\varepsilon } -  \frac{1}{8} \hat{H}_{\alpha \beta }{}^{\delta }
\hat{H}^{\alpha \beta \gamma } \hat{H}_{\gamma }{}^{\epsilon \varepsilon }
\hat{H}_{\delta \epsilon \varepsilon }\nn\\&&  + \frac{1}{144} \hat{H}_{\alpha
\beta \gamma } \hat{H}^{\alpha \beta \gamma } \hat{H}_{\delta \epsilon
\varepsilon } \hat{H}^{\delta \epsilon \varepsilon }+ \hat{H}_{\alpha }{}^{
\gamma \delta } \hat{H}_{\beta \gamma \delta } R^{\alpha
\beta } -  \frac{1}{6} \hat{H}_{\alpha \beta \gamma } \hat{H}^{\alpha \beta
\gamma } R  -
\frac{1}{2} \hat{H}_{\alpha }{}^{\delta \epsilon } \hat{H}^{\alpha \beta
\gamma } R_{\beta \gamma \delta \epsilon }\nn\\&& -
\frac{2}{3} \hat{H}_{\beta \gamma \delta } \hat{H}^{\beta \gamma \delta }
\nabla_{\alpha }\nabla^{\alpha }\Phi + \frac{2}{3} \hat{H}_{\beta
\gamma \delta } \hat{H}^{\beta \gamma \delta } \nabla_{\alpha }\Phi
\nabla^{\alpha }\Phi + 8 R \nabla_{\alpha }\Phi
\nabla^{\alpha }\Phi - 16
R_{\alpha \beta } \nabla^{\alpha }\Phi \nabla^{\beta
}\Phi \nn\\&&+ 16 \nabla_{\alpha }\Phi \nabla^{\alpha
}\Phi \nabla_{\beta }\nabla^{\beta }\Phi - 16 \nabla_{\alpha }\Phi \nabla^{\alpha }\Phi \nabla_{
\beta }\Phi \nabla^{\beta }\Phi + 2 \hat{H}_{\alpha }{}^{\gamma
\delta } \hat{H}_{\beta \gamma \delta } \nabla^{\beta
}\nabla^{\alpha }\Phi \Big]\,,\labell{Mis}
\eeqa
 where $c_1=1/8$ and $R_{GB}^2$ represents the Gauss-Bonnet couplings. Similar couplings exist for the bosonic string theory, in which the Chern-Simons 3-form $\Omega$ is zero and $c_1=1/4$.
For spacetime manifolds with boundaries, the corresponding boundary terms have been found in \cite{Garousi:2021cfc} using T-duality, and they are given by:
\beqa
\prt\!\!\bS^{(1,2,3)}&\!\!\!\!\!=\!\!\!\!\!&-\frac{2 \alpha' c_1}{\kappa^2}\int  d^{9} \sigma\sqrt{|g|} e^{-2\Phi}\Bigg[ Q_2+\frac{4}{3}n^2n^{\alpha } n^{\beta }
\nabla_{\gamma }\nabla^{\gamma }K_{\alpha \beta } - 2n^2 \hat{H}_{\beta }{}^{\delta
\epsilon } \hat{H}_{\gamma \delta \epsilon } n^{\alpha } n^{\beta }
n^{\gamma } \nabla_{\alpha }\Phi\nn\\&& +
\frac{2}{3} \hat{H}_{\beta \gamma \delta } \hat{H}^{\beta \gamma \delta }
n^{\alpha } \nabla_{\alpha }\Phi-\frac{1}{3} \hat{H}_{\beta \gamma \delta } \hat{H}^{\beta
\gamma \delta } K^{\alpha }{}_{\alpha } +  \hat{H}_{\alpha
}{}^{\gamma \delta } \hat{H}_{\beta \gamma \delta } K^{\alpha \beta
} + n^2 \hat{H}_{\alpha
}{}^{\delta \epsilon } \hat{H}_{\beta \delta \epsilon } K^{\gamma
}{}_{\gamma } n^{\alpha } n^{\beta }\nn\\&&- 16 n^2K^{\gamma
}{}_{\gamma } n^{\alpha } n^{\beta } \nabla_{\alpha }\Phi
\nabla_{\beta }\Phi  + 16 K^{\beta }{}_{\beta }
\nabla_{\alpha }\Phi \nabla^{\alpha }\Phi- 16 K_{\alpha \beta } \nabla^{\alpha
}\Phi \nabla^{\beta }\Phi \nn\\&&- 16 n^{\alpha } \nabla_{\alpha
}\Phi \nabla_{\beta }\Phi \nabla^{\beta }\Phi +
\frac{32}{3} n^2 n^{\alpha } n^{\beta } n^{\gamma }
\nabla_{\alpha }\Phi \nabla_{\beta }\Phi \nabla_{\gamma
}\Phi\Bigg]\,.\labell{finalb}
\eeqa
Here, $n^2=n_\mu n^{\mu}$ and $Q_2$ represents the Chern-Simons boundary couplings. The couplings at orders $\alpha'$, $\alpha'^3$, and $\alpha'^5$ are even parity, while the couplings at orders $\alpha'^2$ and $\alpha'^4$ are odd parity.

The effective action in \reef{baction} does not contain any additional odd parity couplings apart from the one mentioned in this action, which involves $\Omega$. The Bianchi identities imply that there are no geometrical odd parity couplings at order $\alpha'$ with a vanishing Chern-Simons form $\Omega$. Therefore, the odd parity coupling in the aforementioned action must be T-duality invariant. A recent study \cite{Garousi:2023pah} explicitly demonstrated the T-duality invariance of the odd parity coupling in \reef{baction} at order $\alpha'$. However, there may exist other even or odd parity couplings at order $\alpha'^2$ that do not involve $\Omega$, in addition to the couplings presented in the previous actions that do involve $\Omega$.
It is worth noting that the even parity coupling $\Omega^2$ in \reef{baction} is inconsistent with T-duality, indicating the presence of other even parity bulk and boundary couplings at this order. For the purpose of this paper, we will not consider the boundary couplings at order $\alpha'^2$. 

A previous study \cite{Garousi:2019cdn} demonstrated the existence of 60 even-parity independent geometrical couplings obtained through the most general field redefinitions allowed only for closed spacetime manifolds. These couplings are scheme-dependent. In a particular scheme,  they are
\beqa
\bS^{(2,3,4,5,6,7,8)}&=&-\frac{2\alpha'^2 }{\kappa^2}\int d^{10} x\sqrt{-G} e^{-2\Phi}\Big[a_1R_{\alpha}{}^{\epsilon}{}_{\gamma}{}^{\zeta} R^{\alpha \beta \gamma \delta} R_{\beta \zeta \delta \epsilon} +a_2 R_{\alpha \beta}{}^{\epsilon \zeta} R^{\alpha \beta \gamma \delta} R_{\gamma \epsilon \delta \zeta}\labell{H6}\\&&+a_3 \hat{H}_{\alpha}{}^{\delta \epsilon} \hat{H}^{\alpha \beta \gamma} \hat{H}_{\beta \delta}{}^{\zeta} \hat{H}_{\gamma}{}^{\iota \kappa} \hat{H}_{\epsilon \iota}{}^{\mu} \hat{H}_{\zeta \kappa \mu}+\cdots+ a_{60} \hat{H}_{\alpha}{}^{\beta \gamma} \hat{H}_{\beta}{}^{\delta \epsilon} \hat{H}_{\gamma}{}^{\zeta \iota} \nabla^{\alpha}\Phi \nabla_{\iota}\hat{H}_{\delta \epsilon \zeta}\Big]\,.\nn
\eeqa
Here, $a_1,\cdots,a_{60}$ are background-independent parameters that cannot be fixed by the gauge symmetries.
The explicit form of all the couplings can be found in \cite{Garousi:2019cdn}. These couplings were obtained for the effective action of the bosonic string theory in \cite{Garousi:2019mca}, where they were required to be invariant under the non-geometrical subgroup of the T-duality group. It is expected that the consistency of the combination of the $\Omega^2$-term and the aforementioned terms at order $\alpha'^2$ with the non-geometrical T-duality will necessitate some of the above couplings to be non-zero in the heterotic theory. 
In \reef{H6}, the couplings at orders $\alpha'^2, \alpha'^4, \alpha'^6,$ and $\alpha'^8$ are even parity, while the couplings at orders $\alpha'^3, \alpha'^5,$ and $\alpha'^7$ are odd parity. It was shown in \cite{Metsaev:1986yb} that, unlike the case in the effective action of the bosonic string theory, the S-matrix method requires the coefficients of the Riemann cubed terms to be zero in the heterotic theory. 
In this paper, we aim to determine all other couplings in the effective action presented above by imposing the requirement that the effective action at order $\alpha'^2$ must be invariant under the non-geometrical subgroup of the T-duality group.

It has come to light that the odd-parity couplings in \reef{Mis} at order $\alpha'^2$ also lack invariance under T-duality. Consequently, it is evident that additional odd-parity couplings at this order must exist, excluding those involving $\Omega$. Remarkably, we have identified 13 independent odd-parity geometrical couplings at order $\alpha'^2$. The constraints imposed by T-duality play a crucial role in determining these couplings.

This paper investigates the Buscher rules for circular reduction and their $\alpha'$-corrections, taking into account the constraint that the generalized Buscher rules must satisfy the $O(1,1,\MZ)$-group and the requirement that the effective actions must be invariant under the generalized Buscher rules. In Section 2, we provide a review of the Buscher rules for circular reduction and study  their $\alpha'$-corrections under the $O(1,1,\MZ)$-transformations. Although there are still undetermined coefficients in the generalized Buscher rules, we demonstrate that these coefficients can be fixed by imposing the constraint that the circular reduction of the effective actions in any scheme must be invariant under $O(1,1,\MZ)$-transformations. We briefly review this constraint for the effective action at order $\alpha'$ in Subsection 2.1 and provide the corrections to the Buscher rules for the effective action in the Meissner scheme. In Subsection 2.2.1, we impose the constraint that the effective action at order $\alpha'^2$ in the minimal scheme (equation \reef{H6} plus $\Omega^2$-term) must also be invariant under $O(1,1,\MZ)$-transformations. This allows us to fix all 60 parameters in the effective action as well as all corresponding corrections to the Buscher rules at order $\alpha'^2$. The resulting effective action is presented in equation \reef{final}. 
Since the expressions for the T-duality transformations are lengthy, we provide them in the Appendix. Our calculations reveal some total derivative terms in the base space. However, since our focus in this paper is solely on closed manifolds with no boundary, we omit these total derivative terms. In Section 2.2.2, we reproduce the calculations pertaining to the odd-parity couplings. Specifically, we observe that the odd-parity couplings at order $\alpha'^2$ in \reef{Mis} do not conform to T-duality. Consequently, we ascertain 13 independent couplings at order $\alpha'^2$ that do not involve $\Omega$. By employing T-duality, we establish that three of these couplings must possess non-zero coefficients. The resulting outcome is presented in \reef{finalodd}. To maintain conciseness, we omit the explicit expressions of the resulting deformations and the terms that amount to total derivatives.
In Section 3, we offer a brief discussion of our findings.

 \section{T-duality  constraint}\label{sec.3}

It has been proved in \cite{Sen:1991zi,Hohm:2014sxa} that if the classical effective action of string theory is compactified on a torus $T^{(d)}$, then the massless fields in the base space should be invariant under the $O(d,d, \MR)$ transformations. These transformations include the geometrical transformations which leave the parent geometrical couplings to be invariant and the nongeometrical transformations which transform a parent coupling to the other couplings. These latter transformations can be used to find the relations between the background-independent geometrical couplings in \reef{H6}.

The proof \cite{Sen:1991zi,Hohm:2014sxa} indicates that the effective actions should be also invariant under the discrete subgroup of $O(d,d, \MR)$, \ie $O(d,d, \MZ)$. These discrete transformations are generated by the inverse transformation and by the shift transformation. The latter transformation involves an antisymmetric matrix of integers. While the shift transformation leaves invariant a geometrical coupling, the inverse transformation connects different geometrical coupling into each other. Hence to simplify the symmetry to the one which has only nongeometrical transformations, we consider the circular reduction that the corresponding discrete group $O(1,1,\MZ)=\MZ_2$ has only the nongeometrical inverse transformations. The discrete transformations at the leading order of $\alpha'$ are  the Buscher rules \cite{Buscher:1987sk,Buscher:1987qj}.

To write the Buscher rules in their simplest form, it is convenient to use the following background for the metric, $B$-field and dilaton \cite{Maharana:1992my}:
  \beqa
G_{\mu\nu}=\left(\matrix{\bg_{ab}+e^{\varphi}g_{a }g_{b }& e^{\varphi}g_{a }&\cr e^{\varphi}g_{b }&e^{\varphi}&}\right),\, B_{\mu\nu}= \left(\matrix{\bb_{ab}+\frac{1}{2}b_{a }g_{b }- \frac{1}{2}b_{b }g_{a }&b_{a }\cr - b_{b }&0&}\right),\,  \Phi=\bar{\phi}+\varphi/4\,,\labell{reduc}\eeqa
where $\bg_{ab}$, $\bb_{ab}$, and $\bphi$ represent the metric, B-field, and dilaton in the base space, respectively. Furthermore, $g_a$ and $b_b$ denote two vectors, while $\vp$ represents a scalar within this space. Inverse of the above metric is 
\beqa
G^{\mu\nu}=\left(\matrix{\bg^{ab} &  -g^{a }&\cr -g^{b }&e^{-\varphi}+g_{c}g^{c}&}\right)\,,\labell{inver}
\eeqa
where $\bg^{ab}$ is the inverse of the base space metric which raises the index of the vectors.
The Buscher rules in this parametrization, are the following transformations:
\beqa
\varphi'= -\varphi
\,\,\,,\,\,g'_{a }= b_{a }\,\,\,,\,\, b'_{a }= g_{a } \,\,\,,\,\,\bg_{ab}'=\bg_{ab} \,\,\,,\,\,\bb_{ab}'=\bb_{ab} \,\,\,,\,\,  \bar{\phi}'= \bar{\phi}\,,\labell{T2}
\eeqa
which abviously form a $\MZ_2$-group, \ie $ (\psi')'= \psi$ where $\psi$ is any field in the base space. 

At the higher orders of $\alpha'$, the above transformations receive higher derivative corrections, \ie
\beqa
\psi'&=&\psi_0'+\sum_{n=1}^{\infty}\frac{\alpha'^n}{n!}\psi_n'\,,
\eeqa
where $\psi_0'$ are the Buscher rules \reef{T2} and $\psi_n'$ is its corrections at order $\alpha'^n$. In terms of different base space fields, the transformations can be written as 
\beqa
&&\varphi'= -\varphi+\sum_{n=1}^{\infty}\frac{\alpha'^n}{n!}\Delta\vp^{(n)}
\,\,\,,\,\,g'_{a }= b_{a }+e^{\vp/2}\sum_{n=1}^{\infty}\frac{\alpha'^n}{n!}\Delta g^{(n)}_a\,\,\,,\,\, b'_{a }= g_{a }+e^{-\vp/2}\sum_{n=1}^{\infty}\frac{\alpha'^n}{n!}\Delta b^{(n)}_a \,\,\,,\,\,\nn\\
&&\bg_{ab}'=\bg_{ab}+\sum_{n=1}^{\infty}\frac{\alpha'^n}{n!}\Delta \bg^{(n)}_{ab} \,\,\,,\,\,\bH_{abc}'=\bH_{abc}+\sum_{n=1}^{\infty}\frac{\alpha'^n}{n!}\Delta\bH^{(n)}_{abc} \,\,\,,\,\,  \bar{\phi}'= \bar{\phi}+\sum_{n=1}^{\infty}\frac{\alpha'^n}{n!}\Delta\bphi^{(n)}\,,\labell{T22}
\eeqa
where $\Delta \vp^{(n)}(\psi), \cdots,\Delta\bphi^{(n)}(\psi)$ contain some contractions of $\nabla\vp,\nabla\bphi, e^{\vp/2}V, e^{-\vp/2}W,\bH,\bR$ and their covariant derivatives at order $\alpha'^n$. In the above equation $\bH_{abc}$ is the torsion in the base space which is defined as $\bH_{abc}=3\prt_{[a}\hat{b}_{bc]}-3g_{[a}W_{bc]}$ where $\hat{b}_{ab}=\bb_{ab}+\frac{1}{2}b_{a }g_{b }- \frac{1}{2}b_{b }g_{a }$. It can be written in terms of $\bb_{ab}$ as 
\beqa
\bH_{abc}&=&3\prt_{[a}\bb_{bc]}-\frac{3}{2}g_{[a}W_{bc]}-\frac{3}{2}b_{[a}V_{bc]}\,.\nn
\eeqa
It satisfies the following Bianchi identity \cite{Kaloper:1997ux}:
\beqa
\prt_{[a} \bH_{bcd]}&=&-\frac{3}{2}V_{[ab}W_{cd]}\,.\labell{anB}
\eeqa
In the above equations, $V_{ab}$ is the field strength of the $U(1)$ gauge field $g_{a}$, \ie $V_{ab}=\prt_{a}g_{b}-\prt_{b}g_{a}$, and $W_{\mu\nu}$ is the field strength of the $U(1)$ gauge field $b_{a}$, \ie $W_{ab}=\prt_{a}b_{b}-\prt_{b}b_{a}$.

 The deformed transformations \reef{T22} must satisfy the $\MZ_2$-group  $ (\psi')'= \psi$. It produces the following relations between the $\alpha'$-corrections of the Buscher rules:
 \beqa
 -\sum_{n=1}^{\infty}\frac{\alpha'^n}{n!}\Delta\vp^{(n)}(\psi)+\sum_{n=1}^{\infty}\frac{\alpha'^n}{n!}\Delta\vp^{(n)}(\psi')&=&0\,,\nn\\
  e^{-\vp/2}\sum_{n=1}^{\infty}\frac{\alpha'^n}{n!}\Delta b_a^{(n)}(\psi)+e^{\vp'/2}\sum_{n=1}^{\infty}\frac{\alpha'^n}{n!}\Delta g_a^{(n)}(\psi')&=&0\,,\nn\\
  e^{\vp/2}\sum_{n=1}^{\infty}\frac{\alpha'^n}{n!}\Delta g_a^{(n)}(\psi)+e^{-\vp'/2}\sum_{n=1}^{\infty}\frac{\alpha'^n}{n!}\Delta b_a^{(n)}(\psi')&=&0\,,\nn\\
 \sum_{n=1}^{\infty}\frac{\alpha'^n}{n!}\Delta \bg_{ab}^{(n)}(\psi)+\sum_{n=1}^{\infty}\frac{\alpha'^n}{n!}\Delta \bg_{ab}^{(n)}(\psi')&=&0\,,\nn\\
 \sum_{n=1}^{\infty}\frac{\alpha'^n}{n!}\Delta \bphi^{(n)}(\psi)+\sum_{n=1}^{\infty}\frac{\alpha'^n}{n!}\Delta \bphi^{(n)}(\psi')&=&0\,.\labell{A1}
 \eeqa
 The corrections $\Delta \vp^{(n)}(\psi), \cdots,\Delta\bphi^{(n)}(\psi)$ involve all contractions of the base space fields at order $\alpha'^{n}$ with arbitrary coeffients that satisfy the above constraints.
 
The constraint $(\psi')'=\psi$ also leads to the following relation:
\beqa
\sum_{n=1}^{\infty}\frac{\alpha'^n}{n!}\Delta \bH_{abc}^{(n)}(\psi)+\sum_{n=1}^{\infty}\frac{\alpha'^n}{n!}\Delta \bH_{abc}^{(n)}(\psi')=0\,.\labell{A2}
\eeqa
However, we cannot conclude that the correction $\Delta \bH^{(n)}_{abc}$ is also all contractions of the base space fields at order $\alpha'^{n+1/2}$ with arbitrary coefficients that satisfy the above constraint. In fact, the corrections $\Delta \bH^{(n)}_{abc}$, $\Delta g_a^{(n)}$, and $\Delta b_a^{(n)}$ satisfy another constraint resulting from the fact that the T-dual transformed fields must satisfy the Bianchi identity \reef{anB}. In terms of forms $db=2W$ and $dg=2V$, the Bianchi identity can be written as:
\beqa
d\bH'=-6 dg'\wedge db'\,.\labell{B1}
\eeqa
Inserting the expansions \reef{T22} into it, one finds
\beqa
\sum_{n=1}^{\infty}\frac{\alpha'^n}{n!}\Delta \bH^{(n)}&\!\!\!\!\!=\!\!\!\!\!&\sum_{n=1}^{\infty}\frac{\alpha'^n}{n!}\tilde{H}^{(n)}-6\sum_{n=1}^{\infty}\frac{\alpha'^n}{n!}db\wedge (e^{-\vp/2}\Delta b^{(n)})-6\sum_{n=1}^{\infty}\frac{\alpha'^n}{n!} (e^{\vp/2}\Delta g^{(n)})\wedge dg\labell{dHbar}\\&&
-3\sum_{n,m=1}^{\infty}\frac{\alpha'^{n+m}}{n!m!} \Big[(e^{\vp/2}\Delta g^{(n)})\wedge d(e^{-\vp/2}\Delta b^{(m)})+d(e^{\vp/2}\Delta g^{(n)})\wedge (e^{-\vp/2}\Delta b^{(m)})\Big]\,,\nn
\eeqa
where the 3-form $\tilde{H}^{(n)}$ contains $U(1)\times U(1)$ gauge-invariant couplings at order $\alpha'^n$ and is a closed 3-form,  \ie $d\tilde{H}^{(n)}=0$. However, the aforementioned corrections must also be expressible in terms of corrections to the base space field $\bb_{ab}$, which imposes constraints on $\tilde{H}^{(n)}$ to be exact, \ie  $\tilde{H}^{(n)}=3d\tilde{B}^{(n)}$. To clarify this point, we first insert the $\alpha'$-expansions in equation \reef{T22} and the following $\alpha'$-expansion of the T-duality of $\bb_{ab}'$:
\beqa
\bb'=\bb+\sum_{n=1}^{\infty}\frac{\alpha'^n}{n!}\Delta \bb^{(n)}\,,
\eeqa
into the  relation between $\bH$ and $\bb$:
\beqa
\bH'=3d\bb'-3g'\wedge db'-3b'\wedge dg'\,.
\eeqa
Then one finds the following relation between $\Delta\bH^{(n)}$ and $\Delta\bb^{(n)}$:
\beqa
\sum_{n=1}^{\infty}\frac{\alpha'^n}{n!}\Delta \bH^{(n)}&\!\!\!\!\!=\!\!\!\!\!&\sum_{n=1}^{\infty}\frac{\alpha'^n}{n!}d\Delta \bb^{(n)}-3\sum_{n=1}^{\infty}\frac{\alpha'^n}{n!}b\wedge d(e^{-\vp/2}\Delta b^{(n)})-3\sum_{n=1}^{\infty}\frac{\alpha'^n}{n!} (e^{\vp/2}\Delta g^{(n)})\wedge dg\nn\\&&-3\sum_{n=1}^{\infty}\frac{\alpha'^n}{n!}g\wedge d(e^{\vp/2}\Delta g^{(n)})-3\sum_{n=1}^{\infty}\frac{\alpha'^n}{n!} (e^{-\vp/2}\Delta b^{(n)})\wedge db\labell{dHbar2}\\&&
-3\sum_{n,m=1}^{\infty}\frac{\alpha'^{n+m}}{n!m!} \Big[(e^{\vp/2}\Delta g^{(n)})\wedge d(e^{-\vp/2}\Delta b^{(m)})+d(e^{\vp/2}\Delta g^{(n)})\wedge (e^{-\vp/2}\Delta b^{(m)})\Big]\,.\nn
\eeqa
Equating the right-hand sides of equations \reef{dHbar} and \reef{dHbar2}, and using the relation $d(A\wedge B)=dA\wedge B-A\wedge dB$ for any two vectors $A$ and $B$, one can derive that if $\tilde{H}$ is exact, then the corrections $\Delta \bb^{(n)}$ can be expressed as follows:
\beqa
\Delta \bb^{(n)}=3\tilde{B}^{(n)}-3b\wedge (e^{-\vp/2}\Delta b^{(n)})-3g\wedge (e^{\vp/2}\Delta g^{(n)})\,.
\eeqa
This relation was first derived in \cite{Kaloper:1997ux} for the case of $n=1$. It is worth noting that the right-hand side of equation \reef{dHbar} is gauge invariant under $U(1) \times U(1)$, whereas the right-hand side of equation \reef{dHbar2} is not. Therefore, we will utilize equation \reef{dHbar} to investigate the T-duality constraint on the effective action. 

Therefore, the relation in equation \reef{dHbar}, in which $\tilde{H}^{(n)}$ is an exact form, relates $\Delta \bH^{(n)}_{abc}$ to $\Delta g_a^{(n)}$ and $\Delta b_a^{(n)}$. The exact form $\tilde{H}^{(n)}$ should also satisfy the constraint in equation \reef{A2}. 
Since the transformed fields $\vp'$ and $\psi'$ have $\alpha'$-expansions, one must first insert their expansions into the constraints in equations \reef{A1} and \reef{A2}, and then Taylor expand the corrections $\Delta\vp^{(n)}(\psi'), \cdots, \Delta\bH^{(n)}(\psi')$ around $\psi_0'$. This yields an expansion in terms of different orders of $\alpha'$. One can then set the terms at each order of $\alpha'$ to be zero to find the appropriate constraints at that order, \ie
\beqa
 -\sum_{n=1}^{\infty}\frac{\alpha'^n}{n!}\Delta\vp^{(n)}(\psi)+\sum_{n=1}^{\infty}\frac{\alpha'^n}{n!}\Delta\vp^{(n)}(\psi_0')+\sum_{n,m=1}^{\infty}\frac{\alpha'^{n+m}}{n!m!}\Delta\vp^{(n,m)}(\psi_0')&=&0\,,\nn\\
 \sum_{n=1}^{\infty}\frac{\alpha'^n}{n!}\Delta b_a^{(n)}(\psi)+\sum_{n=1}^{\infty}\frac{\alpha'^n}{n!}\Delta g_a^{(n)}(\psi_0')+\sum_{n,m=1}^{\infty}\frac{\alpha'^{n+m}}{n!m!}\Delta g_a^{(n,m)}(\psi_0')&=&0\,,\nn\\
 \sum_{n=1}^{\infty}\frac{\alpha'^n}{n!}\Delta g_a^{(n)}(\psi)+\sum_{n=1}^{\infty}\frac{\alpha'^n}{n!}\Delta b_a^{(n)}(\psi_0')+\sum_{n,m=1}^{\infty}\frac{\alpha'^{n+m}}{n!m!}\Delta b_a^{(n,m)}(\psi_0')&=&0\,,\nn\\
 \sum_{n=1}^{\infty}\frac{\alpha'^n}{n!}\Delta \bg_{ab}^{(n)}(\psi)+\sum_{n=1}^{\infty}\frac{\alpha'^n}{n!}\Delta \bg_{ab}^{(n)}(\psi_0')+\sum_{n,m=1}^{\infty}\frac{\alpha'^{n+m}}{n!m!}\Delta  \bg_{ab}^{(n,m)}(\psi_0')&=&0\,,\nn\\
 \sum_{n=1}^{\infty}\frac{\alpha'^n}{n!}\Delta \bphi^{(n)}(\psi)+\sum_{n=1}^{\infty}\frac{\alpha'^n}{n!}\Delta \bphi^{(n)}(\psi_0')+\sum_{n,m=1}^{\infty}\frac{\alpha'^{n+m}}{n!m!}\Delta \bphi^{(n,m)}(\psi_0')&=&0\,,\nn\\
\sum_{n=1}^{\infty}\frac{\alpha'^n}{n!}\Delta \bH_{abc}^{(n)}(\psi)+\sum_{n=1}^{\infty}\frac{\alpha'^n}{n!}\Delta \bH_{abc}^{(n)}(\psi_0')+\sum_{n,m=1}^{\infty}\frac{\alpha'^{n+m}}{n!m!}\Delta \bH_{abc}^{(n,m)}(\psi_0')&=&0\,, \labell{A11}
 \eeqa
where the perturbations $\Delta\vp^{(n,m)}(\psi_0'), \cdots,\Delta\bH_{abc}^{(n,m)}(\psi_0') $  are defined as
\beqa
\Delta\vp^{(n)}(\psi')&=&\Delta\vp^{(n)}(\psi_0')+\sum_{m=1}^{\infty}\frac{\alpha'^m}{m!}\Delta\vp^{(n,m)}(\psi_0')\,,\nn\\
e^{\vp'/2}\Delta g_a^{(n)}(\psi')&=&e^{-\vp/2}\Delta g_a^{(n)}(\psi_0' )+e^{-\vp/2}\sum_{m=1}^{\infty}\frac{\alpha'^m}{m!}\Delta g_a^{(n,m)}(\psi_0')\,,\nn\\
e^{-\vp'/2}\Delta b_a^{(n)}(\psi')&=&e^{\vp/2}\Delta b_a^{(n)}(\psi_0' )+e^{\vp/2}\sum_{m=1}^{\infty}\frac{\alpha'^m}{m!}\Delta b_a^{(n,m)}(\psi_0')\,,\nn\\
\Delta\bg_{ab}^{(n)}(\psi')&=&\Delta\bg_{ab}^{(n)}(\psi_0' )+\sum_{m=1}^{\infty}\frac{\alpha'^m}{m!}\Delta \bg_{ab}^{(n,m)}(\psi_0')\,,\nn\\
\Delta\bphi^{(n)}(\psi')&=&\Delta\bphi^{(n)}(\psi_0' )+\sum_{m=1}^{\infty}\frac{\alpha'^m}{m!}\Delta\bphi^{(n,m)}(\psi_0')\,,\nn\\
\Delta\bH_{abc}^{(n)}(\psi')&=&\Delta\bH_{abc}^{(n)}(\psi_0')+\sum_{m=1}^{\infty}\frac{\alpha'^m}{m!}\Delta\bH_{abc}^{(n,m)}(\psi_0')\,.
\eeqa
Therefore, the corrections to the Buscher rules must satisfy the constraints in equation \reef{A11}. However, these constraints cannot fix all parameters in the corrections. The unfixed parameters should be determined by other constraints, such as the requirement that the effective actions be invariant under T-duality transformations.

Before discussing the T-duality of the effective actions, we would like to point out that the last constraint in equation \reef{A11} can also be expressed in the following form using the relation in equation \reef{dHbar}:
\beqa
&&\sum_{n=1}^{\infty}\frac{\alpha'^n}{n!}\tilde{H}^{(n)}+\sum_{n=1}^{\infty}\frac{\alpha'^n}{n!}\tilde{H}^{(n)}(\psi_0')\nn\\&&-6\sum_{n,m=1}^{\infty}\frac{\alpha'^{n+m}}{n!m!}db\wedge(e^{-\vp/2}\Delta g^{(n+m)}(\psi_0'))-6\sum_{n,m=1}^{\infty}\frac{\alpha'^{n+m}}{n!m!}dg\wedge(e^{\vp/2}\Delta b^{(n+m)}(\psi_0'))\nn\\&&-3\sum_{n,m=1}^{\infty}\frac{\alpha'^{n+m}}{n!m!}\Big[(e^{\vp/2}\Delta g^{(n)})\wedge d(e^{-\vp/2}\Delta b^{(m)})+d(e^{\vp/2}\Delta g^{(n)})\wedge (e^{-\vp/2}\Delta b^{(m)})\Big]\nn\\&&-3\sum_{n,m=1}^{\infty}\frac{\alpha'^{n+m}}{n!m!}\Big[(e^{\vp/2}\Delta g^{(n)}(\psi_0'))\wedge d(e^{-\vp/2}\Delta b^{(m)}(\psi_0'))+d(e^{\vp/2}\Delta g^{(n)}(\psi_0'))\wedge (e^{-\vp/2}\Delta b^{(m)}(\psi_0'))\Big]\nn\\&&
+\sum_{n,m=1}^{\infty}\frac{\alpha'^{n+m}}{n!m!}\Delta \bH_{abc}^{(n,m)}(\psi_0')=0 \,.\labell{A12}
\eeqa
This indicates that the coefficients of the couplings in the exact form $\tilde{H}^{(n)}$ at order $\alpha'^n$ are related to themselves, as well as to the corrections $\Delta g$, $\Delta b$, and $\tilde{H}$ at orders $\alpha'$, $\alpha'^2$, $\cdots$, $\alpha'^{n-1}$.

We now impose the constraint that the effective actions be invariant under the $O(1,1,\mathbb{Z})$-group. To do so, we need to reduce the theory on a circle with a $U(1)$ isometry to obtain the 9-dimensional effective action $S_{\rm eff}(\psi)$, which we then transform under the $O(1,1,\mathbb{Z})$-transformations to produce $S_{\rm eff}(\psi')$. The $O(1,1,\mathbb{Z})$ invariance is given by 
\beqa
S_{\rm eff}(\psi)-S_{\rm eff}(\psi')&=&\int d^{9}x \sqrt{-\bg}\nabla_a\Big[e^{-2\bphi}J^a(\psi)\Big]\,,\labell{TS}
\eeqa
where $J^a$ is an arbitrary covariant vector made up of the 9-dimensional base space fields, with the following $\alpha'$-expansion:
\begin{equation}
J^a=\sum^\infty_{n=0}\frac{\alpha'^n}{n!}J^a_n\,,
\end{equation}
where $J^a_n$ is an arbitrary covariant vector at order $\alpha'^n$. Note that by imposing $(\psi')' = \psi$ on \reef{TS}, it can be deduced that the total derivative terms must satisfy the following relation:
\beqa
 \sqrt{-\bg'}\nabla'_a\Big[e^{-2\bphi'}J^a(\psi')\Big]&=&-\sqrt{-\bg}\nabla_a\Big[e^{-2\bphi}J^a(\psi)\Big]\,.\nn
\eeqa

If the effective action and its circular reduction have the following $\alpha'$-expansions:
\begin{equation}
\bS_{\rm eff}=\sum^\infty_{n=0}\frac{\alpha'^n}{n!}\bS^{(n)},\quad S_{\rm eff}=\sum^\infty_{n=0}\frac{\alpha'^n}{n!}S^{(n)},
\end{equation}
then the constraint in equation \reef{TS} can be expressed as
\begin{equation}
\sum^\infty_{n=0}\frac{\alpha'^n}{n!}S^{(n)}(\psi)-\sum^\infty_{n=0}\frac{\alpha'^n}{n!}S^{(n)}(\psi')=\sum^\infty_{n=0}\frac{\alpha'^n}{n!}\int d^{9}x \sqrt{-\bg}\nabla_a\Big[e^{-2\bphi}J^a_n(\psi)\Big]\,.
\end{equation}
Expanding the second term above around $\psi_0'$, \ie
\begin{equation}
S^{(n)}(\psi')=S^{(n)}(\psi_0')+\sum_{m=1}^{\infty}\frac{\alpha'^m}{m!}S^{(n,m)}(\psi_0')\,,
\end{equation}
yields the constraint in equation \reef{TSn}:
\begin{equation}
\sum^\infty_{n=0}\frac{\alpha'^n}{n!}S^{(n)}-\!\sum^\infty_{n=0}\frac{\alpha'^n}{n!}S^{(n)}(\psi_0')-\!\!\sum^\infty_{n=0,m=1}\frac{\alpha'^{n+m}}{n!m!}S^{(n,m)}(\psi_0')=\!\sum^\infty_{n=0}\frac{\alpha'^n}{n!}\int d^{9}x \sqrt{-\bg}\nabla_a\Big[e^{-2\bphi}J^a_n\Big]. \labell{TSn}
\end{equation}
To find the appropriate constraints on the effective actions, one must set the terms at each order of $\alpha'$ to be zero.

\subsection{T-duality constraint at orders $\alpha'^0,\, \alpha'$}

In this subsection, we review how the effective actions at orders $\alpha'^0$ and $\alpha'$ are found by imposing the nongeometrical T-duality transformations on the effective action. The constraint in equation \reef{TSn} at order $\alpha'^0$ is given by:
\beqa
S^{(0)}-S^{(0)}(\psi'_0)&=& \int d^{9}x \sqrt{-\bg}\nabla_a(e^{-2\bphi}J_0^a)\labell{TS0}
\eeqa
where $J^a_0$ is an arbitrary vector at the leading order of $\alpha'$, and $\psi'_0$ is the Buscher rules \reef{T2}. Reduction of different geometrical couplings in $\bS^{(0)}$ are the following (see \eg \cite{Garousi:2019wgz}):
\beqa
a_1R&=&a_1(\bR-\nabla^a\nabla_a\vp-\frac{1}{2}\nabla_a\vp \nabla^a\vp -\frac{1}{4}e^{\vp}V^2 )\,,\labell{R}\\
a_2\nabla_{\mu}\Phi\nabla^{\mu}\Phi&=&a_2(\nabla_a\bphi\nabla^a \bphi+\frac{1}{2}\nabla_a\bphi\nabla^a\vp+\frac{1}{16}\nabla_a\vp\nabla^a\vp)\,,\nn\\
a_3H^2&=&a_3(\bH_{abc}\bH^{abc}+3e^{-\vp}W^2) \,.\nn
\eeqa
The terms on the left-hand sides of the above equations  are geometrical (gauge invariant) in the 10-dimensional spacetime, while the terms on the right-hand sides are geometrical and gauge invariant in the 9-dimensional base space. However, the terms on the right-hand sides are not invariant under the nongeometrical T-duality transformation in equation \reef{T2}. Requiring them to be invariant under the nongeometrical transformations fixes the coefficients $a_1,a_2,a_3$ up to an overall factor to those in equation \reef{baction}. 
Moreover, there is a total derivative term in the base space that fixes the coefficient of the boundary term to be that in equation \reef{baction}. 

In general, non-geometric T-duality transformations yield two distinct sets of relations among the geometrical couplings. The first set comprises terms involving the zeroth and first partial derivatives of the base space metric, while the second set encompasses terms involving the second partial derivative of the base space metric and higher. Interestingly, a noteworthy observation made in \cite{Garousi:2019mca} is that all the relations derived from the second set are already encompassed by the relations obtained from the first set. Consequently, we can disregard the relations belonging to the second set.
Within the context of the first set of relations, owing to the covariant nature of the formalism, we can select a local frame where the first partial derivative of the metric vanishes. This enables us to simplify the analysis by assuming, for the sake of simplicity, that the base space metric is flat. Henceforth, we consider the base space to be flat in our endeavor to determine the relations between the parameters via T-duality.

The constraint \reef{TSn} at order $\alpha'$ for flat base space is
\beqa
S^{(1)}-S^{(1)}(\psi'_0)-S^{(0,1)}(\psi'_0)&=& \int d^{9}x \,\prt_a[e^{-2\bphi} J_1^a]\,.\labell{TS11}
\eeqa
In the heterotic theory, $S^{(1)}$ has both even and odd parity terms. As a result, $S^{(1)}$ and the corrections to the Buscher rules that appear in $S^{(0,1)}(\psi'_0)$ have both even and odd parity terms. However, as we mentioned earlier, the constraint $(\psi')'=\psi$ on the corrections to the Buscher rules cannot fix the parameters in these corrections. Therefore, we can fix them for effective actions in different schemes, reflecting the fact that the corrections to the Buscher rules depend on the scheme of the effective actions \cite{Garousi:2019wgz}.
The corrections to the Buscher rules at order $\alpha'$ for the effective actions in equations \reef{baction} and \reef{Mis} at order $\alpha'$ have been found in \cite{Meissner:1996sa,Garousi:2023pah} to be:
\beqa
\Delta \bg_{ab}^{(1)}&=&0\,,\nn\\
\Delta\bphi^{(1)}&=&0\,,\nn\\
\Delta\vp^{(1)}&=& c_1\Big(2\prt_a\vp\prt^a\vp+e^\vp V^2+e^{-\vp}W^2\Big) +\frac{1}{4}V^{ab}W_{ab}\,,\nn\\
\Delta g_{a}^{(1)}&=& c_1\Big(2e^{-\vp/2}\prt^b\vp W_{ab}+e^{\vp/2}\bH_{abc} V^{bc} \Big)-\frac{1}{16}\Big(2e^{\vp/2}\prt^b\vp V_{ab}-e^{-\vp/2}\bH_{abc} W^{bc} \Big)\,,\nn\\
\Delta b_{a}^{(1)}&=& c_1\Big(2e^{\vp/2}\prt^b\vp V_{ab}-e^{-\vp/2}\bH_{abc} W^{bc} \Big)-\frac{1}{16}\Big(2e^{-\vp/2}\prt^b\vp W_{ab}+e^{\vp/2}\bH_{abc} V^{bc} \Big)\,,\nn\\
\Delta\bH_{abc}^{(1)}&=&-12c_1 \prt_{[a}(W_b{}^d V_{c]d})-3e^{\vp/2} V_{[ab}\Delta g^{(1)}_{c]}-3e^{-\vp/2} W_{[ab}\Delta b^{(1)}_{c]}\,.\labell{dbH2}
\eeqa
There are also total derivative terms in the base space that fix the geometrical boundary terms at order $\alpha'$ to those in equation \reef{finalb}. Importantly, there are no odd parity corrections for $\Delta \bg_{ab}^{(1)}$ and $\Delta\bphi^{(1)}$. The even parity correction for $\Delta \bg_{ab}^{(1)}$ and $\Delta\bphi^{(1)}$ is zero only for the effective action in the Meissner scheme \cite{Meissner:1996sa}. In this scheme, the corrections involve only the first derivative of the base space fields. This property is required for spacetimes that have boundaries because T-duality should respect the data on the boundary \cite{Garousi:2023pah}. 
There is another scheme in which the corrections involve the first derivative of the base space fields; however, the correction $\Delta\bphi^{(1)}$ is non-zero in that scheme \cite{Garousi:2021yyd}. It is worth noting that the base space dilaton $\bphi$ does not appear in the aforementioned T-duality corrections that have zero $\Delta\bphi^{(1)}$, whereas it appears in the corrections of the T-duality in the scheme \cite{Garousi:2021yyd} in which $\Delta\bphi^{(1)}$ is non-zero.

At higher orders of $\alpha'$, there exist schemes where $\Delta \bg_{ab}^{(n)}=0$. This is due to the fact that if the spacetime has a boundary, the unit vector to the boundary in the string frame must be invariant under T-duality at any order of $\alpha'$. This scheme can also be used for spacetimes that have no boundary.
There may be   schemes where both $\Delta \bg_{ab}^{(n)}$ and $\Delta\bphi^{(n)}$ are zero, meaning that the measure $e^{-2\bphi}\sqrt{-\bg}$ is invariant under T-duality at all orders of $\alpha'$ in those schemes. However, these schemes  have more couplings than the minimal scheme in which the most general field redefinitions are used to find the independent couplings. 
If we consider the geometrical couplings to be in the minimal scheme, such as the scheme in equation \reef{H6}, then the corrections $\Delta \bg_{ab}^{(n)}$ and $\Delta\bphi^{(n)}$ are non-zero. However, if we require the unit vector in the string frame and the Einstein frame to be invariant under the field redefinitions, then the metric and the dilaton remain fixed under these restricted field redefinitions. In that case, there are many other couplings besides those in equation \reef{H6}.
In the ensuing subsection, we investigate closed spacetime manifolds without boundaries, which allow for the most comprehensive field redefinitions. In this context, we encounter a total of 60 independent even-parity and 13 independent odd-parity geometrical couplings.

 \subsection{T-duality constraint at order $  \alpha'^2$}
 
There are both even and odd parity couplings at order $\alpha'^2$ in the actions given by equations \reef{baction} and \reef{Mis}, which involve $\Omega$. Interestingly, neither of these couplings is invariant under T-duality. Consequently, there must exist additional couplings at this order that do not involve $\Omega$. In the subsequent subsection, we identify such couplings that are necessary to ensure the consistency of the even-parity couplings with T-duality. Furthermore, in subsection 2.2.2, we determine the couplings required to establish the consistency of the odd-parity couplings with T-duality.

 \subsubsection{Even-parity couplings}
 
We will now examine the constraint in equation \reef{TSn} in detail at order $\alpha'^2$ to determine the 60 parameters in equation \reef{H6}. This constraint at order $\alpha'^2$ is given by
\begin{equation}
S^{(2)}-S^{(2)}(\psi'_0)-S^{(0,2)}(\psi'_0)-S^{(1,1)}(\psi'_0)= \int d^{9}x \,\prt_a[e^{-2\bphi} J_2^a]\,,
\labell{TS12}
\end{equation}
where $J^a_2$ is an arbitrary vector at order $\alpha'^2$ constructed from the base space fields $\partial\varphi$, $\partial\bphi$, $e^{\varphi/2}V$, $e^{-\varphi/2}W$, and $\bar{H}$. In the heterotic theory, $\bS^{(2)}$ has both even and odd parity terms. However, since we are interested in determining the parameters in equation \reef{H6}, we will only consider the even parity terms at order $\alpha'^2$ in this subsection. Thus, we take $\bS^{(2)}$ to be the $\Omega^2$ term in equation \reef{baction} and the couplings in equation \reef{H6} at order $\alpha'^2$.

The Kaluza-Klein reduction of the frame $e_\mu{}^i$ is given by
\beqa
&&e_\mu{}^i=\left(\matrix{\bar{e}_a{}^\ti & 0 &\cr e^{\varphi/2}g_{a }&e^{\varphi/2}&}\right)\,,
\eeqa
where $\bar{e}_a{}^\ti \bar{e}_b{}^\tj\eta_{\ti\tj}=\bg_{ab}$. This reduction is consistent with the Kaluza-Klein reduction of the metric in equation \reef{reduc}. Using this reduction and the reductions in equation \reef{reduc}, we can determine the circular reduction of the $\Omega^2$-term for flat base space. The calculation can be performed using the "xAct" package \cite{Nutma:2013zea}, \ie
\beqa
\Omega^2&=&\frac{1}{48} e^{3 \vp} V_{a}{}^{c} V^{ab} V_{b}{}^{d} 
V_{c}{}^{e} V_{d}{}^{f} V_{ef} + \frac{1}{48} e^{3 \vp} V_{ab} 
V^{ab} V_{c}{}^{e} V^{cd} V_{d}{}^{f} V_{ef} + \frac{1}{192} e^{3 
\vp} V_{ab} V^{ab} V_{cd} V^{cd} V_{ef} V^{ef}\nn\\&& + \frac{1}{24} 
e^{2 \vp} V_{b}{}^{d} V^{bc} V_{c}{}^{e} V_{de} \prt_{a}\vp \prt^{a}\vp + \frac{1}{48} e^{2 \vp} V_{bc} V^{bc} 
V_{de} V^{de} \prt_{a}\vp \prt^{a}\vp -  \frac{1}{24} 
e^{2 \vp} V_{a}{}^{c} V_{b}{}^{d} V_{c}{}^{e} V_{de} 
\prt^{a}\vp \prt^{b}\vp\nn\\&& + \frac{1}{48} e^{2 \vp} 
V_{a}{}^{c} V_{bc} V_{de} V^{de} \prt^{a}\vp \prt^{b}\vp + \frac{1}{48} e^{\vp} V_{cd} V^{cd} \prt_{a}\vp 
\prt^{a}\vp \prt_{b}\vp \prt^{b}\vp + 
\frac{1}{24} e^{2 \vp} V^{ab} V^{cd} \prt_{b}V_{de} 
\prt_{c}V_{a}{}^{e}\nn\\&& + \frac{1}{48} e^{2 \vp} V_{a}{}^{c} 
V^{ab} \prt_{b}V^{de} \prt_{c}V_{de} -  \frac{1}{12} e^{\Phi 
1} V^{cd} \prt_{a}\vp \prt^{a}\vp \prt^{b}\vp 
\prt_{d}V_{bc} + \frac{1}{24} e^{\vp} V^{ab} V^{cd} 
\prt_{c}\prt_{a}\vp \prt_{d}\prt_{b}\vp\nn\\&& -  
\frac{1}{12} e^{\vp} V_{b}{}^{d} V^{bc} \prt_{a}\vp 
\prt^{a}\vp \prt_{d}\prt_{c}\vp -  \frac{1}{24} 
e^{\vp} \prt^{a}\vp \prt^{b}\vp \prt_{c}V_{bd} 
\prt^{d}V_{a}{}^{c} + \frac{1}{24} e^{\vp} \prt^{a}\vp \prt^{b}\vp \prt_{d}V_{bc} \prt^{d}V_{a}{}^{c}\nn\\&& + 
\frac{1}{12} e^{\vp} V^{bc} \prt^{a}\vp 
\prt_{c}V_{ad} \prt^{d}\prt_{b}\vp -  \frac{1}{12} e^{
\vp} V^{bc} \prt^{a}\vp \prt_{d}V_{ac} 
\prt^{d}\prt_{b}\vp + \frac{1}{24} e^{\vp} 
V_{a}{}^{c} V^{ab} \prt_{d}\prt_{c}\vp 
\prt^{d}\prt_{b}\vp\nn\\&& -  \frac{1}{12} e^{2 \vp} 
V_{b}{}^{d} V^{bc} V_{c}{}^{e} \prt^{a}\vp \prt_{e}V_{ad} - 
 \frac{1}{24} e^{2 \vp} V_{bc} V^{bc} V^{de} \prt^{a}\vp 
\prt_{e}V_{ad} + \frac{1}{12} e^{2 \vp} V_{a}{}^{b} 
V_{c}{}^{e} V^{cd} \prt^{a}\vp \prt_{e}V_{bd}\nn\\&& + 
\frac{1}{12} e^{2 \vp} V_{a}{}^{b} V_{b}{}^{c} V^{de} 
\prt^{a}\vp \prt_{e}V_{cd} -  \frac{1}{12} e^{2 \vp} 
V_{a}{}^{c} V^{ab} V_{b}{}^{d} V_{c}{}^{e} 
\prt_{e}\prt_{d}\vp -  \frac{1}{24} e^{2 \vp} V_{ab} 
V^{ab} V_{c}{}^{e} V^{cd} \prt_{e}\prt_{d}\vp\,.\nn
\eeqa
The reduction of the couplings in equation \reef{H6} that involve only the Riemann curvature, $H$, $\nabla H$, $\nabla\Phi$, and $\nabla\nabla\Phi$, can be found in \cite{Garousi:2019mca}. Therefore, $S^{(2)}$, which is the reduction of the couplings in equation \reef{H6} at order $\alpha'^2$, and the above reduction, can be calculated. Once $S^{(2)}$ is determined, its transformation under the Buscher rules can be calculated to find $S^{(2)}(\psi_0')$.

The circular reduction of the leading order bulk action \reef{baction} is
\beqa
S^{(0)}&=& -\frac{2}{\kappa'^2}\int d^{9}x\sqrt{-\bg}\, e^{-2\bphi}\Big[\bar{R}-\nabla^a\nabla_a\vp-\frac{1}{4}\nabla_a\vp \nabla^a\vp-\frac{1}{4}(e^{\vp}V^2 +e^{-\vp}W^2)\nn\\
&&\qquad\qquad\qquad\qquad\qquad+4\nabla_a\bphi\nabla^a \bphi+2\nabla_a\bphi\nabla^a\vp-\frac{1}{12}\bH_{abc}\bH^{abc}\Big]\,,\labell{R}
\eeqa
where $\kappa'$ is related to the 9-dimensional Newton's constant. 
When we apply the deformed Buscher rules in equation \reef{T22} to the action, we obtain two sets of terms at order $\alpha'^2$. The first set contains two first-order corrections, denoted by $S^{(0,2)}_{(1,1)}$, while the second set contains the second-order corrections, denoted by $S^{(0,2)}_{(2)}$. Specifically, we have:
\beqa
S^{(0,2)}=S^{(0,2)}_{(1,1)}+S^{(0,2)}_{(2)}\,.\labell{S02}
\eeqa
The first one is 
\beqa
S^{(0,2)}_{(1,1)}(\psi_0')&\!\!\!\!\!=\!\!\!\!\!& -\frac{2\alpha'^2}{\kappa'^2}\int d^{9}x  \,  e^{-2\bphi}\Big[ - \frac{1}{12} \Delta \bH^{(1)}{}_{abc} \Delta \bH^{(1)abc} -  
\frac{1}{8} e^{\vp} V_{ab} V^{ab}( \Delta \vp^{(1)})^2\\&& -  
\frac{1}{8} e^{-\vp}W_{ab} W^{ab} (\Delta \vp^{(1)})^2 -  
\frac{1}{4} \prt_{a}\Delta \vp^{(1)} \prt^{a}\Delta \vp^{(1)} -  \frac{1}{2} \Delta b^{(1)a} \prt_{a}\Delta b^{(1)b} 
\prt_{b}\vp \nn\\&&+ \frac{1}{2} \Delta g^{(1)a} \prt_{a}\Delta 
g^{(1)b} \prt_{b}\vp + \frac{1}{8} \Delta b^{(1)a} \Delta 
b^{(1)b} \prt_{a}\vp \prt_{b}\vp + \frac{1}{8} \Delta 
g^{(1)a} \Delta g^{(1)b} \prt_{a}\vp \prt_{b}\vp\nn\\&& + 
\frac{1}{2} \Delta b^{(1)a} \prt_{b}\vp \prt^{b}\Delta 
b^{(1)}{}_{a} -  e^{ \vp/2} V_{ab} \Delta \vp^{(1)} 
\prt^{b}\Delta b^{(1)a} + \frac{1}{2} \prt_{a}\Delta 
b^{(1)}{}_{b} \prt^{b}\Delta b^{(1)a}\nn\\&& -  \frac{1}{2} 
\prt_{b}\Delta b^{(1)}{}_{a} \prt^{b}\Delta b^{(1)a} -  
\frac{1}{2} \Delta g^{(1)a} \prt_{b}\vp \prt^{b}\Delta 
g^{(1)}{}_{a} +e^{- \vp/2}W_{ab} \Delta \vp^{(1)} \prt^{b}\Delta 
g^{(1)a}\nn\\&& + \frac{1}{2} \prt_{a}\Delta 
g^{(1)}{}_{b} \prt^{b}\Delta g^{(1)a} -  \frac{1}{2} 
\prt_{b}\Delta g^{(1)}{}_{a} \prt^{b}\Delta g^{(1)a} + 
\frac{1}{2} e^{ \vp/2} V_{ab} \Delta b^{(1)a} \Delta \vp^{(1)} \prt^{b}\vp \nn\\&&+ \frac{1}{2} e^{- \vp/2}W_{ab} \Delta g^{(1)a} \Delta \vp^{(1)} \prt^{b}\vp - \frac{1}{8} 
\Delta b^{(1)}{}_{a} \Delta b^{(1)a} \prt_{b}\vp \prt^{b}\vp -  \frac{1}{8} \Delta g^{(1)}{}_{a} \Delta g^{(1)a} \prt_{b}\vp \prt^{b}\vp\Big]\,,\nn
\eeqa
where we have  used  the fact that $\Delta \bg_{ab}^{(1)}$ and $\Delta\bphi^{(1)}$ are zero for the first-order corrections in equation \reef{dbH2}. By inserting these first-order corrections into the equation, we obtain $S^{(0,2)}_{(1,1)}(\psi_0')$, which contains no free parameters and includes both even and odd parity terms. However, since we are only interested in the even parity part of equation \reef{TS12} in this subsection, we keep only the even parity terms in the resulting $S^{(0,2)}_{(1,1)}(\psi_0')$.

To find the second term in equation \reef{S02}, we note that for the minimal couplings in equation \reef{H6}, we cannot expect $\Delta \bg_{ab}^{(2)}$ and $\Delta\bphi^{(2)}$ to be zero. Therefore, we need to keep these corrections in the second-order perturbation of equation \reef{R}. The second term in equation \reef{S02} is given by:
\beqa
S^{(0,2)}_{(2)}(\psi_0')&=& -\frac{2\alpha'^2}{\kappa'^2}\int d^{9}x  \,  e^{-2\bphi}\Big[ 
\frac{1}{8} \bH_{a}{}^{cd} \bH_{bcd} \Delta \bg^{(2)ab} - \frac{1}{48} \bH_{bcd} \bH^{bcd} \Delta \bg^{(2)a}{}_{a} \nn\\&&+ \frac{1}{4} e^{\vp} V_{a}{}^{c} V^{ab} \Delta \bg^{(2)}{}_{bc} + 
\frac{1}{4} e^{-\vp}W_{a}{}^{c} W^{ab} \Delta \bg^{(2)}{}_{bc} -  
\frac{1}{16} e^{\vp} V_{ab} V^{ab} \Delta \bg^{(2)c}{}_{c}\nn\\&& -  
\frac{1}{16 }e^{-\vp}W_{ab} W^{ab} \Delta \bg^{(2)c}{}_{c} + \frac{1}{12} 
\bH_{abc} \bH^{abc} \Delta \bphi^{(2)} + \frac{1}{4} e^{\vp} 
V_{ab} V^{ab} \Delta \bphi^{(2)} \nn\\&&+ \frac{1}{4} e^{-\vp} W_{ab} W^{ab} \Delta \bphi^{(2)}+ \frac{1}{8} e^{\vp} V_{ab} V^{ab} 
\Delta \vp^{(2)} -  \frac{1}{8 }
e^{-\vp} W_{ab} W^{ab} \Delta \vp^{(2)}\nn\\&&-  \frac{1}{2} \prt_{a}\prt^{a}\Delta \vp^{(2)} -  \Delta \bphi^{(2)} \prt_{a}\prt^{a}\vp+ \Delta \bg^{(2)b}{}_{b} \prt_{a}\bphi \prt^{a}\bphi- 4 \Delta \bphi^{(2)} \prt_{a}\bphi \prt^{a}\bphi \nn\\&& + 4 \prt_{a}\Delta 
\bphi^{(2)} \prt^{a}\bphi + \prt_{a}\Delta \vp^{(2)} 
\prt^{a}\bphi -  \frac{1}{2} \Delta \bg^{(2)b}{}_{b} 
\prt_{a}\vp\prt^{a}\bphi + 2 \Delta \bphi^{(2)} 
\prt_{a}\vp\prt^{a}\bphi\nn\\&& -  \prt_{a}\vp
\prt^{a}\Delta \bphi^{(2)} + \frac{1}{4} \prt_{a}\vp
\prt^{a}\Delta \vp^{(2)} + \frac{1}{4} \prt_{a}\Delta 
\bg^{(2)b}{}_{b} \prt^{a}\vp-  \frac{1}{16} \Delta 
\bg^{(2)b}{}_{b} \prt_{a}\vp\prt^{a}\vp\nn\\&&+ \frac{1}{4} 
\Delta \bphi^{(2)} \prt_{a}\vp\prt^{a}\vp-  
\frac{1}{2} \prt^{a}\vp\prt_{b}\Delta \bg^{(2)}{}_{a}{}^{b} 
+ \frac{1}{2} \prt_{b}\prt_{a}\Delta \bg^{(2)ab} -  
\frac{1}{2} \Delta \bg^{(2)ab} \prt_{b}\prt_{a}\vp\nn\\&&-  
\frac{1}{2} \prt_{b}\prt^{b}\Delta \bg^{(2)a}{}_{a} + 
\frac{1}{4} \Delta \bg^{(2)a}{}_{a} \prt_{b}\prt^{b}\vp- 2 
\Delta \bg^{(2)}{}_{ab} \prt^{a}\bphi \prt^{b}\bphi + 
\frac{1}{2} e^{ \vp/2} V_{ab} \prt^{b}\Delta b^{(2)a}\nn\\&& + 
\frac{1}{2} e^{-\vp/2} W_{ab} \prt^{b}\Delta g^{(2)a}-  
\frac{1}{4} e^{ \vp/2} V_{ab} \Delta b^{(2)a} \prt^{b}\vp+ \frac{1}{4} e^{- \vp/2}W_{ab} \Delta g^{(2)a} \prt^{b}\vp \nn\\&&+ \Delta \bg^{(2)}{}_{ab} \prt^{a}\bphi \prt^{b}\vp+ 
\frac{1}{8} \Delta \bg^{(2)}{}_{ab} \prt^{a}\vp
\prt^{b}\vp-  
\frac{1}{12} \bH^{abc} \Delta \bH^{(2)}{}_{abc} \Big]\,.\labell{SSS}
\eeqa
The second-order corrections $\Delta \vp^{(2)}, \dots, \Delta\bphi^{(2)}$ contain all contractions of $\partial\vp$, $\partial\bphi$, $e^{\vp/2}V$, $e^{-\vp/2}W$, $\bar{H}$, and their derivatives at order $\alpha'^2$. In this subsection, we only consider the contractions in $\Delta \vp^{(2)}, \Delta\bphi^{(2)}, \Delta\bg^{(2)}_{ab},$ and $\Delta b^{(2)}_a$ that have even parity, and the contractions in $\Delta g^{(2)}_{a}$ and $\Delta \bH^{(2)}_{abc}$ that have odd parity. This is because when these contractions are inserted into equation \reef{SSS}, they produce even parity terms in $S^{(0,2)}_{(2)}$. These corrections must satisfy the constraints in equation \reef{A11} at order $\alpha'^2$, \ie
 \beqa
-\frac{1}{2}\Delta\vp^{(2)}(\psi)+\frac{1}{2}\Delta\vp^{(2)}(\psi_0')+\Delta\vp^{(1,1)}(\psi_0')&=&0\,,\nn\\
\frac{1}{2}\Delta b_a^{(2)}(\psi)+\frac{1}{2}\Delta g_a^{(2)}(\psi'_0)+\Delta g_a^{(1,1)}(\psi_0')&=&0\,,\nn\\
\frac{1}{2}\Delta g_a^{(2)}(\psi)+\frac{1}{2}\Delta b_a^{(2)}(\psi'_0)+\Delta b_a^{(1,1)}(\psi_0')&=&0\,,\nn\\
\frac{1}{2}\Delta \bg_{ab}^{(2)}(\psi)+\frac{1}{2}\Delta \bg_{ab}^{(2)}(\psi'_0)+\Delta \bg_{ab}^{(1,1)}(\psi_0')&=&0\,,\nn\\
\frac{1}{2}\Delta\bphi^{(2)}(\psi)+\frac{1}{2}\Delta\bphi^{(2)}(\psi_0')+\Delta\bphi^{(1,1)}(\psi_0')&=&0\,,\nn\\
\frac{1}{2}\Delta\bH_{abc}^{(2)}(\psi)+\frac{1}{2}\Delta\bH_{abc}^{(2)}(\psi'_0)+\Delta\bH_{abc}^{(1,1)}(\psi_0')&=&0\,,\labell{Z2}
\eeqa
 where 
\beqa
\Delta\bH^{(2)}&=&3d\tilde B^{(2)}-3\Big[W\wedge (e^{-\vp/2}\Delta b^{(2)})+(e^{\vp/2}\Delta g^{(2)})\wedge V\nn\\
&&+2(e^{\vp/2}\Delta g^{(1)})\wedge d(e^{-\vp/2}\Delta b^{(1)})+2d(e^{\vp/2}\Delta g^{(1)})\wedge (e^{-\vp/2}\Delta b^{(1)})\Big]\,.\labell{HH2}
\eeqa 
The 2-form $\tilde{B}^{(2)}$ contains all contractions of $\partial\vp$, $\partial\bphi$, $e^{\vp/2}V$, $e^{-\vp/2}W$, $\bar{H}$, and their derivatives at order $\alpha'^2$ that have odd parity. The constraint in equation \reef{Z2} does not fix all the parameters in the second-order corrections. The unfixed parameters appear in $S^{(0,2)}_{(2)}(\psi_0')$.

The effective action at order $\alpha'$ has both even and odd parity terms. The odd parity term appears in equation \reef{baction}, and the even parity terms are the couplings in equation \reef{Mis} at order $\alpha'$. Therefore, $S^{(1)}$ has both even and odd parity terms,  \ie
\begin{equation}
S^{(1)}=S^{(1)}_o+S^{(1)}_e.
\end{equation}
Since the first-order correction $\Delta\bg^{(1)}_{ab}$ is zero, and we only need the first-order corrections in the perturbation $S^{(1,1)}$ in equation \reef{TS12}, we only need to consider the reduction of these actions when the base space is flat.

The reduction of the odd parity term for flat base space can be found in \cite{Garousi:2023pah}:
 \beqa
S^{(1)}_o&\!\!\!\!\!=\!\!\!\!\!&-\frac{3\alpha'}{\kappa'^2}\int d^{9} x  \,e^{-2\bphi}\Big[\frac{1}{24} e^{\vp} V_{a}{}^{c} V^{ab} V_{b}{}^{d} W_{cd} + 
\frac{1}{48} e^{\vp} V_{ab} V^{ab} V^{cd} W_{cd} \nn\\&&+ \frac{1}{12} e^{\vp} 
\bH_{bcd} V^{ab} \prt_{a}V^{cd} + \frac{1}{24} e^{\vp} 
\bH_{bcd} V_{a}{}^{b} V^{cd} \prt^{a}\vp  + 
\frac{1}{24} V^{bc} W_{bc} \prt_{a}\vp \prt^{a}\vp\nn\\&& -  
\frac{1}{12} W^{bc} \prt^{a}\vp \prt_{c}V_{ab} -  
\frac{1}{12} V^{ab} W_{a}{}^{c} \prt_{c}\prt_{b}\vp+ 
\frac{1}{12} e^{\vp} \bH_{bcd} V^{ab} \prt^{d}V_{a}{}^{c}\Big]\,.\labell{rCS}
\eeqa
For a detailed derivation of the reduction of the couplings in $S^{(1)}_e$, we recommend referring to \cite{Garousi:2019mca}. In this paper, the author provided a comprehensive analysis of the couplings involving the Riemann curvature, $H$, $\nabla H$, $\nabla\Phi$, and $\nabla\nabla\Phi$. The reduction is 
\beqa
S^{(1)}_e&\!\!\!\!\!=\!\!\!\!\!&-\frac{2\alpha' a_1}{\kappa'^2}\int d^{9} x\sqrt{-\bg} e^{-2\bphi}\Big[\frac{1}{144}\bH_{abc}\bH^{abc}\bH_{def}\bH^{def}+\cdots\Big]\,.
\eeqa
The ellipsis (...) denotes numerous additional terms in this reduction. To find $S^{(1,1)}(\psi_0')$, the first-order correction to $S^{(1)}$, we can straightforwardly perturb $S^{(1)}$ to first order. We then insert the first-order corrections in equation \reef{dbH2} and keep only the terms that have even parity. The resulting expression is the desired $S^{(1,1)}(\psi_0')$ in equation \reef{TS12}. Importantly, this expression has no free parameters.

Inserting all the aforementioned ingredients into equation \reef{TS12}, we obtain an equation that involves the 60 parameters in equation \reef{H6}, the unfixed parameters of the second-order corrections $\Delta \vp^{(2)}$, $\Delta\bphi^{(2)}$, $\Delta\bg^{(2)}_{ab}$, $\Delta b^{(2)}_a$, $\Delta g^{(2)}_{a}$, $\tilde{B}^{(2)}$, and the parameters in the current $J_2^a$ of the total derivative terms.
The 60 parameters in equation \reef{H6} are all independent; however, the other parameters in equation \reef{TS12} resulting from the current and the second-order corrections are not independent. Some of the corresponding couplings are related by Bianchi identities, and some of the couplings in equation \reef{TS12} resulting from the second-order corrections are related to the couplings in equation \reef{TS12} resulting from the total derivative terms. Hence, the parameters in the total derivative terms and the second-order corrections can be divided into two parts: one part includes the independent parameters, and the second part includes the dependent parameters. These latter parameters must be removed from equation \reef{TS12}. 
While equation \reef{TS12} can fix the parameters in the former set, it cannot fix the parameters in the latter set. To remove the second set of parameters, we first solve equation \reef{TS12}, and then we set all the unfixed parameters to zero.

The equation \reef{TS12} includes gauge-invariant couplings in flat base space. However, to solve this equation, one must impose various Bianchi identities that  the first derivative of $V$, $W$, and $\bH$ are satisfied. To impose these Bianchi identities, we write the field strengths $V$, $W$, and $\bH$ in terms of the derivatives of $g_a$, $b_a$, and $\bb_{ab}$. In this way, we find an equation in terms of independent couplings that are not gauge-invariant.
The fact that equation \reef{TS12} is gauge-invariant ensures that any terms that have gauge fields $g_a$ or $b_a$ without derivatives must be canceled in the equation.  Hence, to simplify the calculations, we remove the gauge fields $g_a$ and $b_a$ without derivatives from the equation. The coefficients of the remaining independent terms must be zero. This produces a set of algebraic equations that involve all the parameters mentioned above. The resulting equations can easily be solved. After solving them, we set the unfixed parameters of the total derivative terms and the second-order corrections to zero.

We have found that the algebraic equations have no solution if we consider the coupling $\Omega^2$ to be the only coupling at order $\alpha'^2$. However, when we include the 60 couplings in equation \reef{H6} in our calculations, the equations have a solution. Interestingly, we have found that the solution fixes all 60 parameters in terms of the overall number $c_1^2$. The couplings that we have found are as follows:
\beqa
\bS^{(2,3,4,5,6,7,8)}&=&\frac{2\alpha'^2 c_1^2 }{\kappa^2}\int d^{10} x\sqrt{-G} e^{-2\Phi}\Big[\frac{1}{12} \hat{H}_{\alpha }{}^{\delta \epsilon } \hat{H}^{\alpha \beta 
\gamma } \hat{H}_{\beta \delta }{}^{\zeta } \hat{H}_{\gamma }{}^{\iota 
\kappa } \hat{H}_{\epsilon \iota }{}^{\mu } \hat{H}_{\zeta \kappa \mu }\nn\\&& -  
\frac{1}{80} \hat{H}_{\alpha \beta }{}^{\delta } \hat{H}^{\alpha \beta 
\gamma } \hat{H}_{\gamma }{}^{\epsilon \zeta } \hat{H}_{\delta }{}^{\iota 
\kappa } \hat{H}_{\epsilon \zeta }{}^{\mu } \hat{H}_{\iota \kappa \mu } + 
\frac{1}{80} \hat{H}_{\alpha \beta }{}^{\delta } \hat{H}^{\alpha \beta 
\gamma } \hat{H}_{\gamma }{}^{\epsilon \zeta } \hat{H}_{\delta \epsilon 
}{}^{\iota } \hat{H}_{\zeta }{}^{\kappa \mu } \hat{H}_{\iota \kappa \mu }\nn\\&& 
-  \frac{2}{5} \hat{H}_{\alpha }{}^{\delta \epsilon } \hat{H}^{\alpha 
\beta \gamma } \hat{H}_{\beta }{}^{\zeta \iota } \hat{H}_{\delta \zeta 
}{}^{\kappa } R_{\gamma \epsilon \iota \kappa } + 2 \hat{H}_{
\alpha }{}^{\delta \epsilon } \hat{H}^{\alpha \beta \gamma } 
R_{\beta }{}^{\zeta }{}_{\delta }{}^{\iota } 
R_{\gamma \zeta \epsilon \iota }- 2 \hat{H}^{\alpha \beta 
\gamma } \hat{H}^{\delta \epsilon \zeta } R_{\alpha \beta 
\delta }{}^{\iota } R_{\gamma \iota \epsilon \zeta } \nn\\&&-  \frac{1}{40} 
\hat{H}_{\alpha \beta }{}^{\delta } \hat{H}^{\alpha \beta \gamma } 
\hat{H}_{\epsilon \zeta }{}^{\kappa } \hat{H}^{\epsilon \zeta \iota } 
R_{\gamma \iota \delta \kappa }  - 
2 \hat{H}_{\alpha }{}^{\delta \epsilon } \hat{H}^{\alpha \beta \gamma } 
R_{\beta }{}^{\zeta }{}_{\delta }{}^{\iota } 
R_{\gamma \iota \epsilon \zeta } + 2 \hat{H}_{\alpha 
}{}^{\delta \epsilon } \hat{H}^{\alpha \beta \gamma } 
R_{\beta }{}^{\zeta }{}_{\gamma }{}^{\iota } 
R_{\delta \zeta \epsilon \iota }\nn\\&& -  \frac{1}{10} 
\hat{H}_{\alpha \beta }{}^{\delta } \hat{H}^{\alpha \beta \gamma } 
\hat{H}_{\gamma }{}^{\epsilon \zeta } \hat{H}_{\epsilon }{}^{\iota \kappa } 
R_{\delta \iota \zeta \kappa } -  \frac{8}{5} 
\hat{H}_{\alpha }{}^{\delta \epsilon } \hat{H}^{\alpha \beta \gamma } 
\hat{H}_{\beta \delta }{}^{\zeta } \hat{H}_{\gamma }{}^{\iota \kappa } 
R_{\epsilon \iota \zeta \kappa }\nn\\&& -  \frac{1}{20} 
\hat{H}_{\alpha \beta }{}^{\delta } \hat{H}^{\alpha \beta \gamma } 
\hat{H}_{\gamma }{}^{\epsilon \zeta } \hat{H}_{\delta }{}^{\iota \kappa } 
R_{\epsilon \iota \zeta \kappa } -  \frac{1}{20} 
\hat{H}_{\alpha }{}^{\gamma \delta } \hat{H}_{\beta \gamma }{}^{\epsilon } 
\hat{H}_{\delta }{}^{\zeta \iota } \hat{H}_{\epsilon \zeta \iota } 
\nabla^{\beta }\nabla^{\alpha }\Phi\nn\\&& -  \frac{1}{10} \hat{H}_{\gamma 
\delta }{}^{\zeta } \hat{H}^{\gamma \delta \epsilon } 
R_{\alpha \epsilon \beta \zeta } \nabla^{\beta 
}\nabla^{\alpha }\Phi -  \frac{2}{5} \hat{H}_{\alpha }{}^{\gamma 
\delta } \hat{H}_{\gamma }{}^{\epsilon \zeta } R_{\beta 
\epsilon \delta \zeta } \nabla^{\beta }\nabla^{\alpha }\Phi\nn\\&& + 
\frac{1}{5} \hat{H}_{\alpha }{}^{\gamma \delta } \hat{H}_{\beta 
}{}^{\epsilon \zeta } R_{\gamma \epsilon \delta \zeta 
} \nabla^{\beta }\nabla^{\alpha }\Phi -  \frac{1}{10} 
\nabla^{\beta }\nabla^{\alpha }\Phi \nabla_{\epsilon 
}\hat{H}_{\beta \gamma \delta } \nabla^{\epsilon }\hat{H}_{\alpha 
}{}^{\gamma \delta } \nn\\&&-  \frac{1}{20} \hat{H}_{\beta \gamma 
}{}^{\epsilon } \hat{H}^{\beta \gamma \delta } \hat{H}_{\delta }{}^{\zeta 
\iota } \nabla^{\alpha }\Phi \nabla_{\iota }\hat{H}_{\alpha 
\epsilon \zeta } + \frac{1}{40} \hat{H}_{\alpha }{}^{\beta \gamma } 
\hat{H}_{\delta \epsilon }{}^{\iota } \hat{H}^{\delta \epsilon \zeta } 
\nabla^{\alpha }\Phi \nabla_{\iota }\hat{H}_{\beta \gamma \zeta } \nn\\&&- 
 \frac{1}{20} \hat{H}_{\alpha }{}^{\delta \epsilon } \hat{H}^{\alpha \beta 
\gamma } \nabla_{\iota }\hat{H}_{\delta \epsilon \zeta } 
\nabla^{\iota }\hat{H}_{\beta \gamma }{}^{\zeta } -  \frac{1}{5} \hat{H}_{
\alpha }{}^{\delta \epsilon } \hat{H}^{\alpha \beta \gamma } 
\nabla_{\zeta }\hat{H}_{\gamma \epsilon \iota } \nabla^{\iota 
}\hat{H}_{\beta \delta }{}^{\zeta }\nn\\&& + \frac{1}{5} \hat{H}_{\alpha 
}{}^{\delta \epsilon } \hat{H}^{\alpha \beta \gamma } \nabla_{\iota 
}\hat{H}_{\gamma \epsilon \zeta } \nabla^{\iota }\hat{H}_{\beta \delta 
}{}^{\zeta } + \frac{1}{5} \hat{H}_{\alpha \beta }{}^{\delta } 
\hat{H}^{\alpha \beta \gamma } \nabla_{\zeta }\hat{H}_{\delta \epsilon 
\iota } \nabla^{\iota }\hat{H}_{\gamma }{}^{\epsilon \zeta }\nn\\&& -  
\frac{3}{40} \hat{H}_{\alpha \beta }{}^{\delta } \hat{H}^{\alpha \beta 
\gamma } \nabla_{\iota }\hat{H}_{\delta \epsilon \zeta } 
\nabla^{\iota }\hat{H}_{\gamma }{}^{\epsilon \zeta }
\Big]\,.\labell{final}
\eeqa
Note that the coefficients of the Riemann cubed terms become zero, which is consistent with the S-matrix \cite{Metsaev:1986yb}. The algebraic equations also fix the parameters in the second-order corrections to the Buscher rules and the parameters of the total derivative terms at order $\alpha'^2$ in terms of $c_1^2$. However, since these couplings are correct only for closed spacetime manifolds, the total derivative terms are not important, so we do not write them. The solution for the second-order corrections is presented in the Appendix. These corrections are required if one would like to study the T-duality of couplings at order $\alpha'^3$, which is beyond the scope of this paper.

\subsubsection{Odd-parity couplings}

The odd-parity couplings in \reef{Mis}  at order $\alpha'^2$ involving $\Omega$ are listed below:
\beqa
\bS^{(2)}&=&-\frac{3\alpha'^2 c_1}{\kappa^2}\int d^{10} x\sqrt{-G} e^{-2\Phi}\Big[- \frac{1}{2} H_{\gamma \delta }{}^{\epsilon } R^{\alpha 
\beta \gamma \delta } \Omega_{\alpha \beta \epsilon } + 
H_{\gamma }{}^{\delta \epsilon } R^{\alpha \beta }{}_{\alpha 
}{}^{\gamma } \Omega_{\beta \delta \epsilon } -  \frac{1}{3} 
H^{\gamma \delta \epsilon } R^{\alpha \beta }{}_{\alpha \beta 
} \Omega_{\gamma \delta \epsilon } \nn\\&&\qquad\qquad\qquad+ H_{\beta }{}^{\delta 
\epsilon } R^{\alpha \beta }{}_{\alpha }{}^{\gamma } 
\Omega_{\gamma \delta \epsilon } -  \frac{1}{2} H_{\alpha 
\beta }{}^{\epsilon } R^{\alpha \beta \gamma \delta } \Omega_{
\gamma \delta \epsilon } + \frac{1}{6} H_{\alpha }{}^{\delta 
\epsilon } H^{\alpha \beta \gamma } H_{\beta \delta 
}{}^{\varepsilon } \Omega_{\gamma \epsilon \varepsilon } \nn\\&&\qquad\qquad\qquad-  
\frac{1}{2} H_{\alpha \beta }{}^{\delta } H^{\alpha \beta 
\gamma } H_{\gamma }{}^{\epsilon \varepsilon } \Omega_{\delta 
\epsilon \varepsilon } + \frac{1}{36} H_{\alpha \beta \gamma } 
H^{\alpha \beta \gamma } H^{\delta \epsilon \varepsilon } 
\Omega_{\delta \epsilon \varepsilon } -  \frac{4}{3} H^{\beta 
\gamma \delta } \Omega_{\beta \gamma \delta } \nabla_{\alpha 
}\nabla^{\alpha }\Phi \nn\\&&\qquad\qquad\qquad+ \frac{4}{3} H^{\beta \gamma \delta } 
\Omega_{\beta \gamma \delta } \nabla_{\alpha }\Phi 
\nabla^{\alpha }\Phi + 4 H_{\alpha }{}^{\gamma \delta } 
\Omega_{\beta \gamma \delta } \nabla^{\beta }\nabla^{\alpha }
\Phi \Big]\,.\labell{Misodd}
\eeqa
If we reduce the aforementioned couplings on a circle, apply the Buscher rules \reef{T2} to the resulting reduction, insert them into the constraint \reef{TS12}, and account for all the odd-parity terms in $-S^{(0,2)}(\psi'_0)-S^{(1,1)}(\psi'_0)$ that were previously neglected in the subsection 2.2.1, as well as include all odd-parity total derivative terms at order $\alpha'^2$, we would find that they  do not satisfy the constraint  \reef{TS12}. Therefore, there must exist other odd-parity couplings at this order that do not involve $\Omega$.

In accordance with \cite{Garousi:2019cdn}, one can determine all the independent odd-parity geometrical couplings at order $\alpha'^2$. There are a total of 13 independent couplings, listed as follows:
\beqa
\bS_o^{(2)}&=&-\frac{2\alpha'^2 }{\kappa^2}\int d^{10} x\sqrt{-G} e^{-2\Phi}\Big[a_{1}  
   H_{\beta  \gamma  }{}^{\epsilon  } H^{\beta  \gamma  
\delta  } H_{\delta  }{}^{\varepsilon  \mu  } R_{\alpha 
 \varepsilon  \epsilon  \mu  } \nabla^{\alpha  }\Phi  +a_{2}  
   H^{\beta  \gamma  \delta  } R_{\alpha  
}{}^{\epsilon  }{}_{\beta  }{}^{\varepsilon  } R_{\gamma 
 \epsilon  \delta  \varepsilon  } \nabla^{\alpha  }\Phi  \nn\\&&\qquad\qquad\qquad\qquad+   a_{3}  
   H_{\alpha  }{}^{\beta  \gamma  } H_{\beta  }{}^{\delta  
\epsilon  } H_{\delta  }{}^{\varepsilon  \mu  } 
R_{\gamma  \varepsilon  \epsilon  \mu  } 
\nabla^{\alpha  }\Phi  +a_{4}  
   H^{\alpha  \beta  \gamma  } H^{\delta  \epsilon  
\varepsilon  } R_{\gamma  \epsilon  \varepsilon  \mu  } 
\nabla_{\beta  }H_{\alpha  \delta  }{}^{\mu  } \nn\\&&\qquad\qquad\qquad\qquad+a_{5} 
    H_{\alpha  \beta  }{}^{\delta  } H_{\gamma  }{}^{\epsilon  
\varepsilon  } H_{\delta  \epsilon  \varepsilon  } 
\nabla^{\alpha  }\Phi  \nabla^{\gamma  }\nabla^{\beta  }\Phi  
+a_{6}  
   H_{\beta  }{}^{\delta  \epsilon  } R_{\alpha  
\delta  \gamma  \epsilon  } \nabla^{\alpha  }\Phi  
\nabla^{\gamma  }\nabla^{\beta  }\Phi \nn\\&&\qquad\qquad\qquad\qquad +a_{7}  
   H_{\alpha  }{}^{\gamma  \delta  } H_{\gamma  }{}^{\epsilon  
\varepsilon  } \nabla^{\alpha  }\Phi  \nabla^{\beta  }\Phi  
\nabla_{\varepsilon  }H_{\beta  \delta  \epsilon  } +a_{8}  
   H_{\alpha  }{}^{\gamma  \delta  } H_{\gamma  }{}^{\epsilon  
\varepsilon  } \nabla^{\beta  }\nabla^{\alpha  }\Phi  \nabla_{
\varepsilon  }H_{\beta  \delta  \epsilon  }\nn\\&&\qquad\qquad\qquad\qquad +a_{9}  
   H_{\alpha  \beta  }{}^{\delta  } H^{\alpha  \beta  \gamma  
} H_{\gamma  }{}^{\epsilon  \varepsilon  } H_{\epsilon  }{}^{\mu 
 \zeta  } \nabla_{\varepsilon  }H_{\delta  \mu  \zeta  } +
 a_{12}  
   H_{\alpha  \beta  }{}^{\delta  } H^{\alpha  \beta  \gamma  
} H_{\gamma  }{}^{\epsilon  \varepsilon  } H_{\epsilon  }{}^{\mu 
 \zeta  } \nabla_{\zeta  }H_{\delta  \varepsilon  \mu  } \nn\\&&\qquad\qquad\qquad\qquad+a_{13}  
   H_{\alpha  }{}^{\delta  \epsilon  } H^{\alpha  \beta  
\gamma  } H_{\beta  \delta  }{}^{\varepsilon  } H_{\gamma  }{}^{
\mu  \zeta  } \nabla_{\zeta  }H_{\epsilon  \varepsilon  \mu  } 
+a_{10}  
   H_{\alpha  }{}^{\delta  \epsilon  } H^{\alpha  \beta  
\gamma  } R_{\delta  \varepsilon  \epsilon  \mu  } 
\nabla^{\mu  }H_{\beta  \gamma  }{}^{\varepsilon  } \nn\\&&\qquad\qquad\qquad\qquad+a_{11}  
   H_{\alpha  \beta  }{}^{\delta  } H^{\alpha  \beta  \gamma  
} R_{\delta  \mu  \epsilon  \varepsilon  } \nabla^{\mu 
 }H_{\gamma  }{}^{\epsilon  \varepsilon}\Big]\,.\labell{allodd}
\eeqa
Here, $a_1,\cdots,a_{13}$ are background-independent parameters that cannot be fixed by the gauge symmetries.

By incorporating the aforementioned action into the action \reef{Misodd} and following the aforementioned steps, it becomes evident that three parameters in \reef{allodd} exhibit non-zero values. That is 
\beqa
\bS^{(2,3,4,5,6,7)}&=&-\frac{2\alpha'^2 c_1^2}{\kappa^2}\int d^{10} x\sqrt{-G} e^{-2\Phi}\Big[ 4\hat{H}^{\alpha \beta \gamma } \hat{H}^{\delta \epsilon 
\varepsilon } R_{\gamma \epsilon \varepsilon \mu } 
\nabla_{\beta }\hat{H}_{\alpha \delta }{}^{\mu } -  2 
\hat{H}_{\alpha }{}^{\delta \epsilon } \hat{H}^{\alpha \beta \gamma } 
R_{\delta \varepsilon \epsilon \mu } \nabla^{\mu 
}\hat{H}_{\beta \gamma }{}^{\varepsilon }\nn\\&&\qquad\qquad\qquad\qquad\qquad\quad  -  \frac{1}{2} \hat{H}_{
\alpha }{}^{\delta \epsilon }\hat{H}^{\alpha \beta \gamma } 
\hat{H}_{\beta \delta }{}^{\varepsilon } \hat{H}_{\gamma }{}^{\mu \zeta } 
\nabla_{\zeta }\hat{H}_{\epsilon \varepsilon \mu }\Big]\,.\labell{finalodd}
\eeqa
 Furthermore, in the aforementioned context, we have substituted $H$ with $\hat{H}$. In addition, there are corresponding deformations of the Buscher rules similar to the ones outlined in the Appendix, as well as the inclusion of total derivative terms. However, as these details do not provide significant insight, we have chosen not to explicitly present them.
 
\section{Discussion}

In this paper, we have demonstrated that the Chern-Simons $\Omega^2$-term in the effective action of heterotic string theory, at order $\alpha'^2$, does not possess invariance under T-duality transformations. We have examined the Buscher rules, along with their $\alpha'$-corrections, which satisfy the $O(1,1,\mathbb{Z})$-group. However, the requirement that the generalized Buscher rules adhere to the $O(1,1,\mathbb{Z})$-group does not determine the parameters within these rules. Irrespective of the chosen values for these parameters, the $\Omega^2$-term fails to exhibit invariance under $O(1,1,\mathbb{Z})$ transformations.

To address this predicament, we introduce 60 independent NS-NS couplings at order $\alpha'^2$, each with arbitrary coefficients, to the $\Omega^2$-term. We then impose the condition that these even-parity couplings remain invariant under $O(1,1,\mathbb{Z})$ transformations. The investigation also necessitates the inclusion of the effective action at order $\alpha'$. We employ this action within the Miessner scheme, which allows us to determine all 60 parameters, as well as the independent parameters in the $O(1,1,\mathbb{Z})$ transformations, at order $\alpha'^2$. The resulting 60 couplings are presented in equation \reef{final}, while the corrections to the Buscher rules at order $\alpha'^2$ are provided in the Appendix.
In \cite{Garousi:2019cdn}, the 60 independent couplings at order $\alpha'$ were derived using the most general field redefinitions permissible solely for closed spacetime manifolds \cite{Garousi:2021yyd}. Consequently, the couplings presented in equation \reef{final} are only applicable to closed spacetime manifolds. Thus, we have disregarded the total derivative terms associated with ensuring the invariance of the effective action under $O(1,1,\mathbb{Z})$ transformations. 

Additionally, we have observed that the odd-parity couplings in the Miessner action \reef{Mis} at order $\alpha'^2$ are incompatible with T-duality. To resolve this issue, we introduce 13 independent odd-parity couplings, also at order $\alpha'^2$, with arbitrary coefficients to the Miessner action. We then enforce the invariance of these couplings under $O(1,1,\mathbb{Z})$ transformations. This condition allows us to determine all 13 parameters, as well as the independent parameters in the corresponding $O(1,1,\mathbb{Z})$ transformations, at order $\alpha'^2$. The couplings are presented in equation \reef{finalodd}.

Similar calculations to those in subsection 2.2.1 were carried out in \cite{Garousi:2019mca} to find the effective action of the bosonic string theory at order $\alpha'^2$. However, the effective action and the corrections to the Buscher rules at order $\alpha'$ in that paper are different from those in equations \reef{Mis} and \reef{dbH2}. In that work, the effective action in the Metsaev-Tseytlin scheme \cite{Metsaev:1987zx} and its corresponding $O(1,1,\MZ)$-transformations at order $\alpha'$, which were found in \cite{Garousi:2019wgz}, were used. 
It is known that the effective action at order $\alpha'^2$ depends on the scheme used for the effective action at order $\alpha'$ \cite{Bento1990}. If we set the coefficient of $\Omega$ to zero in our calculation, then our results give the effective action for the bosonic string theory at order $\alpha'^2$ when using the effective action at order $\alpha'$ in the Meissner scheme \reef{Mis} and its corresponding $O(1,1,\MZ)$-transformations at order $\alpha'$ given in equation \reef{dbH2}. Our calculations yield the following action for the bosonic string theory:
\beqa
&&\bS^{(2)}=\frac{2\alpha'^2 c_1^2 }{\kappa^2}\int d^{26} x\sqrt{-G} e^{-2\Phi}\Big[\frac{1}{12} H_{\alpha }{}^{\delta \epsilon } H^{\alpha \beta 
\gamma } H_{\beta \delta }{}^{\zeta } H_{\gamma }{}^{\iota 
\kappa } H_{\epsilon \iota }{}^{\mu } H_{\zeta \kappa \mu }\labell{Sb}\\&& -  
\frac{1}{30} H_{\alpha \beta }{}^{\delta } H^{\alpha \beta 
\gamma } H_{\gamma }{}^{\epsilon \zeta } H_{\delta }{}^{\iota 
\kappa } H_{\epsilon \zeta }{}^{\mu } H_{\iota \kappa \mu } -  
\frac{1}{20} H_{\alpha \beta }{}^{\delta } H^{\alpha \beta 
\gamma } H_{\gamma }{}^{\epsilon \zeta } H_{\delta \epsilon 
}{}^{\iota } H_{\zeta }{}^{\kappa \mu } H_{\iota \kappa \mu } \nn\\&&
+ \frac{4}{3} R_{\alpha }{}^{\epsilon }{}_{\gamma }{}^{
\zeta } R^{\alpha \beta \gamma \delta } 
R_{\beta \zeta \delta \epsilon } -  \frac{4}{3} 
R_{\alpha \beta }{}^{\epsilon \zeta } 
R^{\alpha \beta \gamma \delta } R_{\gamma 
\epsilon \delta \zeta } -  \frac{2}{5} H_{\alpha }{}^{\delta 
\epsilon } H^{\alpha \beta \gamma } H_{\beta }{}^{\zeta \iota 
} H_{\delta \zeta }{}^{\kappa } R_{\gamma \epsilon 
\iota \kappa } \nn\\&&+ 2 H_{\alpha }{}^{\delta \epsilon } H^{\alpha 
\beta \gamma } R_{\beta }{}^{\zeta }{}_{\delta 
}{}^{\iota } R_{\gamma \zeta \epsilon \iota } -  
\frac{3}{20} H_{\alpha \beta }{}^{\delta } H^{\alpha \beta 
\gamma } H_{\epsilon \zeta }{}^{\kappa } H^{\epsilon \zeta 
\iota } R_{\gamma \iota \delta \kappa } - 2 H^{\alpha 
\beta \gamma } H^{\delta \epsilon \zeta } R_{\alpha 
\beta \delta }{}^{\iota } R_{\gamma \iota \epsilon 
\zeta }\nn\\&& - 2 H_{\alpha }{}^{\delta \epsilon } H^{\alpha \beta 
\gamma } R_{\beta }{}^{\zeta }{}_{\delta }{}^{\iota } 
R_{\gamma \iota \epsilon \zeta } + 2 H_{\alpha 
}{}^{\delta \epsilon } H^{\alpha \beta \gamma } 
R_{\beta }{}^{\zeta }{}_{\gamma }{}^{\iota } 
R_{\delta \zeta \epsilon \iota } + H_{\alpha \beta 
}{}^{\delta } H^{\alpha \beta \gamma } R_{\gamma 
}{}^{\epsilon \zeta \iota } R_{\delta \zeta \epsilon 
\iota }\nn\\&& -  \frac{3}{5} H_{\alpha \beta }{}^{\delta } H^{\alpha 
\beta \gamma } H_{\gamma }{}^{\epsilon \zeta } H_{\epsilon 
}{}^{\iota \kappa } R_{\delta \iota \zeta \kappa } -  
\frac{8}{5} H_{\alpha }{}^{\delta \epsilon } H^{\alpha \beta 
\gamma } H_{\beta \delta }{}^{\zeta } H_{\gamma }{}^{\iota 
\kappa } R_{\epsilon \iota \zeta \kappa } + 
\frac{1}{5} H_{\alpha \beta }{}^{\delta } H^{\alpha \beta 
\gamma } H_{\gamma }{}^{\epsilon \zeta } H_{\delta }{}^{\iota 
\kappa } R_{\epsilon \iota \zeta \kappa }\nn\\&& -  
\frac{3}{10} H_{\alpha }{}^{\gamma \delta } H_{\beta \gamma 
}{}^{\epsilon } H_{\delta }{}^{\zeta \iota } H_{\epsilon \zeta 
\iota } \nabla^{\beta }\nabla^{\alpha }\Phi -  \frac{3}{5} 
H_{\gamma \delta }{}^{\zeta } H^{\gamma \delta \epsilon } 
R_{\alpha \epsilon \beta \zeta } \nabla^{\beta 
}\nabla^{\alpha }\Phi -  \frac{12}{5} H_{\alpha }{}^{\gamma 
\delta } H_{\gamma }{}^{\epsilon \zeta } R_{\beta 
\epsilon \delta \zeta } \nabla^{\beta }\nabla^{\alpha }\Phi\nn\\&& + 
\frac{6}{5} H_{\alpha }{}^{\gamma \delta } H_{\beta 
}{}^{\epsilon \zeta } R_{\gamma \epsilon \delta \zeta 
} \nabla^{\beta }\nabla^{\alpha }\Phi -  \frac{3}{5} \nabla^{
\beta }\nabla^{\alpha }\Phi \nabla_{\epsilon }H_{\beta \gamma 
\delta } \nabla^{\epsilon }H_{\alpha }{}^{\gamma \delta } -  
\frac{3}{10} H_{\beta \gamma }{}^{\epsilon } H^{\beta \gamma 
\delta } H_{\delta }{}^{\zeta \iota } \nabla^{\alpha }\Phi 
\nabla_{\iota }H_{\alpha \epsilon \zeta }\nn\\&& + \frac{3}{20} 
H_{\alpha }{}^{\beta \gamma } H_{\delta \epsilon }{}^{\iota } 
H^{\delta \epsilon \zeta } \nabla^{\alpha }\Phi \nabla_{\iota 
}H_{\beta \gamma \zeta } -  \frac{1}{20} H_{\alpha }{}^{\delta 
\epsilon } H^{\alpha \beta \gamma } \nabla_{\iota }H_{\delta 
\epsilon \zeta } \nabla^{\iota }H_{\beta \gamma }{}^{\zeta } - 
 \frac{1}{5} H_{\alpha }{}^{\delta \epsilon } H^{\alpha \beta 
\gamma } \nabla_{\zeta }H_{\gamma \epsilon \iota } 
\nabla^{\iota }H_{\beta \delta }{}^{\zeta } \nn\\&&+ \frac{1}{5} 
H_{\alpha }{}^{\delta \epsilon } H^{\alpha \beta \gamma } 
\nabla_{\iota }H_{\gamma \epsilon \zeta } \nabla^{\iota 
}H_{\beta \delta }{}^{\zeta } + \frac{1}{5} H_{\alpha \beta 
}{}^{\delta } H^{\alpha \beta \gamma } \nabla_{\zeta 
}H_{\delta \epsilon \iota } \nabla^{\iota }H_{\gamma 
}{}^{\epsilon \zeta } -  \frac{1}{5} H_{\alpha \beta 
}{}^{\delta } H^{\alpha \beta \gamma } \nabla_{\iota 
}H_{\delta \epsilon \zeta } \nabla^{\iota }H_{\gamma 
}{}^{\epsilon \zeta }\Big]\,.\nn
\eeqa
where $c_1=1/4$. 
Note that the coefficients of the Riemann cubed terms in the action presented above are non-zero. This action should be related to the action found in \cite{Garousi:2019mca} by field redefinitions that transform the Meissner action \reef{Mis} into the Metsaev-Tseytlin action \cite{Metsaev:1987zx}.

The corrections to the Buscher rules at order $\alpha'^2$ that we have derived have non-zero $\Delta \bg_{ab}^{(2)}$ (see the Appendix). However, in the presence of a boundary, one expects the unit vector to the boundary in the string frame and its length to be invariant under T-duality transformations at all orders of $\alpha'$. Therefore, $\Delta \bg_{ab}^{(n)}$ should be zero for spacetime manifolds with a boundary. This implies that the geometrical couplings at order $\alpha'^2$ in the presence of a boundary should incorporate the 60 even-parity couplings in equation \reef{H6}, the 13 odd-parity couplings in \reef{allodd}, as well as additional couplings. These additional couplings are necessary to ensure that the invariance under T-duality results in a vanishing $\Delta \bg_{ab}^{(2)}$. It has been observed in \cite{Garousi:2021yyd} that in the presence of a boundary, one cannot use the most general field redefinitions and must use only restricted field redefinitions, such as those that leave the metric unchanged. If one uses these restricted field redefinitions, then the T-duality transformations should produce zero $\Delta \bg_{ab}^{(n)}$, though they may produce non-zero $\Delta\bphi^{(n)}$. On the other hand, if one requires both the metric and dilaton to be invariant under the restricted field redefinition, \ie  requiring the unit vector in the string frame and in the Einstein frame to be invariant under the field redefinitions, then the measure $e^{-2\bphi}\sqrt{-\bg}$ remains invariant under T-duality at all orders of $\alpha'$, \ie $\Delta\bphi^{(n)}=0$. It would be interesting to find the geometrical couplings when there is only a field redefinition for the $B$-field and use them for the couplings in equation \reef{H6}. In that case, one may find the corrections at order $\alpha'^2$ to have zero $\Delta \bg_{ab}^{(2)}$ and $\Delta\bphi^{(2)}$. The non-zero corrections $\Delta \vp^{(2)},\cdots,\tilde{B}_{ab}^{(2)}$ and the terms of the total derivatives may then be used to find the corresponding boundary couplings at order $\alpha'^2$ by T-duality. We leave the details of this calculation for future work.

\section{Appendix}

In this appendix we write the second order corrections to the Buscher rules corresponding to the effective action \reef{final} in the heterotic string theory. $\Delta\bphi^{(2)}/c_1^2$ is 
\beqa
&&\!\!\!\!\!- \frac{1}{40} e^{\vp} \bH_{b}{}^{de} \bH_{cde} V_{a}{}^{c} V^{ab} 
-  \frac{1}{20} e^{2 \vp} V_{a}{}^{c} V^{ab} V_{b}{}^{d} V_{cd} + 
\frac{159}{80} e^{\vp} \bH_{ab}{}^{e} \bH_{cde} V^{ab} V^{cd} -  
\frac{129}{40} e^{2 \vp} V_{ab} V^{ab} V_{cd} V^{cd} \nn\\&&\!\!\!\!\!+ 
\frac{1}{40} e^{- \vp} \bH_{b}{}^{de} \bH_{cde} W_{a}{}^{c} 
W^{ab} + \frac{1}{20} e^{-2 \vp} W_{a}{}^{c} W^{ab} 
W_{b}{}^{d} W_{cd} -  \frac{159}{80} e^{- \vp} \bH_{ab}{}^{e} 
\bH_{cde} W^{ab} W^{cd}  \nn\\&&\!\!\!\!\!+ \frac{129}{40} e^{-2 \vp} W_{ab} 
W^{ab} W_{cd} W^{cd}-  \frac{13}{64} \bH_{bcd} \bH^{bcd} \prt_{a}
\prt^{a}\vp+ \frac{64}{5} \bH_{acd} V^{bc} W_{b}{}^{d} 
\prt^{a}\bphi -  \frac{117}{20} \bH_{bcd} V_{a}{}^{b} W^{cd} 
\prt^{a}\bphi \nn\\&&\!\!\!\!\!-  \frac{39}{40} \bH^{bcd} \prt_{a}\bH_{bcd} 
\prt^{a}\bphi + \frac{219}{20} e^{\vp} V_{bc} V^{bc} 
\prt_{a}\bphi \prt^{a}\bphi-  \frac{219}{20} e^{- \vp} 
W_{bc} W^{bc} \prt_{a}\bphi \prt^{a}\bphi + \frac{13}{32} 
\bH_{bcd} \bH^{bcd} \prt_{a}\vp \prt^{a}\bphi \nn\\&&\!\!\!\!\! -  
\frac{1583}{80} e^{\vp} V_{bc} V^{bc} \prt_{a}\vp 
\prt^{a}\bphi -  \frac{1583}{80} e^{- \vp} W_{bc} W^{bc} 
\prt_{a}\vp \prt^{a}\bphi + \frac{677}{160} e^{\vp} 
V_{bc} V^{bc} \prt_{a}\vp \prt^{a}\vp \nn\\&&\!\!\!\!\!-  \frac{677}{160} 
e^{- \vp} W_{bc} W^{bc} \prt_{a}\vp \prt^{a}\vp -  
\frac{2323}{20} \prt_{a}\vp \prt^{a}\bphi 
\prt_{b}\prt^{b}\bphi + \frac{2323}{40} 
\prt_{a}\prt^{a}\bphi \prt_{b}\prt^{b}\vp -  
\frac{1313}{20} \prt_{a}\bphi \prt^{a}\bphi 
\prt_{b}\prt^{b}\vp \nn\\&&\!\!\!\!\! -  \frac{4061}{320} \prt_{a}\vp 
\prt^{a}\vp \prt_{b}\prt^{b}\vp+ \frac{4247}{40} 
\prt^{a}\vp \prt_{b}\prt^{b}\prt_{a}\bphi + 
\frac{729}{20} \prt^{a}\bphi 
\prt_{b}\prt^{b}\prt_{a}\vp -  \frac{509}{80} 
\prt_{b}\prt^{b}\prt_{a}\prt^{a}\vp\nn\\&&\!\!\!\!\! + \frac{743}{10} 
e^{\vp} V_{a}{}^{c} V_{bc} \prt^{a}\bphi \prt^{b}\bphi -  
\frac{743}{10} e^{- \vp} W_{a}{}^{c} W_{bc} \prt^{a}\bphi 
\prt^{b}\bphi + \frac{1313}{10} \prt_{a}\bphi 
\prt^{a}\bphi \prt_{b}\vp \prt^{b}\bphi -  
\frac{949}{20} \prt^{a}\bphi \prt_{b}\prt_{a}\vp 
\prt^{b}\bphi \nn\\&&\!\!\!\!\!-  \frac{1}{8} \bH_{a}{}^{cd} \bH_{bcd} 
\prt^{a}\bphi \prt^{b}\vp + \frac{309}{40} e^{\vp} 
V_{a}{}^{c} V_{bc} \prt^{a}\bphi \prt^{b}\vp + 
\frac{309}{40} e^{- \vp} W_{a}{}^{c} W_{bc} \prt^{a}\bphi 
\prt^{b}\vp -  \frac{251}{40} e^{\vp} V_{a}{}^{c} V_{bc} 
\prt^{a}\vp \prt^{b}\vp \nn\\&&\!\!\!\!\!+ \frac{251}{40} e^{- \vp} 
W_{a}{}^{c} W_{bc} \prt^{a}\vp \prt^{b}\vp + 
\frac{6647}{160} \prt_{a}\vp \prt^{a}\bphi 
\prt_{b}\vp \prt^{b}\vp -  \frac{1489}{4} 
\prt^{a}\bphi \prt_{b}\prt_{a}\bphi \prt^{b}\vp -  
\frac{39}{4} \prt^{a}\vp \prt_{b}\prt_{a}\vp 
\prt^{b}\vp \nn\\&&\!\!\!\!\!+ \frac{503}{20} \prt_{b}\prt_{a}\vp 
\prt^{b}\prt^{a}\bphi + \frac{9}{80} \bH_{a}{}^{cd} \bH_{bcd} 
\prt^{b}\prt^{a}\vp + \frac{359}{10} e^{\vp} V^{bc} 
\prt^{a}\bphi \prt_{c}V_{ab} -  \frac{141}{40} e^{\vp} 
V^{bc} \prt^{a}\vp \prt_{c}V_{ab} \nn\\&&\!\!\!\!\!+ \frac{467}{20} 
e^{\vp} \prt_{a}V^{ab} \prt_{c}V_{b}{}^{c} -  
\frac{191}{20} e^{\vp} V_{a}{}^{b} \prt^{a}\bphi 
\prt_{c}V_{b}{}^{c} + \frac{237}{8} e^{\vp} V_{a}{}^{b} 
\prt^{a}\vp \prt_{c}V_{b}{}^{c} -  \frac{359}{10} e^{- 
\vp} W^{bc} \prt^{a}\bphi \prt_{c}W_{ab}\nn\\&&\!\!\!\!\! -  \frac{141}{40} 
e^{- \vp} W^{bc} \prt^{a}\vp \prt_{c}W_{ab} -  
\frac{467}{20} e^{- \vp} \prt_{a}W^{ab} 
\prt_{c}W_{b}{}^{c} + \frac{191}{20} e^{- \vp} W_{a}{}^{b} 
\prt^{a}\bphi \prt_{c}W_{b}{}^{c} + \frac{237}{8} e^{- 
\vp} W_{a}{}^{b} \prt^{a}\vp \prt_{c}W_{b}{}^{c} \nn\\&&\!\!\!\!\!+ 
\frac{196}{5} \bH_{abd} V^{ab} \prt_{c}W^{cd} + \frac{109}{10} 
e^{\vp} V_{a}{}^{c} V^{ab} \prt_{c}\prt_{b}\bphi -  
\frac{109}{10} e^{- \vp} W_{a}{}^{c} W^{ab} 
\prt_{c}\prt_{b}\bphi -  \frac{377}{40} e^{\vp} 
V_{a}{}^{c} V^{ab} \prt_{c}\prt_{b}\vp\nn\\&&\!\!\!\!\! -  \frac{377}{40} 
e^{- \vp} W_{a}{}^{c} W^{ab} \prt_{c}\prt_{b}\vp -  
\frac{219}{40} e^{\vp} V_{ab} V^{ab} \prt_{c}\prt^{c}\bphi 
+ \frac{219}{40} e^{- \vp} W_{ab} W^{ab} 
\prt_{c}\prt^{c}\bphi + \frac{7}{2} e^{\vp} V^{ab} 
\prt_{c}\prt^{c}V_{ab}\nn\\&&\!\!\!\!\! -  \frac{7}{2} e^{- \vp} W^{ab} 
\prt_{c}\prt^{c}W_{ab} + \frac{1137}{160} e^{\vp} V_{ab} 
V^{ab} \prt_{c}\prt^{c}\vp + \frac{1137}{160} e^{- \vp} 
W_{ab} W^{ab} \prt_{c}\prt^{c}\vp + \frac{251}{40} 
e^{\vp} \prt_{c}V_{ab} \prt^{c}V^{ab} \nn\\&&\!\!\!\!\!-  \frac{251}{40} 
e^{- \vp} \prt_{c}W_{ab} \prt^{c}W^{ab} -  \frac{191}{20} 
V^{ab} W^{cd} \prt_{d}\bH_{abc} + \frac{117}{40} \bH^{bcd} 
\prt^{a}\bphi \prt_{d}\bH_{abc} -  \frac{181}{80} \bH^{bcd} 
\prt^{a}\vp \prt_{d}\bH_{abc} \nn\\&&\!\!\!\!\!-  \frac{32}{5} V^{ab} 
W_{a}{}^{c} \prt_{d}\bH_{bc}{}^{d} + \frac{1303}{160} 
\bH_{a}{}^{bc} \prt^{a}\vp \prt_{d}\bH_{bc}{}^{d} -  
\frac{98}{5} \bH^{abc} \prt_{d}\prt_{c}\bH_{ab}{}^{d} + 
\frac{98}{15} \bH^{abc} \prt_{d}\prt^{d}\bH_{abc} \nn\\&&\!\!\!\!\!-  
\frac{191}{40} \prt_{c}\bH_{abd} \prt^{d}\bH^{abc} + 
\frac{191}{120} \prt_{d}\bH_{abc} \prt^{d}\bH^{abc} -  
\frac{227}{4} \bH_{bcd} W^{ab} \prt^{d}V_{a}{}^{c} -  
\frac{433}{20} \bH_{bcd} V^{ab} \prt^{d}W_{a}{}^{c}\,.
\eeqa
Note that there are terms involving third derivatives of the base space, as well as  the terms involving fourth derivatives.  $\Delta\vp^{(2)}/c_1^2$ is 
\beqa
&&\!\!\!\!\!\frac{7}{6} e^{\vp} \bH_{cde} \bH^{cde} V_{ab} V^{ab} + 
\frac{41}{10} e^{\vp} \bH_{b}{}^{de} \bH_{cde} V_{a}{}^{c} V^{ab} + 
\frac{41}{5} e^{2 \vp} V_{a}{}^{c} V^{ab} V_{b}{}^{d} V_{cd} -  
\frac{33}{20} e^{\vp} \bH_{ab}{}^{e} \bH_{cde} V^{ab} V^{cd} \nn\\&&\!\!\!\!\!+ 
\frac{19}{10} e^{2 \vp} V_{ab} V^{ab} V_{cd} V^{cd} + 
\frac{7}{6} e^{- \vp} \bH_{cde} \bH^{cde} W_{ab} W^{ab} + 
\frac{41}{10} e^{- \vp} \bH_{b}{}^{de} \bH_{cde} W_{a}{}^{c} 
W^{ab} + 40 V_{a}{}^{c} V^{ab} W_{b}{}^{d} W_{cd} \nn\\&&\!\!\!\!\!+ \frac{41}{5} 
e^{2 - \vp} W_{a}{}^{c} W^{ab} W_{b}{}^{d} W_{cd} -  
\frac{33}{20} e^{- \vp} \bH_{ab}{}^{e} \bH_{cde} W^{ab} W^{cd} + 
\frac{39}{10} e^{2 - \vp} W_{ab} W^{ab} W_{cd} W^{cd}\nn\\&&\!\!\!\!\! + 
\frac{279}{5} e^{\vp} V_{bc} V^{bc} \prt_{a}\bphi 
\prt^{a}\bphi + \frac{279}{5} e^{- \vp} W_{bc} W^{bc} 
\prt_{a}\bphi \prt^{a}\bphi -  \frac{129}{20} \bH_{bcd} 
V_{a}{}^{b} W^{cd} \prt^{a}\vp -  \frac{289}{120} \bH^{bcd} 
\prt_{a}\bH_{bcd} \prt^{a}\vp\nn\\&&\!\!\!\!\! -  \frac{341}{80} e^{\vp} 
V_{bc} V^{bc} \prt_{a}\vp \prt^{a}\vp + \frac{619}{80} 
e^{- \vp} W_{bc} W^{bc} \prt_{a}\vp \prt^{a}\vp -  
\frac{579}{2} \prt_{a}\vp \prt^{a}\vp 
\prt_{b}\prt^{b}\bphi + \frac{37}{5} 
\prt_{a}\prt^{a}\vp \prt_{b}\prt^{b}\vp\nn\\&&\!\!\!\!\! -  
\frac{811}{20} \prt_{a}\vp \prt^{a}\bphi 
\prt_{b}\prt^{b}\vp + \frac{1394}{5} \prt^{a}\bphi 
\prt_{b}\prt^{b}\prt_{a}\bphi + \frac{76}{5} 
\prt^{a}\vp \prt_{b}\prt^{b}\prt_{a}\vp -  
\frac{91}{5} \prt_{b}\prt^{b}\prt_{a}\prt^{a}\bphi + 
\frac{1}{5} \bH_{a}{}^{cd} \bH_{bcd} \prt^{a}\bphi 
\prt^{b}\bphi \nn\\&&\!\!\!\!\!+ \frac{272}{5} e^{\vp} V_{a}{}^{c} V_{bc} 
\prt^{a}\bphi \prt^{b}\bphi + \frac{272}{5} e^{- \vp} 
W_{a}{}^{c} W_{bc} \prt^{a}\bphi \prt^{b}\bphi + 
\frac{103}{2} \prt_{a}\vp \prt^{a}\bphi \prt_{b}\vp 
\prt^{b}\bphi -  \frac{2424}{5} \prt^{a}\bphi 
\prt_{b}\prt_{a}\bphi \prt^{b}\bphi\nn\\&&\!\!\!\!\! -  \frac{17}{2} 
e^{\vp} V_{a}{}^{c} V_{bc} \prt^{a}\bphi \prt^{b}\vp + 
\frac{17}{2} e^{- \vp} W_{a}{}^{c} W_{bc} \prt^{a}\bphi 
\prt^{b}\vp -  \frac{599}{40} \bH_{a}{}^{cd} \bH_{bcd} 
\prt^{a}\vp \prt^{b}\vp -  \frac{273}{10} e^{\vp} 
V_{a}{}^{c} V_{bc} \prt^{a}\vp \prt^{b}\vp\nn\\&&\!\!\!\!\! -  
\frac{273}{10} e^{- \vp} W_{a}{}^{c} W_{bc} \prt^{a}\vp 
\prt^{b}\vp + \frac{1737}{5} \prt_{a}\bphi 
\prt^{a}\bphi \prt_{b}\vp \prt^{b}\vp + \frac{3}{2} 
\prt_{a}\vp \prt^{a}\vp \prt_{b}\vp \prt^{b}\vp + 
\frac{343}{5} \prt^{a}\vp \prt_{b}\prt_{a}\bphi 
\prt^{b}\vp \nn\\&&\!\!\!\!\!-  \frac{262}{5} \prt^{a}\bphi 
\prt_{b}\prt_{a}\vp \prt^{b}\vp - 16 \prt^{a}\vp 
\prt_{b}\prt_{a}\vp \prt^{b}\vp -  \frac{1}{10} 
\bH_{a}{}^{cd} \bH_{bcd} \prt^{b}\prt^{a}\bphi + \frac{328}{5} 
\prt_{b}\prt_{a}\bphi \prt^{b}\prt^{a}\bphi + 
\frac{2}{5} \prt_{b}\prt_{a}\vp \prt^{b}\prt^{a}\vp 
\nn\\&&\!\!\!\!\!+ \frac{144}{5} e^{\vp} V^{bc} \prt^{a}\bphi 
\prt_{c}V_{ab} + \frac{679}{10} e^{\vp} V^{bc} 
\prt^{a}\vp \prt_{c}V_{ab} + \frac{41}{5} e^{\vp} 
\prt_{a}V^{ab} \prt_{c}V_{b}{}^{c} + \frac{54}{5} e^{\vp} 
V_{a}{}^{b} \prt^{a}\bphi \prt_{c}V_{b}{}^{c}\nn\\&&\!\!\!\!\! + 
\frac{351}{20} e^{\vp} V_{a}{}^{b} \prt^{a}\vp 
\prt_{c}V_{b}{}^{c} + \frac{144}{5} e^{- \vp} W^{bc} 
\prt^{a}\bphi \prt_{c}W_{ab} -  \frac{199}{10} e^{- \vp} 
W^{bc} \prt^{a}\vp \prt_{c}W_{ab} + \frac{11}{5} \bH_{bcd} V^{ab} 
\prt^{d}W_{a}{}^{c}\nn\\&&\!\!\!\!\!+ \frac{41}{5} e^{- 
\vp} \prt_{a}W^{ab} \prt_{c}W_{b}{}^{c} + \frac{54}{5} 
e^{- \vp} W_{a}{}^{b} \prt^{a}\bphi \prt_{c}W_{b}{}^{c} -  
\frac{351}{20} e^{- \vp} W_{a}{}^{b} \prt^{a}\vp 
\prt_{c}W_{b}{}^{c} - 31 e^{\vp} V_{a}{}^{c} V^{ab} 
\prt_{c}\prt_{b}\bphi \nn\\&&\!\!\!\!\!- 31 e^{- \vp} W_{a}{}^{c} W^{ab} 
\prt_{c}\prt_{b}\bphi -  \frac{17}{2} e^{\vp} V_{a}{}^{c} 
V^{ab} \prt_{c}\prt_{b}\vp -  \frac{15}{2} e^{- \vp} 
W_{a}{}^{c} W^{ab} \prt_{c}\prt_{b}\vp -  \frac{559}{10} 
e^{\vp} V_{ab} V^{ab} \prt_{c}\prt^{c}\bphi\nn\\&&\!\!\!\!\! -  
\frac{559}{10} e^{- \vp} W_{ab} W^{ab} 
\prt_{c}\prt^{c}\bphi -  \frac{37}{10} e^{\vp} V^{ab} 
\prt_{c}\prt^{c}V_{ab} -  \frac{37}{10} e^{- \vp} W^{ab} 
\prt_{c}\prt^{c}W_{ab} -  \frac{217}{40} e^{\vp} V_{ab} 
V^{ab} \prt_{c}\prt^{c}\vp\nn\\&&\!\!\!\!\! + \frac{217}{40} e^{- \vp} 
W_{ab} W^{ab} \prt_{c}\prt^{c}\vp -  \frac{17}{5} e^{\vp} 
\prt_{c}V_{ab} \prt^{c}V^{ab} -  \frac{17}{5} e^{- \vp} 
\prt_{c}W_{ab} \prt^{c}W^{ab} - 8 V^{ab} W^{cd} 
\prt_{d}\bH_{abc} \nn\\&&\!\!\!\!\!-  \frac{39}{5} \bH^{bcd} \prt^{a}\bphi 
\prt_{d}\bH_{abc} + \frac{289}{40} \bH^{bcd} \prt^{a}\vp 
\prt_{d}\bH_{abc} + \frac{1}{10} \prt_{a}\bH^{abc} 
\prt_{d}\bH_{bc}{}^{d} -  \frac{3}{10} \bH_{a}{}^{bc} \prt^{a}\bphi \prt_{d}\bH_{bc}{}^{d} \nn\\&&\!\!\!\!\!-  \frac{71}{10} \bH^{abc} 
\prt_{d}\prt_{c}\bH_{ab}{}^{d} -  \frac{163}{10} 
\prt_{c}\bH_{abd} \prt^{d}\bH^{abc} + \frac{91}{30} 
\prt_{d}\bH_{abc} \prt^{d}\bH^{abc} -  \frac{69}{5} \bH_{bcd} 
W^{ab} \prt^{d}V_{a}{}^{c}\,. 
\eeqa
In addition to  first and  second derivatives, the above expression  also involves third and fourth derivatives of the base space. To simplify our notation for witting $\Delta\bg_{ab}^{(2)}$, we introduce a symmetric tensor $S^{ab}$ and multiply it with $\Delta\bg_{ab}^{(2)}$. This allows us to write $\Delta\bg_{ab}^{(2)}$ in a simpler form.  $S^{ab}\Delta\bg_{ab}^{(2)}/c_1^2$   is 
\beqa
&&\!\!\!\!\!\frac{1}{10} e^{\vp} S^{ab} \bH_{c}{}^{ef} \bH_{def} V_{a}{}^{c} 
V_{b}{}^{d} + \frac{2}{5} e^{\vp} S^{ab} \bH_{b}{}^{ef} \bH_{def} 
V_{a}{}^{c} V_{c}{}^{d} + \frac{1}{5} e^{\vp} S^{ab} 
\bH_{ad}{}^{f} \bH_{bef} V_{c}{}^{e} V^{cd} \nn\\&&\!\!\!\!\!-  \frac{1}{5} e^{2 
\vp} S^{ab} V_{a}{}^{c} V_{b}{}^{d} V_{c}{}^{e} V_{de} -  
\frac{33}{5} e^{\vp} S^{ab} \bH_{bc}{}^{f} \bH_{def} V_{a}{}^{c} 
V^{de} + \frac{73}{20} e^{\vp} S^{a}{}_{a} \bH_{bc}{}^{f} 
\bH_{def} V^{bc} V^{de}\nn\\&&\!\!\!\!\! -  \frac{79}{10} e^{2 \vp} S^{ab} 
V_{a}{}^{c} V_{bc} V_{de} V^{de} -  \frac{2}{5} e^{\vp} S^{ab} 
\bH_{acd} \bH_{bef} V^{cd} V^{ef} -  \frac{1}{10} e^{- \vp} 
S^{ab} \bH_{c}{}^{ef} \bH_{def} W_{a}{}^{c} W_{b}{}^{d}\nn\\&&\!\!\!\!\! -  
\frac{2}{5} e^{- \vp} S^{ab} \bH_{b}{}^{ef} \bH_{def} 
W_{a}{}^{c} W_{c}{}^{d} -  \frac{1}{5} e^{- \vp} S^{ab} 
\bH_{ad}{}^{f} \bH_{bef} W_{c}{}^{e} W^{cd} + \frac{1}{5} e^{2 - 
\vp} S^{ab} W_{a}{}^{c} W_{b}{}^{d} W_{c}{}^{e} W_{de}\nn\\&&\!\!\!\!\! + 
\frac{33}{5} e^{- \vp} S^{ab} \bH_{bc}{}^{f} \bH_{def} 
W_{a}{}^{c} W^{de} -  \frac{73}{20} e^{- \vp} S^{a}{}_{a} 
\bH_{bc}{}^{f} \bH_{def} W^{bc} W^{de} + \frac{79}{10} e^{2 - 
\vp} S^{ab} W_{a}{}^{c} W_{bc} W_{de} W^{de} \nn\\&&\!\!\!\!\!+ \frac{2}{5} 
e^{- \vp} S^{ab} \bH_{acd} \bH_{bef} W^{cd} W^{ef} -  
\frac{156}{5} S^{bc} \bH_{cde} V_{b}{}^{d} W_{a}{}^{e} \prt^{a}\bphi + \frac{104}{5} S^{bc} \bH_{cde} V_{a}{}^{d} W_{b}{}^{e} 
\prt^{a}\bphi \nn\\&&\!\!\!\!\!+ \frac{64}{5} S^{bc} \bH_{ade} V_{b}{}^{d} 
W_{c}{}^{e} \prt^{a}\bphi + \frac{64}{5} S^{b}{}_{b} \bH_{ade} 
V^{cd} W_{c}{}^{e} \prt^{a}\bphi - 26 S^{bc} \bH_{cde} V_{ab} 
W^{de} \prt^{a}\bphi\nn\\&&\!\!\!\!\! + \frac{81}{5} S^{b}{}_{b} \bH_{cde} 
V_{a}{}^{c} W^{de} \prt^{a}\bphi -  \frac{118}{5} S_{a}{}^{b} 
\bH_{cde} V_{b}{}^{c} W^{de} \prt^{a}\bphi - 13 S^{bc} 
\bH_{b}{}^{de} \prt_{a}\bH_{cde} \prt^{a}\bphi \nn\\&&\!\!\!\!\!+ \frac{27}{10} 
S^{b}{}_{b} \bH^{cde} \prt_{a}\bH_{cde} \prt^{a}\bphi -  
\frac{2}{5} e^{\vp} S^{bc} V_{b}{}^{d} V_{cd} \prt_{a}\bphi 
\prt^{a}\bphi + \frac{2}{5} e^{- \vp} S^{bc} W_{b}{}^{d} 
W_{cd} \prt_{a}\bphi \prt^{a}\bphi \nn\\&&\!\!\!\!\!+ 30 e^{\vp} S^{bc} 
V_{b}{}^{d} \prt_{a}V_{cd} \prt^{a}\bphi - 30 e^{- \vp} 
S^{bc} W_{b}{}^{d} \prt_{a}W_{cd} \prt^{a}\bphi + 
\frac{29}{8} S^{bc} \bH_{b}{}^{de} \bH_{cde} \prt_{a}\vp 
\prt^{a}\bphi \nn\\&&\!\!\!\!\!-  \frac{169}{20} e^{\vp} S^{bc} V_{b}{}^{d} 
V_{cd} \prt_{a}\vp \prt^{a}\bphi -  \frac{169}{20} e^{- 
\vp} S^{bc} W_{b}{}^{d} W_{cd} \prt_{a}\vp \prt^{a}\bphi 
+ 28 S^{bc} \bH_{cde} V_{b}{}^{d} W_{a}{}^{e} \prt^{a}\vp \nn\\&&\!\!\!\!\!- 28 
S^{bc} \bH_{cde} V_{a}{}^{d} W_{b}{}^{e} \prt^{a}\vp - 45 
e^{\vp} S^{bc} V_{b}{}^{d} \prt_{a}V_{cd} \prt^{a}\vp - 45 
e^{- \vp} S^{bc} W_{b}{}^{d} \prt_{a}W_{cd} \prt^{a}\vp 
\nn\\&&\!\!\!\!\!-  \frac{303}{20} e^{\vp} S^{bc} V_{b}{}^{d} V_{cd} 
\prt_{a}\vp \prt^{a}\vp + \frac{303}{20} e^{- \vp} 
S^{bc} W_{b}{}^{d} W_{cd} \prt_{a}\vp \prt^{a}\vp - 32 
S^{ab} V_{a}{}^{c} W^{de} \prt_{b}\bH_{cde}\nn\\&&\!\!\!\!\! -  \frac{16}{3} 
S^{ab} \prt_{a}\bH^{cde} \prt_{b}\bH_{cde} -  \frac{59}{15} 
S_{a}{}^{b} \bH^{cde} \prt^{a}\bphi \prt_{b}\bH_{cde} + 
\frac{238}{5} S^{ab} \bH_{cde} V^{cd} \prt_{b}W_{a}{}^{e} + 
\frac{39}{5} S^{ab} \bH_{cde} V_{a}{}^{c} \prt_{b}W^{de} \nn\\&&\!\!\!\!\!-  
\frac{37}{5} S^{a}{}_{a} \bH_{cde} V^{bc} \prt_{b}W^{de} + 
\frac{119}{15} S^{ab} \bH^{cde} \prt_{b}\prt_{a}\bH_{cde} -  
\frac{443}{40} e^{\vp} S^{ab} V_{cd} V^{cd} 
\prt_{b}\prt_{a}\vp\nn\\&&\!\!\!\!\! -  \frac{443}{40} e^{- \vp} S^{ab} 
W_{cd} W^{cd} \prt_{b}\prt_{a}\vp + \frac{556}{5} e^{\vp} 
S^{cd} V_{ac} V_{bd} \prt^{a}\bphi \prt^{b}\bphi + 26 
e^{\vp} S^{c}{}_{c} V_{a}{}^{d} V_{bd} \prt^{a}\bphi 
\prt^{b}\bphi \nn\\&&\!\!\!\!\!+ \frac{236}{5} e^{\vp} S_{a}{}^{c} 
V_{b}{}^{d} V_{cd} \prt^{a}\bphi \prt^{b}\bphi -  
\frac{556}{5} e^{- \vp} S^{cd} W_{ac} W_{bd} \prt^{a}\bphi 
\prt^{b}\bphi - 26 e^{- \vp} S^{c}{}_{c} W_{a}{}^{d} W_{bd} 
\prt^{a}\bphi \prt^{b}\bphi \nn\\&&\!\!\!\!\!-  \frac{236}{5} e^{- \vp} 
S_{a}{}^{c} W_{b}{}^{d} W_{cd} \prt^{a}\bphi \prt^{b}\bphi + 
\frac{888}{5} S^{c}{}_{c} \prt_{a}\bphi \prt^{a}\bphi 
\prt_{b}\vp \prt^{b}\bphi - 15 S^{c}{}_{c} 
\prt^{a}\bphi \prt_{b}\prt_{a}\vp \prt^{b}\bphi\nn\\&&\!\!\!\!\! -  
\frac{1}{5} S_{a}{}^{c} \bH_{b}{}^{de} \bH_{cde} \prt^{a}\bphi 
\prt^{b}\vp + \frac{41}{5} e^{\vp} S^{cd} V_{ac} V_{bd} 
\prt^{a}\bphi \prt^{b}\vp + \frac{157}{10} e^{\vp} 
S^{c}{}_{c} V_{a}{}^{d} V_{bd} \prt^{a}\bphi \prt^{b}\vp\nn\\&&\!\!\!\!\! + 
61 e^{\vp} S_{b}{}^{c} V_{a}{}^{d} V_{cd} \prt^{a}\bphi 
\prt^{b}\vp - 24 e^{\vp} S_{a}{}^{c} V_{b}{}^{d} V_{cd} 
\prt^{a}\bphi \prt^{b}\vp + \frac{431}{20} e^{\vp} 
S_{ab} V_{cd} V^{cd} \prt^{a}\bphi \prt^{b}\vp\nn\\&&\!\!\!\!\! + 
\frac{41}{5} e^{- \vp} S^{cd} W_{ac} W_{bd} \prt^{a}\bphi 
\prt^{b}\vp + \frac{157}{10} e^{- \vp} S^{c}{}_{c} 
W_{a}{}^{d} W_{bd} \prt^{a}\bphi \prt^{b}\vp + 61 e^{- 
\vp} S_{b}{}^{c} W_{a}{}^{d} W_{cd} \prt^{a}\bphi 
\prt^{b}\vp \nn\\&&\!\!\!\!\!- 24 e^{- \vp} S_{a}{}^{c} W_{b}{}^{d} W_{cd} 
\prt^{a}\bphi \prt^{b}\vp + \frac{431}{20} e^{- \vp} 
S_{ab} W_{cd} W^{cd} \prt^{a}\bphi \prt^{b}\vp -  
\frac{319}{10} e^{\vp} S^{cd} V_{ac} V_{bd} \prt^{a}\vp 
\prt^{b}\vp \nn\\&&\!\!\!\!\!+ \frac{36}{5} e^{\vp} S^{c}{}_{c} V_{a}{}^{d} 
V_{bd} \prt^{a}\vp \prt^{b}\vp -  \frac{303}{10} 
e^{\vp} S_{a}{}^{c} V_{b}{}^{d} V_{cd} \prt^{a}\vp 
\prt^{b}\vp + \frac{319}{10} e^{- \vp} S^{cd} W_{ac} 
W_{bd} \prt^{a}\vp \prt^{b}\vp \nn\\&&\!\!\!\!\!-  \frac{36}{5} e^{- 
\vp} S^{c}{}_{c} W_{a}{}^{d} W_{bd} \prt^{a}\vp 
\prt^{b}\vp + \frac{303}{10} e^{- \vp} S_{a}{}^{c} 
W_{b}{}^{d} W_{cd} \prt^{a}\vp \prt^{b}\vp + 
\frac{727}{20} S^{c}{}_{c} \prt_{a}\vp \prt^{a}\bphi 
\prt_{b}\vp \prt^{b}\vp\nn\\&&\!\!\!\!\! -  \frac{2713}{5} S^{c}{}_{c} 
\prt^{a}\bphi \prt_{b}\prt_{a}\bphi \prt^{b}\vp -  
\frac{74}{5} S^{c}{}_{c} \prt^{a}\vp 
\prt_{b}\prt_{a}\vp \prt^{b}\vp + 13 S^{bc} 
\bH_{b}{}^{de} \prt^{a}\bphi \prt_{c}\bH_{ade} -  
\frac{143}{20} S^{bc} \bH_{b}{}^{de} \prt^{a}\vp 
\prt_{c}\bH_{ade}\nn\\&&\!\!\!\!\! + \frac{464}{5} e^{\vp} S^{bc} V_{a}{}^{d} 
\prt^{a}\bphi \prt_{c}V_{bd} -  \frac{232}{5} e^{\vp} 
S^{bc} V_{a}{}^{d} \prt^{a}\vp \prt_{c}V_{bd} + 32 S^{ab} 
\bH_{bde} V^{cd} \prt_{c}W_{a}{}^{e}\nn\\&&\!\!\!\!\! -  \frac{464}{5} e^{- 
\vp} S^{bc} W_{a}{}^{d} \prt^{a}\bphi \prt_{c}W_{bd} -  
\frac{232}{5} e^{- \vp} S^{bc} W_{a}{}^{d} \prt^{a}\vp 
\prt_{c}W_{bd} + \frac{121}{5} S^{ab} \bH_{bde} V_{a}{}^{c} 
\prt_{c}W^{de}\nn\\&&\!\!\!\!\! -  \frac{5168}{5} S_{b}{}^{c} \prt^{a}\bphi 
\prt^{b}\vp \prt_{c}\prt_{a}\bphi -  \frac{362}{5} 
S^{bc} \prt_{a}\vp \prt^{a}\bphi 
\prt_{c}\prt_{b}\bphi -  \frac{211}{5} S_{a}{}^{c} 
\prt^{a}\bphi \prt^{b}\vp \prt_{c}\prt_{b}\bphi -  
\frac{2586}{5} S^{bc} \prt_{a}\bphi \prt^{a}\bphi 
\prt_{c}\prt_{b}\vp\nn\\&&\!\!\!\!\! -  \frac{431}{20} S^{bc} 
\prt_{a}\vp \prt^{a}\vp \prt_{c}\prt_{b}\vp -  
\frac{219}{5} S_{a}{}^{c} \prt^{a}\bphi \prt^{b}\bphi 
\prt_{c}\prt_{b}\vp - 66 S_{a}{}^{c} \prt^{a}\vp 
\prt^{b}\vp \prt_{c}\prt_{b}\vp + \frac{358}{5} 
S^{bc} \prt^{a}\vp \prt_{c}\prt_{b}\prt_{a}\bphi \nn\\&&\!\!\!\!\!- 
148 S^{b}{}_{b} \prt_{a}\vp \prt^{a}\bphi 
\prt_{c}\prt^{c}\bphi + 431 S^{ab} 
\prt_{b}\prt_{a}\vp \prt_{c}\prt^{c}\bphi - 862 
S_{ab} \prt^{a}\bphi \prt^{b}\vp 
\prt_{c}\prt^{c}\bphi -  \frac{29}{16} S^{ab} \bH_{a}{}^{de} 
\bH_{bde} \prt_{c}\prt^{c}\vp\nn\\&&\!\!\!\!\! -  \frac{444}{5} S^{b}{}_{b} 
\prt_{a}\bphi \prt^{a}\bphi \prt_{c}\prt^{c}\vp -  
\frac{37}{5} S^{b}{}_{b} \prt_{a}\vp \prt^{a}\vp 
\prt_{c}\prt^{c}\vp + \frac{181}{5} S^{ab} 
\prt_{b}\prt_{a}\bphi \prt_{c}\prt^{c}\vp + 74 
S^{a}{}_{a} \prt_{b}\prt^{b}\bphi \prt_{c}\prt^{c}\vp 
\nn\\&&\!\!\!\!\!-  \frac{2537}{80} S_{ab} \prt^{a}\vp \prt^{b}\vp 
\prt_{c}\prt^{c}\vp + \frac{311}{2} S^{b}{}_{b} 
\prt^{a}\vp \prt_{c}\prt^{c}\prt_{a}\bphi + 15 
S^{b}{}_{b} \prt^{a}\bphi \prt_{c}\prt^{c}\prt_{a}\vp 
+ \frac{2149}{5} S_{a}{}^{b} \prt^{a}\vp 
\prt_{c}\prt^{c}\prt_{b}\bphi\nn\\&&\!\!\!\!\! + \frac{219}{10} 
S_{a}{}^{b} \prt^{a}\bphi \prt_{c}\prt^{c}\prt_{b}\vp 
-  \frac{15}{4} S^{a}{}_{a} 
\prt_{c}\prt^{c}\prt_{b}\prt^{b}\vp + \frac{5172}{5} 
S_{bc} \prt_{a}\bphi \prt^{a}\bphi \prt^{b}\bphi 
\prt^{c}\vp + \frac{509}{8} S_{bc} \prt_{a}\vp 
\prt^{a}\bphi \prt^{b}\vp \prt^{c}\vp\nn\\&&\!\!\!\!\! + 
\frac{429}{10} S_{ac} \prt^{a}\bphi \prt_{b}\vp 
\prt^{b}\vp \prt^{c}\vp + \frac{217}{5} S^{ab} 
\prt_{c}\prt_{b}\vp \prt^{c}\prt_{a}\bphi + 
\frac{1}{10} S^{ab} \bH_{b}{}^{de} \bH_{cde} 
\prt^{c}\prt_{a}\vp + 15 S^{a}{}_{a} 
\prt_{c}\prt_{b}\vp \prt^{c}\prt^{b}\bphi \nn\\&&\!\!\!\!\!+ 
\frac{82}{5} S^{ab} V^{cd} W_{a}{}^{e} \prt_{d}\bH_{bce} + 
\frac{144}{5} e^{\vp} S^{bc} V_{b}{}^{d} \prt^{a}\bphi 
\prt_{d}V_{ac} -  \frac{72}{5} e^{\vp} S^{bc} V_{b}{}^{d} 
\prt^{a}\vp \prt_{d}V_{ac} + \frac{37}{5} e^{\vp} 
S^{b}{}_{b} V^{cd} \prt^{a}\vp \prt_{d}V_{ac} \nn\\&&\!\!\!\!\!+ 
\frac{599}{10} e^{\vp} S_{a}{}^{b} V^{cd} \prt^{a}\vp 
\prt_{d}V_{bc} - 16 e^{\vp} S^{ab} \prt_{c}V_{a}{}^{c} 
\prt_{d}V_{b}{}^{d} + \frac{118}{5} e^{\vp} S^{bc} V_{ab} 
\prt^{a}\bphi \prt_{d}V_{c}{}^{d} -  \frac{89}{5} e^{\vp} 
S^{b}{}_{b} V_{a}{}^{c} \prt^{a}\bphi \prt_{d}V_{c}{}^{d}\nn\\&&\!\!\!\!\! + 
\frac{118}{5} e^{\vp} S_{a}{}^{b} V_{b}{}^{c} \prt^{a}\bphi 
\prt_{d}V_{c}{}^{d} + \frac{479}{10} e^{\vp} S^{bc} V_{ab} 
\prt^{a}\vp \prt_{d}V_{c}{}^{d} + \frac{41}{5} e^{\vp} 
S^{b}{}_{b} V_{a}{}^{c} \prt^{a}\vp \prt_{d}V_{c}{}^{d}\nn\\&&\!\!\!\!\! + 
\frac{303}{10} e^{\vp} S_{a}{}^{b} V_{b}{}^{c} \prt^{a}\vp 
\prt_{d}V_{c}{}^{d} + \frac{232}{5} e^{\vp} S^{ab} 
\prt_{b}V_{a}{}^{c} \prt_{d}V_{c}{}^{d} + \frac{77}{5} 
e^{\vp} S^{a}{}_{a} \prt_{b}V^{bc} \prt_{d}V_{c}{}^{d} -  
\frac{92}{5} S^{ab} \bH_{bce} W_{a}{}^{c} \prt_{d}V^{de}\nn\\&&\!\!\!\!\! -  
\frac{187}{10} S^{a}{}_{a} \bH_{bce} W^{bc} \prt_{d}V^{de} -  
\frac{144}{5} e^{- \vp} S^{bc} W_{b}{}^{d} \prt^{a}\bphi 
\prt_{d}W_{ac} -  \frac{72}{5} e^{- \vp} S^{bc} W_{b}{}^{d} 
\prt^{a}\vp \prt_{d}W_{ac} \nn\\&&\!\!\!\!\!+ \frac{37}{5} e^{- \vp} 
S^{b}{}_{b} W^{cd} \prt^{a}\vp \prt_{d}W_{ac} + 
\frac{599}{10} e^{- \vp} S_{a}{}^{b} W^{cd} \prt^{a}\vp 
\prt_{d}W_{bc} + 16 e^{- \vp} S^{ab} \prt_{c}W_{a}{}^{c} 
\prt_{d}W_{b}{}^{d} \nn\\&&\!\!\!\!\!-  \frac{118}{5} e^{- \vp} S^{bc} 
W_{ab} \prt^{a}\bphi \prt_{d}W_{c}{}^{d} + \frac{89}{5} 
e^{- \vp} S^{b}{}_{b} W_{a}{}^{c} \prt^{a}\bphi 
\prt_{d}W_{c}{}^{d} -  \frac{118}{5} e^{- \vp} S_{a}{}^{b} 
W_{b}{}^{c} \prt^{a}\bphi \prt_{d}W_{c}{}^{d} \nn\\&&\!\!\!\!\!+ 
\frac{479}{10} e^{- \vp} S^{bc} W_{ab} \prt^{a}\vp 
\prt_{d}W_{c}{}^{d} + \frac{41}{5} e^{- \vp} S^{b}{}_{b} 
W_{a}{}^{c} \prt^{a}\vp \prt_{d}W_{c}{}^{d} + 
\frac{303}{10} e^{- \vp} S_{a}{}^{b} W_{b}{}^{c} 
\prt^{a}\vp \prt_{d}W_{c}{}^{d}\nn\\&&\!\!\!\!\! -  \frac{232}{5} e^{- 
\vp} S^{ab} \prt_{b}W_{a}{}^{c} \prt_{d}W_{c}{}^{d} -  
\frac{77}{5} e^{- \vp} S^{a}{}_{a} \prt_{b}W^{bc} 
\prt_{d}W_{c}{}^{d} -  \frac{68}{5} S^{ab} \bH_{bce} V_{a}{}^{c} 
\prt_{d}W^{de} \nn\\&&\!\!\!\!\!+ \frac{113}{10} S^{a}{}_{a} \bH_{bce} V^{bc} 
\prt_{d}W^{de} -  \frac{146}{5} e^{\vp} S^{ab} V_{a}{}^{c} 
V_{c}{}^{d} \prt_{d}\prt_{b}\bphi + \frac{146}{5} e^{- 
\vp} S^{ab} W_{a}{}^{c} W_{c}{}^{d} \prt_{d}\prt_{b}\bphi\nn\\&&\!\!\!\!\! + 
\frac{307}{10} e^{\vp} S^{ab} V_{a}{}^{c} V_{c}{}^{d} 
\prt_{d}\prt_{b}\vp + \frac{307}{10} e^{- \vp} S^{ab} 
W_{a}{}^{c} W_{c}{}^{d} \prt_{d}\prt_{b}\vp + \frac{2}{5} 
e^{\vp} S^{ab} V_{a}{}^{c} V_{b}{}^{d} 
\prt_{d}\prt_{c}\bphi \nn\\&&\!\!\!\!\!+ \frac{74}{5} e^{\vp} S^{a}{}_{a} 
V_{b}{}^{d} V^{bc} \prt_{d}\prt_{c}\bphi -  \frac{2}{5} 
e^{- \vp} S^{ab} W_{a}{}^{c} W_{b}{}^{d} \prt_{d}\prt_{c}\bphi -  \frac{74}{5} e^{- \vp} S^{a}{}_{a} W_{b}{}^{d} W^{bc} 
\prt_{d}\prt_{c}\bphi\nn\\&&\!\!\!\!\! + \frac{72}{5} e^{\vp} S^{ab} 
V_{a}{}^{c} \prt_{d}\prt_{c}V_{b}{}^{d} -  \frac{72}{5} 
e^{- \vp} S^{ab} W_{a}{}^{c} \prt_{d}\prt_{c}W_{b}{}^{d} - 
 \frac{163}{10} e^{\vp} S^{ab} V_{a}{}^{c} V_{b}{}^{d} 
\prt_{d}\prt_{c}\vp\nn\\&&\!\!\!\!\! -  \frac{37}{5} e^{\vp} S^{a}{}_{a} 
V_{b}{}^{d} V^{bc} \prt_{d}\prt_{c}\vp -  \frac{163}{10} 
e^{- \vp} S^{ab} W_{a}{}^{c} W_{b}{}^{d} 
\prt_{d}\prt_{c}\vp -  \frac{37}{5} e^{- \vp} 
S^{a}{}_{a} W_{b}{}^{d} W^{bc} \prt_{d}\prt_{c}\vp\nn\\&&\!\!\!\!\! + 
\frac{1}{5} e^{\vp} S^{ab} V_{a}{}^{c} V_{bc} 
\prt_{d}\prt^{d}\bphi -  \frac{1}{5} e^{- \vp} S^{ab} 
W_{a}{}^{c} W_{bc} \prt_{d}\prt^{d}\bphi -  \frac{74}{5} e^{
\vp} S^{ab} V_{a}{}^{c} \prt_{d}\prt^{d}V_{bc} \nn\\&&\!\!\!\!\!+ 
\frac{74}{5} e^{- \vp} S^{ab} W_{a}{}^{c} 
\prt_{d}\prt^{d}W_{bc} + \frac{177}{40} e^{\vp} S^{ab} 
V_{a}{}^{c} V_{bc} \prt_{d}\prt^{d}\vp + \frac{177}{40} 
e^{- \vp} S^{ab} W_{a}{}^{c} W_{bc} \prt_{d}\prt^{d}\vp 
\nn\\&&\!\!\!\!\!+ \frac{37}{10} e^{\vp} S^{a}{}_{a} \prt_{d}V_{bc} 
\prt^{d}V^{bc} -  \frac{37}{10} e^{- \vp} S^{a}{}_{a} 
\prt_{d}W_{bc} \prt^{d}W^{bc} + \frac{1}{5} S^{ab} 
\bH_{ac}{}^{e} \bH_{bde} \prt^{d}\prt^{c}\vp + 26 S^{bc} 
\bH_{b}{}^{de} \prt^{a}\bphi \prt_{e}\bH_{acd}\nn\\&&\!\!\!\!\! -  
\frac{81}{10} S^{b}{}_{b} \bH^{cde} \prt^{a}\bphi 
\prt_{e}\bH_{acd} -  \frac{143}{10} S^{bc} \bH_{b}{}^{de} 
\prt^{a}\vp \prt_{e}\bH_{acd} -  \frac{82}{5} S^{ab} V^{cd} 
W_{a}{}^{e} \prt_{e}\bH_{bcd} -  \frac{82}{5} S^{ab} V_{a}{}^{c} 
W^{de} \prt_{e}\bH_{bcd}\nn\\&&\!\!\!\!\! -  \frac{37}{5} S^{a}{}_{a} V^{bc} 
W^{de} \prt_{e}\bH_{bcd} + \frac{59}{5} S_{a}{}^{b} \bH^{cde} 
\prt^{a}\bphi \prt_{e}\bH_{bcd} + \frac{1}{4} S_{a}{}^{b} 
\bH^{cde} \prt^{a}\vp \prt_{e}\bH_{bcd} -  \frac{32}{5} 
S^{ab} V_{a}{}^{c} W_{b}{}^{d} \prt_{e}\bH_{cd}{}^{e} \nn\\&&\!\!\!\!\!-  
\frac{32}{5} S^{a}{}_{a} V^{bc} W_{b}{}^{d} 
\prt_{e}\bH_{cd}{}^{e} + \frac{431}{40} S^{b}{}_{b} 
\bH_{a}{}^{cd} \prt^{a}\vp \prt_{e}\bH_{cd}{}^{e} + 
\frac{1}{10} S_{a}{}^{b} \bH_{b}{}^{cd} \prt^{a}\vp 
\prt_{e}\bH_{cd}{}^{e}\nn\\&&\!\!\!\!\! + \frac{117}{5} S^{ab} \bH_{bcd} W^{cd} 
\prt_{e}V_{a}{}^{e} -  \frac{37}{5} S^{ab} \bH_{bcd} V^{cd} 
\prt_{e}W_{a}{}^{e} -  \frac{119}{5} S^{ab} \bH^{cde} 
\prt_{e}\prt_{b}\bH_{acd} + 8 S^{ab} \bH_{a}{}^{cd} 
\prt_{e}\prt_{b}\bH_{cd}{}^{e}\nn\\&&\!\!\!\!\! + 16 S^{ab} \bH_{a}{}^{cd} 
\prt_{e}\prt_{d}\bH_{bc}{}^{e} + \frac{37}{10} S^{a}{}_{a} 
\bH^{bcd} \prt_{e}\prt_{d}\bH_{bc}{}^{e} - 8 S^{ab} 
\bH_{a}{}^{cd} \prt_{e}\prt^{e}\bH_{bcd} -  \frac{37}{30} 
S^{a}{}_{a} \bH^{bcd} \prt_{e}\prt^{e}\bH_{bcd} \nn\\&&\!\!\!\!\!+ \frac{39}{5} 
S^{ab} \prt_{b}\bH_{cde} \prt^{e}\bH_{a}{}^{cd} -  
\frac{82}{5} S^{ab} \prt_{d}\bH_{bce} \prt^{e}\bH_{a}{}^{cd} + 
\frac{41}{5} S^{ab} \prt_{e}\bH_{bcd} \prt^{e}\bH_{a}{}^{cd} - 
 \frac{37}{10} S^{a}{}_{a} \prt_{d}\bH_{bce} \prt^{e}\bH^{bcd} 
\nn\\&&\!\!\!\!\!+ \frac{37}{30} S^{a}{}_{a} \prt_{e}\bH_{bcd} 
\prt^{e}\bH^{bcd} -  \frac{398}{5} S^{ab} \bH_{cde} W_{a}{}^{c} 
\prt^{e}V_{b}{}^{d} + \frac{82}{5} S^{ab} \bH_{bde} W_{a}{}^{c} 
\prt^{e}V_{c}{}^{d}\,.
\eeqa
The expression for $\Delta\bg_{ab}^{(2)}$ also  involves various terms with third and fourth derivatives of the base space. $\Delta b_{a}^{(2)}/c_1^2$    is 
\beqa
&&\!\!\!\!\!\!\!\!\!\!\frac{31}{20} e^{\vp/2} \bH_{ade} V^{bc} V^{de} W_{bc} -  
\frac{2}{5} e^{\vp/2} \bH_{ace} V^{bc} V^{de} W_{bd} + 4 e^{\vp/2} \bH_{cde} V_{a}{}^{b} V^{cd} W_{b}{}^{e} -  \frac{2}{5} 
e^{-\vp/2} \bH_{a}{}^{de} \bH_{bd}{}^{f} \bH_{cef} W^{bc} \nn\\&&\!\!\!\!\!\!\!\!\!\!+ 2 e^{-\vp/2} \bH_{a}{}^{de} \bH_{bc}{}^{f} \bH_{def} W^{bc} + \frac{2}{5} 
e^{-\vp/2} \bH_{ab}{}^{d} \bH_{c}{}^{ef} \bH_{def} W^{bc}+ 
\frac{9}{5} e^{\vp/2} \bH_{ade} V_{b}{}^{d} V^{bc} W_{c}{}^{e}\nn\\&&\!\!\!\!\!\!\!\!\!\! - 
 \frac{49}{20} e^{\vp/2} \bH_{ade} V_{bc} V^{bc} W^{de} -  
4e^{-3 \vp/2} \bH_{cde} W_{a}{}^{b} W_{b}{}^{c} W^{de} -  
4 e^{-3 \vp/2}\bH_{ade} W_{bc} W^{bc} W^{de} \nn\\&&\!\!\!\!\!\!\!\!\!\!+ 4 e^{\vp/2} \bH_{a}{}^{de} \bH_{cde} \prt_{b}V^{bc} - 14 e^{\vp/2} 
\bH_{b}{}^{de} \bH_{cde} V_{a}{}^{c} \prt^{b}\bphi+ 14 e^{\vp/2} \bH_{a}{}^{de} \bH_{cde} V_{b}{}^{c} \prt^{b}\bphi- 55 
e^{3/2 \vp} V_{a}{}^{c} V_{b}{}^{d} V_{cd} \prt^{b}\bphi\nn\\&&\!\!\!\!\!\!\!\!\!\!-  
\frac{8}{5} e^{\vp/2} \bH_{ac}{}^{e} \bH_{bde} V^{cd} 
\prt^{b}\bphi-  \frac{4}{5} e^{-\vp/2} V^{cd} W_{ac} W_{bd} 
\prt^{b}\bphi-  \frac{138}{5} e^{-\vp/2} \bH_{bcd} W^{cd} 
\prt_{a}\bphi\prt^{b}\bphi\nn\\&&\!\!\!\!\!\!\!\!\!\!-  \frac{367}{5} e^{-\vp/2} 
\bH_{acd} W^{cd} \prt_{b}\bphi\prt^{b}\bphi+ \frac{51}{5} 
e^{-\vp/2} \bH_{acd} W^{cd} \prt_{b}\vp \prt^{b}\bphi-  
\frac{19}{5} e^{\vp/2} \bH_{b}{}^{de} \bH_{cde} V_{a}{}^{c} 
\prt^{b}\vp \nn\\&&\!\!\!\!\!\!\!\!\!\!+ 8 e^{\vp/2} \bH_{a}{}^{de} \bH_{cde} V_{b}{}^{c} 
\prt^{b}\vp -  \frac{63}{4} e^{3 \vp/2} V_{a}{}^{c} 
V_{b}{}^{d} V_{cd} \prt^{b}\vp + \frac{4}{5} e^{\vp/2} 
\bH_{ac}{}^{e} \bH_{bde} V^{cd} \prt^{b}\vp - 4 e^{\vp/2} 
\bH_{ab}{}^{e} \bH_{cde} V^{cd} \prt^{b}\nn\\&&\!\!\!\!\!\!\!\!\!\!- 4 e^{3 \vp/2} 
V_{ab} V_{cd} V^{cd} \prt^{b}\vp -  \frac{12}{5} e^{-\vp/2} 
V^{cd} W_{ac} W_{bd} \prt^{b}\vp + 8 e^{-\vp/2} V^{cd} W_{ab} 
W_{cd} \prt^{b}\vp \nn\\&&\!\!\!\!\!\!\!\!\!\!+ 16 e^{-\vp/2} V_{b}{}^{c} W_{a}{}^{d} 
W_{cd} \prt^{b}\vp - 16 e^{-\vp/2} V_{a}{}^{c} W_{b}{}^{d} 
W_{cd} \prt^{b}\vp + 12 e^{-\vp/2} V_{ab} W_{cd} W^{cd} 
\prt^{b}\vp \nn\\&&\!\!\!\!\!\!\!\!\!\!-  \frac{27}{2} e^{-\vp/2} \bH_{bcd} W^{cd} 
\prt_{a}\bphi\prt^{b}\vp + \frac{15}{8} e^{-\vp/2} 
\bH_{bcd} W^{cd} \prt_{a}\vp \prt^{b}\vp -  \frac{133}{20} 
e^{-\vp/2} \bH_{acd} W^{cd} \prt_{b}\vp \prt^{b}\vp\nn\\&&\!\!\!\!\!\!\!\!\!\!- 66 
e^{\vp/2} \prt_{b}\prt^{b}\bphi\prt_{c}V_{a}{}^{c} + 
\frac{15}{4} e^{\vp/2} \prt_{b}\prt^{b}\vp 
\prt_{c}V_{a}{}^{c} + 68 e^{\vp/2} \prt_{b}\bphi
\prt^{b}\bphi\prt_{c}V_{a}{}^{c} + \frac{63}{10} e^{\vp/2} \prt_{b}\vp \prt^{b}\bphi\prt_{c}V_{a}{}^{c} \nn\\&&\!\!\!\!\!\!\!\!\!\!+ 4 
e^{\vp/2} \prt_{b}\vp \prt^{b}\vp 
\prt_{c}V_{a}{}^{c} - 27 e^{\vp/2} \prt_{a}\vp 
\prt^{b}\bphi\prt_{c}V_{b}{}^{c} + 27 e^{\vp/2} 
\prt_{a}\bphi\prt^{b}\vp \prt_{c}V_{b}{}^{c} -  
\frac{39}{10} e^{\vp/2} \prt_{a}\vp \prt^{b}\vp 
\prt_{c}V_{b}{}^{c}\nn\\&&\!\!\!\!\!\!\!\!\!\! -  \frac{49}{5} e^{\vp/2} 
\prt^{b}\prt_{a}\bphi\prt_{c}V_{b}{}^{c} + 
\frac{231}{20} e^{\vp/2} \prt^{b}\prt_{a}\vp 
\prt_{c}V_{b}{}^{c} + 16 e^{-\vp/2} V^{bc} W_{b}{}^{d} 
\prt_{c}W_{ad}-  \frac{64}{5} e^{-\vp/2} \bH_{abd} 
\prt^{b}\bphi\prt_{c}W^{cd}\nn\\&&\!\!\!\!\!\!\!\!\!\! - 8 e^{-\vp/2} \bH_{abd} 
\prt^{b}\vp \prt_{c}W^{cd} - 112 e^{\vp/2} V_{b}{}^{c} 
\prt^{b}\bphi\prt_{c}\prt_{a}\bphi-  \frac{353}{10} 
e^{\vp/2} V_{b}{}^{c} \prt^{b}\vp 
\prt_{c}\prt_{a}\bphi-  \frac{283}{10} e^{\vp/2} 
V_{b}{}^{c} \prt^{b}\bphi\prt_{c}\prt_{a}\vp \nn\\&&\!\!\!\!\!\!\!\!\!\!+ 
\frac{1}{5} e^{\vp/2} V_{b}{}^{c} \prt^{b}\vp 
\prt_{c}\prt_{a}\vp -  \frac{944}{5} e^{\vp/2} 
V_{a}{}^{c} \prt^{b}\bphi\prt_{c}\prt_{b}\bphi+ 
\frac{52}{5} e^{\vp/2} V_{a}{}^{c} \prt^{b}\vp \prt_{c}
\prt_{b}\bphi-  \frac{67}{5} e^{\vp/2} \prt^{b}\bphi
\prt_{c}\prt_{b}V_{a}{}^{c} \nn\\&&\!\!\!\!\!\!\!\!\!\!+ \frac{231}{20} e^{\vp/2} 
\prt^{b}\vp \prt_{c}\prt_{b}V_{a}{}^{c} + \frac{57}{2} 
e^{\vp/2} V_{a}{}^{c} \prt^{b}\bphi
\prt_{c}\prt_{b}\vp -  \frac{78}{5} e^{\vp/2} 
V_{a}{}^{c} \prt^{b}\vp \prt_{c}\prt_{b}\vp- 306 e^{\vp/2} V_{ab} \prt^{b}\bphi\prt_{c}\prt^{c}\bphi\nn\\&&\!\!\!\!\!\!\!\!\!\!-  
\frac{111}{2} e^{\vp/2} V_{ab} \prt^{b}\vp 
\prt_{c}\prt^{c}\bphi+ \frac{207}{5} e^{\vp/2} 
\prt^{b}\bphi\prt_{c}\prt^{c}V_{ab}-  \frac{233}{20} 
e^{\vp/2} \prt^{b}\vp \prt_{c}\prt^{c}V_{ab} + 
\frac{67}{20} e^{\vp/2} V_{ab} \prt^{b}\bphi
\prt_{c}\prt^{c}\vp \nn\\&&\!\!\!\!\!\!\!\!\!\!+ \frac{321}{40} e^{\vp/2} V_{ab} 
\prt^{b}\vp \prt_{c}\prt^{c}\vp + 75 e^{\vp/2} 
V_{a}{}^{b} \prt_{c}\prt^{c}\prt_{b}\bphi-  \frac{7}{20} 
e^{\vp/2} V_{a}{}^{b} \prt_{c}\prt^{c}\prt_{b}\vp + 
\frac{126}{5} e^{-\vp/2} \bH_{acd} W_{b}{}^{d} \prt^{b}\bphi
\prt^{c}\bphi\nn\\&&\!\!\!\!\!\!\!\!\!\!+ \frac{1948}{5} e^{\vp/2} V_{ac} 
\prt_{b}\bphi\prt^{b}\bphi\prt^{c}\bphi-  
\frac{343}{10} e^{\vp/2} V_{ac} \prt_{b}\vp 
\prt^{b}\bphi\prt^{c}\bphi- 56 e^{\vp/2} 
\prt^{b}\bphi\prt_{c}V_{ab} \prt^{c}\bphi-  
\frac{38}{5} e^{\vp/2} \bH_{b}{}^{de} \bH_{cde} 
\prt^{c}V_{a}{}^{b} \nn\\&&\!\!\!\!\!\!\!\!\!\!-  \frac{16}{5} e^{-\vp/2} \bH_{bcd} 
W_{a}{}^{d} \prt^{b}\bphi\prt^{c}\vp + 16 e^{-\vp/2} 
\bH_{acd} W_{b}{}^{d} \prt^{b}\bphi\prt^{c}\vp+ 
\frac{29}{10} e^{-\vp/2} \bH_{abd} W_{c}{}^{d} \prt^{b}\bphi
\prt^{c}\vp \nn\\&&\!\!\!\!\!\!\!\!\!\!+ 54 e^{\vp/2} V_{bc} \prt_{a}\bphi
\prt^{b}\bphi\prt^{c}\vp -  \frac{129}{10} e^{\vp/2} 
V_{bc} \prt_{a}\vp \prt^{b}\bphi\prt^{c}\vp + 
\frac{277}{5} e^{\vp/2} V_{ac} \prt_{b}\bphi
\prt^{b}\bphi\prt^{c}\vp -  \frac{1}{10} e^{\vp/2} 
\prt_{b}V_{ac} \prt^{b}\bphi\prt^{c}\vp \nn\\&&\!\!\!\!\!\!\!\!\!\!-  
\frac{161}{20} e^{\vp/2} V_{ac} \prt_{b}\vp 
\prt^{b}\bphi\prt^{c}\vp + \frac{23}{5} e^{-\vp/2} 
\bH_{acd} W_{b}{}^{d} \prt^{b}\vp \prt^{c}\vp + 
\frac{143}{40} e^{\vp/2} V_{ac} \prt_{b}\vp 
\prt^{b}\vp \prt^{c}\vp \nn\\&&\!\!\!\!\!\!\!\!\!\!+ \frac{283}{10} e^{\vp/2} 
\prt^{b}\bphi\prt_{c}V_{ab} \prt^{c}\vp- 8 e^{\vp/2} \prt^{b}\vp \prt_{c}V_{ab} \prt^{c}\vp + 
\frac{139}{10} e^{\vp/2} V_{ab} \prt^{b}\bphi
\prt_{c}\vp \prt^{c}\vp + \frac{318}{5} e^{\vp/2} 
\prt_{c}V_{ab} \prt^{c}\prt^{b}\bphi\nn\\&&\!\!\!\!\!\!\!\!\!\!-  \frac{1}{10} 
e^{\vp/2} \prt_{c}V_{ab} \prt^{c}\prt^{b}\vp + 
\frac{211}{5} e^{-\vp/2} W^{cd} \prt^{b}\bphi
\prt_{d}\bH_{abc} -  \frac{1}{20} e^{-\vp/2} W^{cd} 
\prt^{b}\vp \prt_{d}\bH_{abc}-  \frac{32}{5} e^{-\vp/2} 
\prt_{b}W^{bc} \prt_{d}\bH_{ac}{}^{d} \nn\\&&\!\!\!\!\!\!\!\!\!\!+ \frac{63}{5} e^{-\vp/2} W_{b}{}^{c} \prt^{b}\bphi\prt_{d}\bH_{ac}{}^{d} + 
\frac{29}{20} e^{-\vp/2} W_{b}{}^{c} \prt^{b}\vp 
\prt_{d}\bH_{ac}{}^{d} + \frac{69}{5} e^{-\vp/2} W^{bc} 
\prt_{a}\bphi\prt_{d}\bH_{bc}{}^{d} \nn\\&&\!\!\!\!\!\!\!\!\!\!-  \frac{1}{10} e^{-\vp/2} W^{bc} \prt_{a}\vp \prt_{d}\bH_{bc}{}^{d} + 
\frac{8}{5} e^{-\vp/2} W_{a}{}^{c} \prt^{b}\vp 
\prt_{d}\bH_{bc}{}^{d}- 7 e^{-\vp/2} \prt^{c}W_{a}{}^{b} 
\prt_{d}\bH_{bc}{}^{d} -  \frac{72}{5} e^{3/2 \vp} V_{b}{}^{d} 
V^{bc} \prt_{d}V_{ac} \nn\\&&\!\!\!\!\!\!\!\!\!\!- 4 e^{-\vp/2} W_{bc} W^{bc} 
\prt_{d}V_{a}{}^{d} + 16 e^{3/2 \vp} V_{a}{}^{b} V^{cd} 
\prt_{d}V_{bc} + \frac{119}{5} e^{-\vp/2} W_{a}{}^{b} W^{cd} 
\prt_{d}V_{bc} + \frac{1}{5} e^{3/2 \vp} V_{a}{}^{b} 
V_{b}{}^{c} \prt_{d}V_{c}{}^{d}\nn\\&&\!\!\!\!\!\!\!\!\!\!- 8 e^{-\vp/2} W_{a}{}^{b} 
W_{b}{}^{c} \prt_{d}V_{c}{}^{d} + 2 e^{-\vp/2} V^{bc} 
W_{b}{}^{d} \prt_{d}W_{ac} -  \frac{149}{10} e^{-\vp/2} 
V^{bc} W_{bc} \prt_{d}W_{a}{}^{d} -  \frac{49}{10} e^{-\vp/2} 
V^{bc} W_{a}{}^{d} \prt_{d}W_{bc} \nn\\&&\!\!\!\!\!\!\!\!\!\!+ 32 e^{-\vp/2} V_{a}{}^{b} 
W^{cd} \prt_{d}W_{bc} -  \frac{2}{5} e^{-\vp/2} V^{bc} W_{ab} 
\prt_{d}W_{c}{}^{d}- 24 e^{-\vp/2} V_{a}{}^{b} W_{b}{}^{c} 
\prt_{d}W_{c}{}^{d} -  \frac{81}{10} e^{-\vp/2} W^{bc} 
\prt_{d}\prt_{a}\bH_{bc}{}^{d}\nn\\&&\!\!\!\!\!\!\!\!\!\! -  \frac{44}{5} e^{-\vp/2} 
W^{bc} \prt_{d}\prt_{c}\bH_{ab}{}^{d} + \frac{79}{10} e^{-\vp/2} W^{bc} \prt_{d}\prt^{d}\bH_{abc} + \frac{659}{10} 
e^{-\vp/2} \bH_{abc} W^{bc} \prt_{d}\prt^{d}\bphi\nn\\&&\!\!\!\!\!\!\!\!\!\!-  
\frac{31}{20} e^{-\vp/2} \bH_{abc} W^{bc} 
\prt_{d}\prt^{d}\vp+ \frac{18}{5} e^{\vp/2} 
\bH_{ab}{}^{e} \bH_{cde} \prt^{d}V^{bc} -  \frac{67}{5} e^{-\vp/2} \bH_{bcd} \prt^{b}\bphi\prt^{d}W_{a}{}^{c} \nn\\&&\!\!\!\!\!\!\!\!\!\!+ 
\frac{1}{20} e^{-\vp/2} \bH_{bcd} \prt^{b}\vp 
\prt^{d}W_{a}{}^{c} + \frac{138}{5} e^{-\vp/2} \bH_{acd} 
\prt^{b}\bphi\prt^{d}W_{b}{}^{c} -  \frac{31}{5} e^{-\vp/2} \bH_{acd} \prt^{b}\vp \prt^{d}W_{b}{}^{c} \nn\\&&\!\!\!\!\!\!\!\!\!\!-  
\frac{4}{5} e^{-\vp/2} \bH_{bcd} W^{bc} 
\prt^{d}\prt_{a}\bphi+ \frac{1}{20} e^{-\vp/2} \bH_{bcd} 
W^{bc} \prt^{d}\prt_{a}\vp + \frac{91}{5} e^{-\vp/2} 
\bH_{acd} W^{bc} \prt^{d}\prt_{b}\bphi+ \frac{1}{5} e^{-\vp/2} \bH_{acd} \prt^{d}\prt_{b}W^{bc}\nn\\&&\!\!\!\!\!\!\!\!\!\! -  \frac{47}{20} 
e^{-\vp/2} \bH_{acd} W^{bc} \prt^{d}\prt_{b}\vp -  
\frac{4}{5} e^{\vp/2} \bH_{b}{}^{de} V^{bc} \prt_{e}\bH_{acd} 
+ 3 e^{\vp/2} \bH_{a}{}^{de} V^{bc} \prt_{e}\bH_{bcd} + 
\frac{41}{10} e^{\vp/2} \bH_{b}{}^{cd} V_{a}{}^{b} 
\prt_{e}\bH_{cd}{}^{e} \nn\\&&\!\!\!\!\!\!\!\!\!\!+ \frac{4}{5} e^{\vp/2} \bH_{ab}{}^{d} 
V^{bc} \prt_{e}\bH_{cd}{}^{e}\,.
\eeqa
It  has the  third  derivatives of the base space at various terms. $\Delta g_{a}^{(2)}/c_1^2$    is 
\beqa
&&\!\!\!\!\!\frac{2}{5} e^{\vp/2} \bH_{a}{}^{de} \bH_{bd}{}^{f} \bH_{cef} 
V^{bc} - 2 e^{\vp/2} \bH_{a}{}^{de} \bH_{bc}{}^{f} \bH_{def} V^{bc} 
-  \frac{2}{5} e^{\vp/2} \bH_{ab}{}^{d} \bH_{c}{}^{ef} \bH_{def} 
V^{bc} \nn\\&&\!\!\!\!\!+ \frac{2}{5} e^{-\vp/2} \bH_{ade} V^{bc} W_{b}{}^{d} 
W_{c}{}^{e} - 4 e^{-\vp/2} \bH_{a}{}^{de} \bH_{cde} 
\prt_{b}W^{bc} + 14 e^{-\vp/2} \bH_{b}{}^{de} \bH_{cde} 
W_{a}{}^{c} \prt^{b}\bphi \nn\\&&\!\!\!\!\!- 14 e^{-\vp/2} \bH_{a}{}^{de} 
\bH_{cde} W_{b}{}^{c} \prt^{b}\bphi + \frac{4}{5} e^{\vp/2} 
V_{a}{}^{c} V_{b}{}^{d} W_{cd} \prt^{b}\bphi + 55 e^{-3\vp/2}
W_{a}{}^{c} W_{b}{}^{d} W_{cd} \prt^{b}\bphi \nn\\&&\!\!\!\!\!+ 
\frac{8}{5} e^{-\vp/2} \bH_{ac}{}^{e} \bH_{bde} W^{cd} 
\prt^{b}\bphi + \frac{138}{5} e^{\vp/2} \bH_{bcd} V^{cd} 
\prt_{a}\bphi \prt^{b}\bphi + \frac{367}{5} e^{\vp/2} 
\bH_{acd} V^{cd} \prt_{b}\bphi \prt^{b}\bphi \nn\\&&\!\!\!\!\!+ \frac{51}{5} 
e^{\vp/2} \bH_{acd} V^{cd} \prt_{b}\vp \prt^{b}\bphi -  
\frac{19}{5} e^{-\vp/2} \bH_{b}{}^{de} \bH_{cde} W_{a}{}^{c} 
\prt^{b}\vp + 8 e^{-\vp/2} \bH_{a}{}^{de} \bH_{cde} W_{b}{}^{c} 
\prt^{b}\vp \nn\\&&\!\!\!\!\!-  \frac{12}{5} e^{\vp/2} V_{a}{}^{c} 
V_{b}{}^{d} W_{cd} \prt^{b}\vp -  \frac{63}{4 }e^{-3\vp/2} W_{a}{}^{c} 
W_{b}{}^{d} W_{cd} \prt^{b}\vp + \frac{4}{5} 
e^{-\vp/2} \bH_{ac}{}^{e} \bH_{bde} W^{cd} \prt^{b}\vp \nn\\&&\!\!\!\!\!-  
\frac{27}{2} e^{\vp/2} \bH_{bcd} V^{cd} \prt_{a}\bphi 
\prt^{b}\vp + \frac{17}{8} e^{\vp/2} \bH_{bcd} V^{cd} 
\prt_{a}\vp \prt^{b}\vp -  \frac{27}{20} e^{\vp/2} 
\bH_{acd} V^{cd} \prt_{b}\vp \prt^{b}\vp \nn\\&&\!\!\!\!\!+ \frac{9}{10} 
e^{-\vp/2} \bH_{a}{}^{de} W^{bc} \prt_{c}\bH_{bde} - 2 e^{\vp/2} V^{bc} W_{b}{}^{d} \prt_{c}V_{ad} + \frac{64}{5} e^{\vp/2} \bH_{abd} \prt^{b}\bphi \prt_{c}V^{cd}\nn\\&&\!\!\!\!\! - 8 e^{\vp/2} 
\bH_{abd} \prt^{b}\vp \prt_{c}V^{cd} + 66 e^{-\vp/2} 
\prt_{b}\prt^{b}\bphi \prt_{c}W_{a}{}^{c} + \frac{15}{4} 
e^{-\vp/2} \prt_{b}\prt^{b}\vp \prt_{c}W_{a}{}^{c} - 68 
e^{-\vp/2} \prt_{b}\bphi \prt^{b}\bphi 
\prt_{c}W_{a}{}^{c} \nn\\&&\!\!\!\!\!+ \frac{63}{10} e^{-\vp/2} 
\prt_{b}\vp \prt^{b}\bphi \prt_{c}W_{a}{}^{c} - 4 e^{-\vp/2} \prt_{b}\vp \prt^{b}\vp \prt_{c}W_{a}{}^{c} - 27 
e^{-\vp/2} \prt_{a}\vp \prt^{b}\bphi 
\prt_{c}W_{b}{}^{c} + 27 e^{-\vp/2} \prt_{a}\bphi 
\prt^{b}\vp \prt_{c}W_{b}{}^{c} \nn\\&&\!\!\!\!\!+ \frac{39}{10} e^{-\vp/2} \prt_{a}\vp \prt^{b}\vp \prt_{c}W_{b}{}^{c} + 
\frac{49}{5} e^{-\vp/2} \prt^{b}\prt_{a}\bphi 
\prt_{c}W_{b}{}^{c} + \frac{231}{20} e^{-\vp/2} 
\prt^{b}\prt_{a}\vp \prt_{c}W_{b}{}^{c} + 112 e^{-\vp/2} W_{b}{}^{c} \prt^{b}\bphi \prt_{c}\prt_{a}\bphi\nn\\&&\!\!\!\!\! -  
\frac{353}{10} e^{-\vp/2} W_{b}{}^{c} \prt^{b}\vp 
\prt_{c}\prt_{a}\bphi -  \frac{283}{10} e^{-\vp/2} 
W_{b}{}^{c} \prt^{b}\bphi \prt_{c}\prt_{a}\vp + 
\frac{39}{5} e^{-\vp/2} W_{b}{}^{c} \prt^{b}\vp \prt_{c}
\prt_{a}\vp \nn\\&&\!\!\!\!\!+ \frac{944}{5} e^{-\vp/2} W_{a}{}^{c} 
\prt^{b}\bphi \prt_{c}\prt_{b}\bphi + \frac{52}{5} 
e^{-\vp/2} W_{a}{}^{c} \prt^{b}\vp 
\prt_{c}\prt_{b}\bphi + \frac{67}{5} e^{-\vp/2} 
\prt^{b}\bphi \prt_{c}\prt_{b}W_{a}{}^{c} + 
\frac{231}{20} e^{-\vp/2} \prt^{b}\vp 
\prt_{c}\prt_{b}W_{a}{}^{c} \nn\\&&\!\!\!\!\!+ \frac{57}{2} e^{-\vp/2} 
W_{a}{}^{c} \prt^{b}\bphi \prt_{c}\prt_{b}\vp -  
\frac{42}{5} e^{-\vp/2} W_{a}{}^{c} \prt^{b}\vp \prt_{c}
\prt_{b}\vp + 306 e^{-\vp/2} W_{ab} \prt^{b}\bphi 
\prt_{c}\prt^{c}\bphi \nn\\&&\!\!\!\!\!-  \frac{111}{2} e^{-\vp/2} W_{ab} 
\prt^{b}\vp \prt_{c}\prt^{c}\bphi -  \frac{207}{5} 
e^{-\vp/2} \prt^{b}\bphi \prt_{c}\prt^{c}W_{ab} -  
\frac{233}{20} e^{-\vp/2} \prt^{b}\vp 
\prt_{c}\prt^{c}W_{ab}\nn\\&&\!\!\!\!\! + \frac{67}{20} e^{-\vp/2} W_{ab} 
\prt^{b}\bphi \prt_{c}\prt^{c}\vp -  \frac{321}{40} 
e^{-\vp/2} W_{ab} \prt^{b}\vp \prt_{c}\prt^{c}\vp - 
75 e^{-\vp/2} W_{a}{}^{b} \prt_{c}\prt^{c}\prt_{b}\bphi 
-  \frac{7}{20} e^{-\vp/2} W_{a}{}^{b} 
\prt_{c}\prt^{c}\prt_{b}\vp\nn\\&&\!\!\!\!\! -  \frac{126}{5} e^{\vp/2} \bH_{acd} V_{b}{}^{d} \prt^{b}\bphi \prt^{c}\bphi -  
\frac{1948}{5} e^{-\vp/2} W_{ac} \prt_{b}\bphi 
\prt^{b}\bphi \prt^{c}\bphi -  \frac{343}{10} e^{-\vp/2} 
W_{ac} \prt_{b}\vp \prt^{b}\bphi \prt^{c}\bphi \nn\\&&\!\!\!\!\!+ 56 
e^{-\vp/2} \prt^{b}\bphi \prt_{c}W_{ab} \prt^{c}\bphi + 
\frac{38}{5} e^{-\vp/2} \bH_{b}{}^{de} \bH_{cde} 
\prt^{c}W_{a}{}^{b} -  \frac{16}{5} e^{\vp/2} \bH_{bcd} 
V_{a}{}^{d} \prt^{b}\bphi \prt^{c}\vp \nn\\&&\!\!\!\!\!+ 16 e^{\vp/2} 
\bH_{acd} V_{b}{}^{d} \prt^{b}\bphi \prt^{c}\vp + 
\frac{29}{10} e^{\vp/2} \bH_{abd} V_{c}{}^{d} \prt^{b}\bphi 
\prt^{c}\vp + 54 e^{-\vp/2} W_{bc} \prt_{a}\bphi 
\prt^{b}\bphi \prt^{c}\vp \nn\\&&\!\!\!\!\!+ \frac{129}{10} e^{-\vp/2} 
W_{bc} \prt_{a}\vp \prt^{b}\bphi \prt^{c}\vp + 
\frac{277}{5} e^{-\vp/2} W_{ac} \prt_{b}\bphi 
\prt^{b}\bphi \prt^{c}\vp -  \frac{1}{10} e^{-\vp/2} 
\prt_{b}W_{ac} \prt^{b}\bphi \prt^{c}\vp \nn\\&&\!\!\!\!\!+ 
\frac{161}{20} e^{-\vp/2} W_{ac} \prt_{b}\vp 
\prt^{b}\bphi \prt^{c}\vp -  \frac{23}{5} e^{\vp/2} 
\bH_{acd} V_{b}{}^{d} \prt^{b}\vp \prt^{c}\vp + 
\frac{143}{40} e^{-\vp/2} W_{ac} \prt_{b}\vp 
\prt^{b}\vp \prt^{c}\vp \nn\\&&\!\!\!\!\!+ \frac{283}{10} e^{-\vp/2} 
\prt^{b}\bphi \prt_{c}W_{ab} \prt^{c}\vp -  
\frac{139}{10} e^{-\vp/2} W_{ab} \prt^{b}\bphi 
\prt_{c}\vp \prt^{c}\vp -  \frac{318}{5} e^{-\vp/2} 
\prt_{c}W_{ab} \prt^{c}\prt^{b}\bphi\nn\\&&\!\!\!\!\! -  \frac{1}{10} 
e^{-\vp/2} \prt_{c}W_{ab} \prt^{c}\prt^{b}\vp -  
\frac{211}{5} e^{\vp/2} V^{cd} \prt^{b}\bphi 
\prt_{d}\bH_{abc} + \frac{159}{20} e^{\vp/2} V^{cd} 
\prt^{b}\vp \prt_{d}\bH_{abc} + \frac{32}{5} e^{\vp/2} 
\prt_{b}V^{bc} \prt_{d}\bH_{ac}{}^{d} \nn\\&&\!\!\!\!\!-  \frac{63}{5} e^{\vp/2} V_{b}{}^{c} \prt^{b}\bphi \prt_{d}\bH_{ac}{}^{d} + 
\frac{29}{20} e^{\vp/2} V_{b}{}^{c} \prt^{b}\vp 
\prt_{d}\bH_{ac}{}^{d} -  \frac{69}{5} e^{\vp/2} V^{bc} 
\prt_{a}\bphi \prt_{d}\bH_{bc}{}^{d} -  \frac{1}{10} e^{\vp/2} V^{bc} \prt_{a}\vp \prt_{d}\bH_{bc}{}^{d} \nn\\&&\!\!\!\!\!+ 
\frac{8}{5} e^{\vp/2} V_{a}{}^{c} \prt^{b}\vp 
\prt_{d}\bH_{bc}{}^{d} + 7 e^{\vp/2} \prt^{c}V_{a}{}^{b} 
\prt_{d}\bH_{bc}{}^{d} + \frac{109}{10} e^{\vp/2} V^{bc} 
W_{bc} \prt_{d}V_{a}{}^{d} -  \frac{49}{5} e^{\vp/2} 
V_{a}{}^{b} W^{cd} \prt_{d}V_{bc} \nn\\&&\!\!\!\!\!+ 16 e^{\vp/2} V^{bc} W_{ab} 
\prt_{d}V_{c}{}^{d} + \frac{2}{5} e^{\vp/2} V_{a}{}^{b} 
W_{b}{}^{c} \prt_{d}V_{c}{}^{d} + \frac{72}{5} e^{-3\vp/2} W_{b}{}^{d} W^{bc} 
\prt_{d}W_{ac} -  \frac{119}{5} e^{\vp/2} 
V_{a}{}^{b} V^{cd} \prt_{d}W_{bc}\nn\\&&\!\!\!\!\! -  \frac{1}{5}e^{-3\vp/2}W_{a}{}^{b} 
W_{b}{}^{c} \prt_{d}W_{c}{}^{d} + 
\frac{41}{10} e^{\vp/2} V^{bc} 
\prt_{d}\prt_{a}\bH_{bc}{}^{d} + \frac{4}{5} e^{\vp/2} 
V^{bc} \prt_{d}\prt_{c}\bH_{ab}{}^{d} -  \frac{39}{10} e^{\vp/2} V^{bc} \prt_{d}\prt^{d}\bH_{abc} \nn\\&&\!\!\!\!\!-  \frac{659}{10} 
e^{\vp/2} \bH_{abc} V^{bc} \prt_{d}\prt^{d}\bphi -  
\frac{31}{20} e^{\vp/2} \bH_{abc} V^{bc} 
\prt_{d}\prt^{d}\vp + \frac{67}{5} e^{\vp/2} \bH_{bcd} 
\prt^{b}\bphi \prt^{d}V_{a}{}^{c} -  \frac{159}{20} e^{\vp/2} \bH_{bcd} \prt^{b}\vp \prt^{d}V_{a}{}^{c} \nn\\&&\!\!\!\!\! -  
\frac{138}{5} e^{\vp/2} \bH_{acd} \prt^{b}\bphi 
\prt^{d}V_{b}{}^{c} + \frac{49}{5} e^{\vp/2} \bH_{acd} 
\prt^{b}\vp \prt^{d}V_{b}{}^{c} + \frac{22}{5} e^{-\vp/2} \bH_{ab}{}^{e} \bH_{cde} \prt^{d}W^{bc} + \frac{4}{5} 
e^{\vp/2} \bH_{bcd} V^{bc} \prt^{d}\prt_{a}\bphi  \nn\\&&\!\!\!\!\!+ 
\frac{1}{20} e^{\vp/2} \bH_{bcd} V^{bc} 
\prt^{d}\prt_{a}\vp -  \frac{91}{5} e^{\vp/2} \bH_{acd} 
V^{bc} \prt^{d}\prt_{b}\bphi -  \frac{1}{5} e^{\vp/2} 
\bH_{acd} \prt^{d}\prt_{b}V^{bc} + \frac{113}{20} e^{\vp/2} \bH_{acd} V^{bc} \prt^{d}\prt_{b}\vp  \nn\\&&\!\!\!\!\!+ \frac{4}{5} 
e^{-\vp/2} \bH_{b}{}^{de} W^{bc} \prt_{e}\bH_{acd} + 
\frac{19}{10} e^{-\vp/2} \bH_{a}{}^{de} W^{bc} 
\prt_{e}\bH_{bcd} -  \frac{41}{10} e^{-\vp/2} \bH_{b}{}^{cd} 
W_{a}{}^{b} \prt_{e}\bH_{cd}{}^{e} \nn\\&&\!\!\!\!\! -  \frac{4}{5} e^{-\vp/2} 
\bH_{ab}{}^{d} W^{bc} \prt_{e}\bH_{cd}{}^{e}\,.
\eeqa
The above expression also  involves various terms with third derivatives of the base space. To simplify the expression for   $\tilde{B}_{ab}^{(2)}$, we introduce an antisymmetric tensor $A^{ab}$ and multiply it with $\tilde{B}_{ab}^{(2)}$. This allows us to write $\tilde{B}_{ab}^{(2)}$ in a simpler form.  $A^{ab}\tilde{B}_{ab}^{(2)}/c_1^2$   is 
\beqa
&&\!\!\!\!\!\frac{36}{5}  A^{ab} \bH_{c}{}^{ef} \bH_{def} V_{a}{}^{c} 
W_{b}{}^{d} -  \frac{48}{5}  A^{ab} \bH_{bd}{}^{f} \bH_{cef} 
V_{a}{}^{c} W^{de} + \frac{24}{5}  A^{ab} \bH_{bc}{}^{f} \bH_{def} 
V_{a}{}^{c} W^{de} \nn\\&&\!\!\!\!\!-  \frac{48}{5}  e^{\vp} A^{bc} \bH_{cde} 
V_{a}{}^{d} V_{b}{}^{e} \prt^{a}\bphi -  \frac{12}{5}  
e^{\vp} A^{bc} \bH_{ace} V_{b}{}^{d} V_{d}{}^{e} \prt^{a}\bphi + 
\frac{48}{5}  e^{- \vp} A^{bc} \bH_{cde} W_{a}{}^{d} 
W_{b}{}^{e} \prt^{a}\bphi \nn\\&&\!\!\!\!\!+ \frac{12}{5}  e^{- \vp} A^{bc} 
\bH_{ace} W_{b}{}^{d} W_{d}{}^{e} \prt^{a}\bphi -  
\frac{2304}{5}  A^{bc} V_{b}{}^{d} W_{cd} \prt_{a}\bphi 
\prt^{a}\bphi + \frac{36}{5}  A^{bc} V_{b}{}^{d} 
\prt_{a}W_{cd} \prt^{a}\bphi \nn\\&&\!\!\!\!\!+ \frac{24}{5}  e^{\vp} 
A^{bc} \bH_{cde} V_{a}{}^{d} V_{b}{}^{e} \prt^{a}\vp -  
\frac{24}{5}  e^{\vp} A^{bc} \bH_{ade} V_{b}{}^{d} V_{c}{}^{e} 
\prt^{a}\vp + 6  e^{\vp} A^{bc} \bH_{ace} V_{b}{}^{d} 
V_{d}{}^{e} \prt^{a}\vp \nn\\&&\!\!\!\!\!+ \frac{24}{5}  e^{- \vp} A^{bc} 
\bH_{cde} W_{a}{}^{d} W_{b}{}^{e} \prt^{a}\vp -  \frac{24}{5}  
e^{- \vp} A^{bc} \bH_{ade} W_{b}{}^{d} W_{c}{}^{e} 
\prt^{a}\vp + 6  e^{- \vp} A^{bc} \bH_{ace} W_{b}{}^{d} 
W_{d}{}^{e} \prt^{a}\vp \nn\\&&\!\!\!\!\!- 24  e^{- \vp} A^{bc} \bH_{cde} 
W_{ab} W^{de} \prt^{a}\vp + \frac{432}{5}  A^{bc} V_{b}{}^{d} 
\prt_{a}W_{cd} \prt^{a}\vp -  \frac{96}{5}  A^{bc} 
V_{b}{}^{d} W_{cd} \prt_{a}\vp \prt^{a}\vp \nn\\&&\!\!\!\!\!+ \frac{12}{5} 
 A^{ab} \bH_{a}{}^{cd} \bH_{c}{}^{ef} \prt_{b}\bH_{def} -  
\frac{96}{5}  A^{cd} V_{ac} W_{bd} \prt^{a}\bphi 
\prt^{b}\bphi -  \frac{96}{5}  A_{b}{}^{c} V_{c}{}^{d} W_{ad} 
\prt^{a}\bphi \prt^{b}\vp -  \frac{96}{5}  A_{b}{}^{c} 
V_{a}{}^{d} W_{cd} \prt^{a}\bphi \prt^{b}\vp  \nn\\&&\!\!\!\!\!+ 
\frac{57}{10}  A^{cd} \prt_{a}\bH_{bcd} \prt^{a}\bphi 
\prt^{b}\vp-  \frac{72}{5}  A^{cd} V_{ac} W_{bd} \prt^{a}
\vp \prt^{b}\vp + \frac{12}{5}  A_{a}{}^{c} V_{c}{}^{d} 
W_{bd} \prt^{a}\vp \prt^{b}\vp -  \frac{12}{5}  
A_{a}{}^{c} V_{b}{}^{d} W_{cd} \prt^{a}\vp \prt^{b}\vp  \nn\\&&\!\!\!\!\!- 72 
 A^{cd} \prt^{a}\bphi \prt_{b}\bH_{acd} \prt^{b}\vp - 24 
 e^{- \vp} A^{ab} W_{a}{}^{c} W^{de} \prt_{c}\bH_{bde} + 
\frac{36}{5}  e^{\vp} A^{ab} \bH_{bde} V^{cd} 
\prt_{c}V_{a}{}^{e} \nn\\&&\!\!\!\!\! -  \frac{36}{5}  e^{- \vp} A^{ab} 
\bH_{bde} W^{cd} \prt_{c}W_{a}{}^{e} + \frac{663}{40}  A^{ab} 
\prt_{c}\prt^{c}\vp \prt_{d}\bH_{ab}{}^{d} -  
\frac{663}{20}  A^{bc} \prt_{a}\vp \prt^{a}\bphi 
\prt_{d}\bH_{bc}{}^{d} \nn\\&&\!\!\!\!\! -  \frac{414}{5}  A^{ab} 
\prt^{c}\prt_{a}\bphi \prt_{d}\bH_{bc}{}^{d} + 
\frac{651}{10}  A^{ab} \prt^{c}\prt_{a}\vp 
\prt_{d}\bH_{bc}{}^{d} -  \frac{12}{5}  A^{ab} \bH_{a}{}^{cd} 
\bH_{c}{}^{ef} \prt_{d}\bH_{bef} + \frac{468}{5}  A^{bc} W_{ab} 
\prt^{a}\bphi \prt_{d}V_{c}{}^{d} \nn\\&&\!\!\!\!\! - 72  A^{bc} W_{ab} 
\prt^{a}\vp \prt_{d}V_{c}{}^{d} -  \frac{48}{5}  
A_{a}{}^{b} W_{b}{}^{c} \prt^{a}\vp \prt_{d}V_{c}{}^{d} -  
\frac{24}{5}  e^{\vp} A^{ab} \bH_{bce} V_{a}{}^{c} 
\prt_{d}V^{de} -  \frac{6}{5}  A^{ab} \prt_{c}V^{cd} 
\prt_{d}W_{ab}  \nn\\&&\!\!\!\!\!-  \frac{792}{5}  A^{bc} V_{b}{}^{d} 
\prt^{a}\bphi \prt_{d}W_{ac} + \frac{672}{5}  A^{bc} 
V_{b}{}^{d} \prt^{a}\vp \prt_{d}W_{ac} + \frac{414}{5}  
A^{bc} V_{bc} \prt^{a}\bphi \prt_{d}W_{a}{}^{d} + 
\frac{432}{5}  A^{bc} V_{a}{}^{d} \prt^{a}\bphi 
\prt_{d}W_{bc} \nn\\&&\!\!\!\!\! -  \frac{336}{5}  A^{bc} V_{a}{}^{d} \prt^{a}
\vp \prt_{d}W_{bc} - 6  A^{ab} \prt_{c}V_{a}{}^{c} 
\prt_{d}W_{b}{}^{d} + 72  A^{bc} V_{ab} \prt^{a}\bphi 
\prt_{d}W_{c}{}^{d} - 72  A^{bc} V_{ab} \prt^{a}\vp 
\prt_{d}W_{c}{}^{d} \nn\\&&\!\!\!\!\! -  \frac{48}{5}  A_{a}{}^{b} V_{b}{}^{c} 
\prt^{a}\vp \prt_{d}W_{c}{}^{d} -  \frac{12}{5}  A^{ab} 
\prt_{b}V_{a}{}^{c} \prt_{d}W_{c}{}^{d} + \frac{24}{5}  
e^{- \vp} A^{ab} \bH_{bce} W_{a}{}^{c} \prt_{d}W^{de} + 
\frac{207}{5}  A^{bc} \prt^{a}\bphi 
\prt_{d}\prt_{a}\bH_{bc}{}^{d}  \nn\\&&\!\!\!\!\!-  \frac{36}{5}  A^{ab} 
V_{a}{}^{c} W_{c}{}^{d} \prt_{d}\prt_{b}\bphi + 24  A^{ab} 
V^{cd} W_{ac} \prt_{d}\prt_{b}\vp -  \frac{312}{5}  A^{ab} 
V_{a}{}^{c} W_{c}{}^{d} \prt_{d}\prt_{b}\vp + \frac{444}{5} 
 A^{ab} V_{a}{}^{c} W_{b}{}^{d} \prt_{d}\prt_{c}\bphi  \nn\\&&\!\!\!\!\!- 48  
A^{ab} V_{a}{}^{c} W_{b}{}^{d} \prt_{d}\prt_{c}\vp -  
\frac{207}{5}  A^{bc} \prt^{a}\bphi 
\prt_{d}\prt^{d}\bH_{abc} + \frac{663}{20}  A^{bc} 
\prt^{a}\vp \prt_{d}\prt^{d}\bH_{abc} + 384  A^{ab} 
V_{a}{}^{c} W_{bc} \prt_{d}\prt^{d}\bphi  \nn\\&&\!\!\!\!\!+ 42  A^{ab} 
W_{a}{}^{c} \prt_{d}\prt^{d}V_{bc} - 42  A^{ab} V_{a}{}^{c} 
\prt_{d}\prt^{d}W_{bc} -  \frac{57}{5}  A^{bc} \bH_{acd} 
\prt^{a}\vp \prt^{d}\prt_{b}\bphi + \frac{69}{5}  
A^{bc} \bH_{acd} \prt^{a}\bphi \prt^{d}\prt_{b}\vp  \nn\\&&\!\!\!\!\!+ 
\frac{3}{5}  A^{ab} \prt_{d}\bH_{abc} 
\prt^{d}\prt^{c}\vp -  \frac{9}{5}  e^{\vp} A^{ab} 
V_{c}{}^{e} V^{cd} \prt_{e}\bH_{abd} + \frac{9}{5}  e^{- 
\vp} A^{ab} W_{c}{}^{e} W^{cd} \prt_{e}\bH_{abd} -  
\frac{48}{5}  e^{\vp} A^{ab} V_{a}{}^{c} V^{de} 
\prt_{e}\bH_{bcd}  \nn\\&&\!\!\!\!\!+ \frac{48}{5}  e^{- \vp} A^{ab} 
W_{a}{}^{c} W^{de} \prt_{e}\bH_{bcd} + \frac{6}{5}  e^{\vp} 
A^{ab} V_{a}{}^{c} V_{c}{}^{d} \prt_{e}\bH_{bd}{}^{e} -  
\frac{6}{5}  e^{- \vp} A^{ab} W_{a}{}^{c} W_{c}{}^{d} 
\prt_{e}\bH_{bd}{}^{e}  \nn\\&&\!\!\!\!\!+ \frac{6}{5}  e^{\vp} A^{ab} \bH_{bde} 
V^{cd} \prt^{e}V_{ac} - 6  e^{\vp} A^{ab} \bH_{bde} V_{a}{}^{c} 
\prt^{e}V_{c}{}^{d} - 12  e^{- \vp} A^{ab} \bH_{cde} W^{cd} 
\prt^{e}W_{ab} \nn\\&&\!\!\!\!\! -  \frac{6}{5}  e^{- \vp} A^{ab} \bH_{bde} 
W^{cd} \prt^{e}W_{ac} + 54  e^{- \vp} A^{ab} \bH_{bde} 
W_{a}{}^{c} \prt^{e}W_{c}{}^{d} -  \frac{24}{5}  A^{ab} 
\bH_{a}{}^{cd} \bH_{c}{}^{ef} \prt_{f}\bH_{bde}\,.
\eeqa
The above corrections at order $\alpha'^2$ are required if one would like to study the effective action at order $\alpha'^3$ for closed spacetime manifolds.

\end{document}